\documentclass{article}

\usepackage{jcappub}
\usepackage{amsmath}
\usepackage{amssymb}
\usepackage{physics}
\usepackage{placeins}
\usepackage[caption=false]{subfig}
\usepackage{tikz}
\usepackage{dcolumn}
\usepackage[capitalize]{cleveref}

\usetikzlibrary{patterns, arrows,shadows,positioning}
\tikzset{rectangle/.append style={draw=black, thick, fill=white, drop shadow}}
\DeclareMathOperator{\e}{e}
\DeclareMathOperator{\smoothstep}{SStep}
\DeclareMathOperator{\smoothpl}{SPL}
\DeclareMathOperator{\cosi}{Ci}
\DeclareMathOperator{\sini}{Si}

\newcommand{\Refa}[1]{Ref.~{\cite{#1}}}
\newcommand{\Refs}[1]{Refs.~{\cite{#1}}}

\newcommand{\hypergauss}[4]{\phantom{}_{_2}\mathrm{F}\!_{_1}\!\left(#1,#2;#3;#4\right)}

\newcommand{\tdec}{R} 
\newcommand{\tgauss}{D_{\mathrm{A}+}} 

\newcommand{\vrms}{\overline{v}}
\newcommand{\vrmsst}{\overline{v}_*}
\newcommand{\vsweep}{v_\mathrm{sw}}
\newcommand{\vdc}{v_\mathrm{dc}}

\newcommand{\Psdv}{P_v} 
\newcommand{\Psv}{\mathcal{P}_v} 
\newcommand{\UETCv}{P_v} 

\newcommand{\side}{\mathrm{s}}
\newcommand{\vonK}{von K\'arm\'an}

\newcommand{\tdevel}{\tau_*} 
\newcommand{\tend}{\tau_\mathrm{end}} 
\newcommand{\tfin}{\tau_\mathrm{fin}} 
\newcommand{\tini}{\tau_\mathrm{ini}} 
\newcommand{\tgro}{\tau_\mathrm{gro}} 
\newcommand{\tuetc}{\tau_\mathrm{ref}} 
\newcommand{\deltuetc}{\Delta \tau_\mathrm{UETC}} 

\newcommand{\tshock}{\tau_\mathrm{sh}}
\newcommand{\tmid}{\tau_\mathrm{mid}}
\newcommand{\tdiff}{\tau_\mathrm{diff}}
\newcommand{\tauxist}{\tau_{\xi_*}}

\newcommand{\ndecay}{\mathcal{N}_\mathrm{e}}
\newcommand{\ncut}{\mathcal{N}_\mathrm{cut}}

\usepackage{xspace}

\makeatletter
\DeclareRobustCommand\onedot{\futurelet\@let@token\@onedot}
\def\@onedot{\ifx\@let@token.\else.\null\fi\xspace}

\def\eg{\emph{e.g}\onedot} 
\def\ie{\emph{i.e}\onedot} 
\makeatother

\title{Generation of gravitational waves from freely decaying turbulence}
\subheader{HIP-2021-35/TH}

\author[a]{Pierre Auclair}
\author[b,c]{Chiara Caprini}
\author[d]{Daniel Cutting}
\author[d,e]{Mark Hindmarsh}
\author[d]{Kari Rummukainen}
\author[f]{Dani\`ele A.~Steer}
\author[d]{David J.~Weir}

\affiliation[a]{Cosmology, Universe and Relativity at Louvain, Institute of Mathematics and Physics, Louvain University, 2 Chemin du Cyclotron, 1348 Louvain-la-Neuve, Belgium}
\affiliation[b]{D\'epartement de Physique Th\'eorique, Université de Gen\`eve, CH-1211 Gen\`eve, Switzerland}
\affiliation[c]{Theoretical Physics Department, CERN, CH-1211 Gen\`eve, Switzerland}
\affiliation[d]{Department  of  Physics  and  Helsinki  Institute  of  Physics,  PL  64,  FI-00014  University  of  Helsinki,  Finland}
\affiliation[e]{Department  of  Physics  and  Astronomy,  University  of  Sussex,  Falmer,  Brighton  BN1  9QH,  U.K}
\affiliation[f]{Universit\'e Paris Cit\'e, CNRS, Astroparticule et Cosmologie, F-75013 Paris, France}

\emailAdd{pierre.auclair@uclouvain.be}
\emailAdd{chiara.caprini@cern.ch}
\emailAdd{daniel.cutting@helsinki.fi}
\emailAdd{mark.hindmarsh@helsinki.fi}
\emailAdd{kari.rummukainen@helsinki.fi}
\emailAdd{steer@apc.univ-paris7.fr}
\emailAdd{david.weir@helsinki.fi}

\abstract{
	We study the stochastic gravitational wave background (SGWB) produced by freely decaying vortical turbulence in the early Universe. We thoroughly investigate the time correlation of the velocity field, and hence of the anisotropic stresses producing the gravitational waves. With hydrodynamical simulations, we show that the unequal time correlation function (UETC) of the Fourier components of the velocity field is Gaussian in the time difference, as predicted by the "sweeping" decorrelation model. We introduce a decorrelation model that can be extended to wavelengths around the integral scale of the flow. Supplemented with the evolution laws of the kinetic energy and of the integral scale, this provides a new model UETC of the turbulent velocity field consistent with the simulations. We discuss the UETC as a positive definite kernel, and propose to use the Gibbs kernel for the velocity UETC as a natural way to ensure positive definiteness of the SGWB. The SGWB is given by a 4-dimensional integration of the resulting anisotropic stress UETC with the gravitational wave Green's function. We perform this integration using a Monte Carlo algorithm based on importance sampling, and find that the result matches that of the simulations. Furthermore, the SGWB obtained from the numerical integration and from the simulations show close agreement with a model in which the source is constant in time and abruptly turns off after a few eddy turnover times. Based on this assumption, we provide an approximate analytical form for the SGWB spectrum and its scaling with the initial kinetic energy and integral scale. Finally, we use our model and numerical integration algorithm to show that including an initial growth phase for the turbulent flow heavily influences the spectral shape of the SGWB. This highlights the importance of a complete understanding of the turbulence generation mechanism.
}

\begin{document}

\maketitle

\section{Introduction}
\label{sec:orgd6ac989}

Gravitational wave (GW) signals from the early Universe have the
potential to provide a new observational perspective
on high energy physics phenomena. In this context, first order cosmological phase transitions
provide a compelling source of GWs. This was first proposed many years
ago (see
\Refs{Witten:1984rs,hogan,Turner:1990rc,Kosowsky:1991ua,Kosowsky:1992vn,Kamionkowski:1993fg}),
before it became clear that the electroweak (EW) symmetry breaking
proceeds as a crossover in the Standard Model
\cite{Kajantie:1996mn}. However, it has since emerged that many
scenarios beyond the Standard Model (BSM) lead to first order phase
transitions at -- and beyond -- the EW scale, reopening the case for the
study of GW production from first order phase transitions (for a GW-oriented review, see Ref.~\cite{Caprini:2019egz}).

This is particularly interesting in the context of the Laser
Interferometer Space Antenna (LISA), which is sensitive to a frequency window
in the millihertz range~\cite{Audley:2017drz}. In the context of primordial GW-sourcing processes
that are localized in time, such as a first order phase transition, the characteristic frequency of the GWs today can be connected to the characteristic time/length scale of the source anisotropic stresses $L_*$, via
\begin{equation}
	f\simeq 16.5 \cdot 10^{-3} \,\mathrm{mHz} \,\frac{1}{L_*\mathcal{H}_*} \frac{T_*}{100\,\mathrm{GeV}} \left(\frac{g_*}{100}\right)^{1/6}.
\end{equation}
Here a subscript $*$ denotes the epoch at which the phase transition occurs.
This shows that LISA can potentially detect the GW signal from
transitions in the window 100~GeV -- 1~TeV, if the
characteristic time/length scale of the anisotropic stresses is of the
order $L_*\sim 10^{-2}$ to $10^{-3}$ of the Hubble scale $\mathcal{H}_*^{-1}$ at
the time of the phase transition. These can be considered as typical values for $L_*\mathcal{H}_*$, given that $L_*$ is related to the mean bubble separation $R_*$ (see \eg~\cite{Caprini:2015zlo,Caprini:2019egz} and references therein).
LISA will therefore offer a new way to probe BSM physics, complementary to the Large Hadron Collider (see \eg~\Refa{Caldwell:2022qsj}).

There are several processes possibly leading to sizable anisotropic
stresses in connection with a first order phase transition. This
rich phenomenology renders phase transitions particularly appealing as primordial GW sources. Bubble percolation, with the consequent breaking of spherical symmetry, is the most direct one~\cite{Turner:1990rc,Kosowsky:1991ua,Kosowsky:1992vn}.
The GW generation by bubble collision has been analysed both with simulations~\cite{Huber:2008hg,Weir:2016tov,Konstandin:2017sat,Cutting:2018tjt,Cutting:2020nla,Lewicki:2020jiv,Lewicki:2020azd,Gould:2021dpm} and analytical approaches~\cite{Caprini:2007xq,Jinno:2016vai,Jinno:2017fby}.

The first 3-dimensional (3D) simulations of the coupled system of a scalar field and a relativistic fluid have shown that sound waves, produced in the fluid by expanding bubbles, are also a promising source of GWs \cite{Hindmarsh:2013xza,Hindmarsh:2015qta,Hindmarsh:2017gnf}.
These  simulations have been performed using the code SCOTTS (Simulations of COsmic Thermal TransitionS) on which we will also rely in the present work.
Analytical insight into the  generation of GWs by sound waves has been developed in \Refs{Hindmarsh:2016lnk,Hindmarsh:2019phv}, and a new simulation technique for computing the GW spectra in the sound shell limit has been proposed in \Refa{Jinno:2020eqg}.

Refs.~\cite{Hindmarsh:2013xza,Hindmarsh:2015qta,Hindmarsh:2017gnf} performed simulations of first order phase transitions of weak to intermediate strength, \eg~$\alpha\lesssim 0.1$, where $\alpha$ is the ratio of the trace anomaly of the energy momentum tensor and the thermal energy. It was found that the bulk fluid motion consisted primarily of compressional modes, which correspond to sound waves. In this case, sound waves are expected to be the dominant GW source. However, shocks will develop in the fluid and are expected to convert the acoustic phase into a turbulent one \cite{Pen:2015qta,Dahl:2021wyk}.
The characteristic time of shock formation can be estimated as $\tshock\sim R_*/\vrms_{\parallel}$, where $\vrms_{\parallel}$ denotes the root-mean-squared (rms) velocity of the acoustic motion. For stronger transitions, in which the initial value of $\vrms_{\parallel}$ is typically larger, the onset of turbulence is expected to occur earlier and give a more substantial contribution to the total GW signal. Furthermore, recent simulations of transitions with $\alpha\sim 0.5$ found that, if the bubble wall velocity is subsonic, flows with substantial vorticity are already generated during the transition itself when fluid shells and bubble walls collide~\cite{Cutting:2019zws}.

In the present work, we study the GW signal generated by a
hypothetical turbulent phase in the aftermath of a first order
phase transition. The first analyses of the GW signal from turbulence have relied on analytical modelling of the turbulent flow, and semi-analytical estimates of the GW signal~\cite{Kamionkowski:1993fg,Caprini:2006jb,Gogoberidze:2007an,Kahniashvili:2008pe,Kahniashvili:2008er,Caprini:2009yp,Caprini:2009pr}.
In particular, Ref.~\cite{Niksa:2018ofa}  evaluates the GW signal from all components (compressional, vortical, magnetic field) of both standard, and helical, freely-decaying magneto-hydrodynamic (MHD) turbulence. Various spectral shapes and scalings of the GW power spectrum per logarithmic wave number interval have been found, depending on the relative amplitude of the compressional and vortical components, and on the presence of helicity.
In \Refa{Niksa:2018ofa}, it was also argued that the correct auto-correlation time to be used in the equations describing GW production by turbulence is the Eulerian eddy turnover time (see Ref.~\cite{kaneda_lagrangian_1993}).
Far in the inertial range, the Eulerian eddy turnover time becomes  $\tau_E(k)\sim (k \vsweep)^{-1}$, where $\vsweep\simeq \vrms/3$ is a locally uniform velocity field sweeping the vortices, in accordance with the Kraichnan random sweeping approximation~\cite{kraichnan:1964}, and $ \vrms$ is the rms velocity of the turbulence.
Earlier work used the Lagrangian eddy turnover time $\tau_{\ell}(k)\sim (k v_{\ell})^{-1}$, where $v_{\ell}$ denotes the characteristic velocity on the scale $\ell\sim 1/k$ \cite{Gogoberidze:2007an,Kahniashvili:2008pe,Caprini:2009yp}.
Our simulations confirm that the Eulerian eddy turnover time is the correct option.

Simulations of both non-helical and helical MHD turbulence and the resulting GW generation have been carried out in \Refs{Pol:2018pao,Pol:2019yex,Kahniashvili:2020jgm,Brandenburg:2021bvg}.
In these works there is an initial phase in which the MHD turbulence develops, starting from a nearly monochromatic electromotive force, or a kinetic forcing.
They find that the spectral shape of the GW signal varies, depending on whether a fully-developed turbulent magnetic and/or velocity field is present in the initial conditions, or whether it is sourced by the forcing.

In this paper, we develop a semi-analytical model of GW generation by freely-decaying kinetic, vortical, non-helical turbulence, which we support with the results of simulations. We rely on the relativistic hydrodynamic code SCOTTS, developed by the authors of \Refs{Hindmarsh:2013xza,Hindmarsh:2015qta,Hindmarsh:2017gnf,Cutting:2019zws}, but we use it to study the evolution of vortical fluid motion, excluding the dynamics of the scalar field undergoing the transition (see \cref{sec:numerical}). We have chosen to over-simplify the turbulence model (no compressional modes, no helicity, no magnetic field, no initial forcing phase) in order to have full control of our analytical understanding of the GW production.
Inclusion of these additional features will be tackled in future works.

The layout of the paper is as follows. We begin with a short review of stochastic gravitational wave backgrounds in the context of vortical turbulence in \cref{sec:sgwbreview}. We first introduce the equations of motion for the metric perturbations in an FLRW background, and write their solutions in terms of Green's functions and the anisotropic shear stress. The gravitational wave energy density power spectrum in the radiation era is then constructed, assuming a statistically homogeneous and isotropic signal. The solution for the gravitational wave energy density power spectrum is given in terms of an integral over the Green's function and the unequal time correlator (UETC) of the anisotropic stress. The section concludes by reviewing the computation of the anisotropic stress UETC in the context of vortical turbulence, decomposing the anisotropic stress UETC into a kernel which contains the fluid velocity UETC.

We continue by modelling freely decaying turbulence in \cref{sec:free-decay-model}. In \cref{sec:free-decay-velps}, we introduce the velocity power spectrum for a divergence free velocity field, along with important quantities such as the integral scale of the flow $\xi$, and the rms velocity $\vrms$. In our modelling, we assume that the turbulent flow is described by a \vonK~spectrum, which has a Kolmogorov power law in the inertial range and a causal slope of $k^5$ at large scales, where $k$ refers to the power spectrum wavenumber. The evolution of the velocity field in decaying turbulence is considered in \cref{sec:free-decay-evol}.
Here we review the self-similarity properties of the turbulent power spectrum, starting from dimensional arguments and the assumption that there is a single length scale in the flow.
On further assuming that the eddy turnover time, $\tau_\xi$, is the only principal timescale in the flow, one arrives at exponential decay laws for the kinetic energy and integral scale, where the decay exponents of both quantities are in principle related. We conclude with \cref{sec:UETCvelocity}, where we introduce a model for the velocity UETC. To model the decorrelation in time, we follow \Refa{Niksa:2018ofa} and adopt the exponential decorrelation provided by the short-time analysis of  \Refa{kaneda_lagrangian_1993}. This consistently reproduces the random sweeping model \cite{kraichnan:1964} in the inertial range, which is supported by numerical works simulating decaying isotropic turbulence (see \eg Ref.~\cite{he_computation_2004}).

In \cref{sec:PosKer} we expand on the discussion of UETCs by considering them in the context of positive kernels and Mercer's condition.
Indeed, any two-point correlation function of a random variable satisfies a set of inequalities known as Mercer's condition.
Positive kernels are defined as the functions satisfying Mercer's condition, thus they are candidates to model UETCs.
Earlier works, such as \Refa{Caprini:2009yp}, encountered the question of how to define a valid correlation function for a non-stationary process. Here, we consistently address this issue in the context of the theory of positive kernels, and propose to model the unequal time correlator (UETC) of the turbulent velocity field as a  Gibbs Kernel. This provides a way to symmetrize the unequal time correlator and, most importantly, it guarantees that the anisotropic stress two-point function is a valid correlation function. As a consequence, this leads to a GW energy density power spectrum which is always positive, as it should be.

After discussing the theory, we move on to describe our simulations in \cref{sec:numerical}. We introduce the Minkowski space simulation code, SCOTTS, and discuss how it has been modified to study relativistic vortical turbulence. To prepare the initial conditions, we initialize the velocity field in momentum space, where each mode is randomly drawn using Gaussian statistics from a \vonK~power spectrum. We also project the velocity field so that it is purely vortical. After outlining our suite of simulation runs, we specify the methodology for outputting the velocity power spectra and UETC, as well as the procedure for evolving the metric perturbations and obtaining the gravitational wave power spectrum in the simulations.

In the following two sections we present our results. First, we focus on the evolution of the velocity field in the simulations, in \cref{sec:results-vel}.
We study the evolution of the kinetic energy and of the integral scale of the flow in \cref{sec:ResETC}, and link it to the model of \cref{sec:free-decay-evol}. In the simulations, the velocity power spectrum displays self-similar scaling towards the end of the simulation, but
the theoretical relation between the decay law exponents and the velocity spectrum power law at large scales is not satisfied. We also test our model for the velocity UETC in \cref{sec:results-uetc}. The combination of the Gibbs Kernel model of the UETC with the decorrelation of \Refa{kaneda_lagrangian_1993} performs well for both the inertial range and for large scale modes. We compare our model to that of \Refa{Niksa:2018ofa}, and find that our model has a closer agreement with the simulation data.

Our results for the gravitational wave power spectrum are given in \cref{sec:sgwbresults}.
There are three different cases. We begin by evaluating the gravitational wave power spectrum under the assumption that the turbulence is stationary, in \cref{sec:stationary}.
Next we cover the case of instantaneously generated turbulence in \cref{sec:gw-instant}, which contains our main results regarding the SGWB spectrum.
Here our initial conditions are a fully developed turbulent power spectrum.
We first show the SGWB extracted from the simulations.
We then calculate the gravitational wave signal using our model of freely decaying turbulence outlined in \cref{sec:free-decay-model,sec:PosKer}. In this case, to calculate the gravitational wave power spectrum, we use Monte Carlo integration with importance sampling to compute the integrals found in \cref{sec:sgwbreview}. Using this technique, we confirm that the precise values of the decay law exponents are unimportant for the final spectral shape. We also describe the constant source approximation, in which the source of gravitational waves is taken to be constant in time, before it is turned off abruptly~\cite{RoperPol:2022iel}.
We obtain in this context a ready-to-use formula for the SGWB spectrum from hydrodynamic turbulence.
We compare the gravitational wave power spectrum under all of these methods, finding good agreement in their mutual regions of applicability. The final case we consider in \cref{sec:GWcontinuous} is the inclusion of a simplified turbulence growth phase. Using our turbulence model and
Monte Carlo integration, we calculate the gravitational wave power spectrum in the case of linear growth for both a $\mathcal{C}^0$ and a $\mathcal{C}^{1}$ {growth phase}. The inclusion of a growth phase substantially affects the shape of the gravitational wave power spectrum, underlining the importance of studying the creation mechanism of turbulence in the early Universe.
Finally, we summarize our results and conclusions in \cref{sec:discussion}.

To help the reader, in \cref{tab:notation} we summarize some timescales and temporal parameters which appear in different sections of this paper.
\begin{table}
  \centering
  \begin{tabular}{ p{2cm}p{7cm}p{4,5cm}  }
    \multicolumn{3}{c}{\emph{Simulations (SCOTTS code) (\cref{sec:numerical})}} \\
    \hline
    time variable & description                            & properties                          \\
    \hline
    $\tdevel$     & initial time in the simulation         &                                     \\
    $\tend$       & final time in the simulation           & $k (\tend-\tdevel) \gtrsim 10 $     \\
    $\tuetc$      & reference time at which UETCs measured & $\tdevel < \tuetc < \tend$          \\
  \end{tabular}

  \vspace{1cm}

  \begin{tabular}{cll}
    \multicolumn{3}{c}{\emph{Semi-analytical evaluation of GW spectrum (\cref{sec:sgwbresults})}} \\
    \hline
    time variable & description & properties \\
    \hline
    $\tini$       & initial integration time & turbulence starts \\
    $\tgro$       & \parbox[t]{7cm}{timescale over which turbulence is sourced \\
      (see Section \ref{sec:GWcontinuous})} & \\
    $\tdevel$     & start of turbulence free decay & $\tdevel=\tini+\tgro \equiv \mathcal{H}_*^{-1}$ \\
    $\tauxist$    & initial eddy turnover time & $\tauxist = \xi_* / \vrmsst$ \\
    $\ndecay$     & characteristic decay time of the source in units of  $\tauxist$ & dimensionless, $\order{1}$ \\
    $\ncut$       & \parbox[t]{7cm}{ duration of the source in units of  $\tauxist$ \\
    (used in the stationary source approximation of \cref{sec:stationary}, and in the constant source approximation of \cref{sec:const})} & dimensionless, $\order{1}$ \\
  \end{tabular}

  \caption{
    Important timescales and temporal parameters used in this article.
    A $*$ subscript indicates quantities evaluated at $\tdevel$.
    In the above, $k$ refers to the magnitude of the non-zero wavenumbers on the discrete momentum space lattice in the simulations.}
  \label{tab:notation}
\end{table}

\section{Stochastic background of gravitational waves from vortical turbulence}
\label{sec:sgwbreview}
\subsection{Generation of gravitational waves}
\label{sec:gen-gw}

In the cosmological context, GWs are described by transverse and traceless tensor spatial perturbations $h_{ij}$ of the background FLRW metric for a homogeneous and isotropic Universe:
\begin{equation}
	\dd{s}^2 = a^2(\tau) \qty[ - \dd{\tau}^2 + (\delta_{ij} + 2 h_{ij}) \dd{x}^i \dd{x}^j ],
	\label{eq:metric}
\end{equation}
with $\partial_i h_{ij}=h_{ii}=0$.
Here $a(\tau)$ is the scale factor, and $\tau$ the conformal time.
It follows from the linearized Einstein equations that $h_{ij}\qty(\vb{k}, \tau)$ satisfies
\begin{equation}
	\ddot{h}_{ij} + 2\mathcal{H} \dot{h}_{ij} + k^2 h_{ij} = 8 \pi G a^2(\tau) \Pi^{(TT)}_{ij}\qty(\vb{k}, \tau)
	\label{eq:gw-propagation}
\end{equation}
where $\mathcal{H}=\dot{a}/{a}$ is the comoving Hubble parameter, $\cdot = \dv*{\tau}$, $\vb{k}$ denotes a comoving wave-vector, and $\Pi_{ij}^{(TT)}$ is the transverse traceless part of the anisotropic stress tensor. The anisotropic stress is defined by
\begin{equation}
	\Pi_{ij}=T_{ij} - p(\delta_{ij}+2h_{ij})\text,
	\label{eq:Piij}
\end{equation}
with $T_{\mu\nu}$ the fluid energy momentum tensor, and $p$ its pressure.
The transverse traceless component of the anisotropic stress is extracted by means of the projector
\begin{align}
	\Pi_{ij}^{(TT)}(\vb{k},\tau)            & =\Lambda_{ij\ell m}(\vu{k})\Pi_{\ell m}(\vb{k},\tau),
	\label{eq:Piij_proj}                                                                                                                            \\
	\text{with}~~\Lambda_{ij\ell m}(\vu{k}) & = \bot_{i\ell }(\vu{k}) \bot_{jm}(\vu{k}) - \frac{1}{2} \bot_{ij}(\vu{k}) \bot_{\ell m}(\vu{k})\text,
	\label{eq:lambda_proj}                                                                                                                          \\
	\text{and}~~\bot_{ij}(\vu{k})           & = \delta_{ij} - \hat{k}_i \hat{k}_j\text. \label{eq:PSsim}
\end{align}
We work in the radiation era, and therefore we further define  $\Pi_{ij}^{(TT)} = (4\rho/3) \tilde{\Pi}_{ij}$ with $\rho$ the energy density in the early, radiation dominated Universe, and $p=\rho /3$.
Using Friedmann's equation, \cref{eq:gw-propagation} can be rewritten as
\begin{equation}
	\ddot{h}_{ij} + 2\mathcal{H} \dot{h}_{ij} + k^2 h_{ij} = 4 \mathcal{H}^2 \tilde{\Pi}_{ij}\qty(\vb{k}, \tau). \label{eq:gw-propagation-rad}
\end{equation}
Assuming that there were no changes in the relativistic degrees of freedom, $\mathcal{H} = \tau^{-1}$, and
changing variable from $h_{ij}(\vb{k}, \tau)$ to $\tau h_{ij}(\vb{k}, \tau)$,  \cref{eq:gw-propagation-rad} becomes the equation for a forced harmonic oscillator whose Green's function is known (see \Refa{Caprini:2018mtu} for a review).
The solution and its time derivative are
\begin{align}
	h_{ij}\qty(\vb{k}, \tau)       & =
	4\int_{\tini}^\tau\frac{\dd \zeta}{\zeta} \frac{\sin k(\tau-\zeta)}{k \tau} \tilde{\Pi}_{ij}(\vb{k}, \zeta) ,
	\label{eq:strain}                  \\
	\dot{h}_{ij}\qty(\vb{k}, \tau) & =
	4\int_{\tini}^{\tau}\frac{\dd \zeta}{\zeta} \qty[\frac{\cos k(\tau-\zeta)}{\tau}-\frac{\sin k(\tau-\zeta)}{k\tau^2}]\tilde{\Pi}_{ij}(\vb{k}, \zeta) .
	\label{eq:hdot-short}
\end{align}
where $\tini$ is the conformal time at which the source turns on.

\subsection{GW energy density power spectrum}
\label{sec:gw-energy-ps}

We assume that the GWs are generated by a stochastic process (specifically vortical turbulence) and consist of signals coming from patches in the sky that were causally disconnected at the moment of their emission in the very early Universe.
Assuming statistical homogeneity and isotropy, the two-point correlation function of $\dot{h}_{ij}$ can be written as
\begin{equation}
	\ev{\dot{h}^*_{ij}(\vb{k}, \tau) \dot{h}_{ij}(\vb{q}, \tau)} \equiv
	(2\pi)^3 \delta(\vb{q}-\vb{k}) P_{\dot{h}}(k, \tau),
\end{equation}
which defines the spectral density $P_{\dot{h}}$.
The fractional GW energy density power spectrum at conformal time $\tau$ in the radiation era is then
\begin{equation}
	\eval{\dv{\Omega_\mathrm{gw}}{\ln k}}_\tau =
	\frac{k^3P_{\dot{h}}(k, \tau)}{2 (2\pi)^3 G a^2(\tau) \rho(\tau)}.
	\label{eq:OmGW}
\end{equation}
The relevant statistical object of the GW source is the unequal time correlation function (UETC) of the anisotropic stress, defined by
\begin{equation}
	\ev{\tilde{\Pi}_{ij}(\vb{k}, \tau) \tilde{\Pi}^*_{ij}(\vb{q}, \zeta)} \equiv (2\pi)^3\delta(\vb{k-q}) P_{\tilde{\Pi}}(k, \tau,\zeta).
	\label{eq:Pi}
\end{equation}
From \cref{eq:hdot-short}, one then obtains
\begin{equation}
	P_{\dot{h}}(k,\tau) = \frac{16}{\tau^2} \iint_{\tini}^{\tau}\frac{\dd{\eta}}{\eta} \frac{\dd{\zeta}}{\zeta} \,
	\mathcal{G}(k,\tau, \eta,\zeta)
	P_{\tilde{\Pi}}(k, \eta,\zeta),
	\label{eq:h-square}
\end{equation}
with
\begin{equation}
	\mathcal{G}(k,\tau, \eta,\zeta)=
	\cos k(\tau - \eta)\cos k(\tau - \zeta)-2\frac{\cos k(\tau - \eta)\sin k(\tau - \zeta)}{k\tau}+\frac{\sin k(\tau - \eta)\sin k(\tau - \zeta)}{(k\tau)^2}\,.
	\label{eq:green_with_all_terms}
\end{equation}
The GW energy density fraction \cref{eq:OmGW} becomes then
\begin{equation}
	\eval{\dv{\Omega_\mathrm{gw}}{\ln k}}_{\tau} = \frac{8}{3 \pi^2}
	k^3
	\iint_{\tini}^{\tau} \mathcal{G}(k,\tau, \eta,\zeta) P_{\tilde{\Pi}}(k, \eta,\zeta)\frac{\dd{\eta}}{\eta} \frac{\dd{\zeta}}{\zeta},
	\label{eq:sbgw}
\end{equation}
where we have used the Friedmann equation $3 \mathcal{H}^2 = 8 a^2 \pi G \rho$.
As can be appreciated from \cref{eq:sbgw}, the GW signal is  sourced by the UETC of the anisotropic stress energy of the turbulent fluid, $P_{\tilde{\Pi}}(k,\tau, \zeta)$.

\subsection{The unequal time anisotropic stress correlator}
\label{sec:uetc-stress}

We now review the computation of the UETC of the turbulent fluid anisotropic stress, defined in \cref{eq:Pi}.
The starting point is the spatial, off-diagonal part of the energy momentum tensor of a relativistic fluid.
In order to simplify the computation, we neglect the spatial dependence of the fluid enthalpy density, and set the Lorentz factor $\gamma = 1$.
This amounts to assuming non-relativistic turbulence.
As we shall see, we find an excellent agreement between the semi-analytical results based on the non-relativistic source model and the relativistic simulations, also in what concerns the GW signal.
Therefore,
the assumption of non-relativistic turbulence does not significantly affect our final results, within the tested range of initial rms velocity.
We then approximate
\begin{equation}
	T_{ij} (\vb{x}, \tau) \approx (\rho + p)  v_i(\vb{x}, \tau) v_j(\vb{x}, \tau), \quad \text{for } i\neq j.
	\label{eq:Tlm}
\end{equation}
On using \cref{eq:Piij,eq:Piij_proj},
the two point correlation of the normalized anisotropic stress becomes
\begin{multline}
	\ev{\tilde{\Pi}_{ij}(\vb{k}, \tau) \tilde{\Pi}^*_{ij}(\vb{k'}, \zeta)} = \\
	\Lambda_{ij\ell m}(\vu{k}) \Lambda_{ijrs}(\vu{k'})\int \frac{\dd[3]{p}}{(2\pi)^3} \frac{\dd[3]{h}}{(2\pi)^3}
	\ev{v_\ell(\vb{p}, \tau) v_m (\vb{k-p}, \tau) v_r^*(\vb{h}, \zeta) v_s^* (\vb{k'-h}, \zeta) }.
\end{multline}
We assume Gaussianity of the velocity field to rewrite the four-point velocity correlator in terms of two-point correlators using Wick's theorem:
\begin{equation}
\begin{split}
  \ev{\tilde{\Pi}_{ij}(\vb{k}, \tau) \tilde{\Pi}^*_{ij}(\vb{k'}, \zeta)}  = &
	  \Lambda_{ij\ell m}(\vu{k}) \Lambda_{ijrs}(\vu{k'})
	\int \frac{\dd[3]{p}}{(2\pi)^3} \frac{\dd[3]{h}}{(2\pi)^3} \\
	& \quad [ \ev{v_\ell(\vb{p}, \tau) v_m (\vb{k-p}, \tau)} \ev{v_r^*(\vb{h}, \zeta) v_s^* (\vb{k'-h}, \zeta) } \\
	&  \qquad + \ev{v_\ell(\vb{p}, \tau) v_r^*(\vb{h}, \zeta) } \ev{v_m (\vb{k-p}, \tau)  v_s^* (\vb{k'-h}, \zeta) } \\
	 & \qquad + \ev{v_\ell(\vb{p}, \tau) v_s^* (\vb{k'-h}, \zeta} \ev{v_m (\vb{k-p}, \tau) v_r^*(\vb{h}, \zeta)  }]. \label{eq:Piquasinorm}
\end{split}
\end{equation}
Similarly to the anisotropic stress UETC of \cref{eq:Pi}, we can define the turbulent velocity UETC.
Under the further assumption that the velocity field is divergence-free, statistically isotropic and homogeneous, the velocity UETC is
\begin{equation}
	\ev{v_i(\vb{k}, \tau) v_j^*(\vb{q}, \zeta)} \equiv (2\pi)^3 \delta\qty(\vb{k} - \vb{q}) \bot_{ij}(\vu k) \UETCv(k, \tau, \zeta).
	\label{eq:velocity-ps}
\end{equation}
Then on using $\Lambda_{ij\ell m}(\vu{k}) \Lambda_{ijrs}(\vu{k}) = \Lambda_{\ell m r s}(\vu{k})$, \cref{eq:Piquasinorm} becomes
\begin{multline}
	\ev{\tilde{\Pi}_{ij}(\vb{k}, \tau) \tilde{\Pi}^*_{ij}(\vb{k'}, \zeta)} = \\
	\Lambda_{\ell m r s}(\vu{k}) \delta(\vb{k-k'})
	\int \dd[3]{p}
	\UETCv(p, \tau, \zeta) \UETCv(q, \tau, \zeta)\qty[  \bot_{\ell r}(\vu{p}) \bot_{m s}(\vu{q})
		+ \bot_{\ell s}(\vu{p})\bot_{m r}(\vu{q})  ],
	\label{eq:stress-full}
\end{multline}
where $\vb{q} = \vb{k} - \vb{p}$ (note that
the first term in \cref{eq:Piquasinorm} does not contribute).
The contraction over indices can be carried out explicitly using \cref{eq:Piij_proj,eq:lambda_proj,eq:PSsim},
\begin{align}
	2 \Lambda_{\ell m r s}(\vu{k}) \bot_{\ell r}(\vu{p}) \bot_{m s}(\vu{q}) & = 1 + 2 \qty[(\vu{k}\vdot \vu{p} )^2 + (\vu{k}\vdot \vu{q})^2] + (\vu{k}\vdot \vu{p} )^2  (\vu{k}\vdot \vu{q})^2 \label{eq:projection-niksa} \\
	2 \Lambda_{\ell m r s}(\vu{k}) \bot_{\ell s}(\vu{p}) \bot_{m r}(\vu{q}) & = 1 + (\vu{k}\vdot \vu{p} )^2  (\vu{k}\vdot \vu{q})^2 \label{eq:projection-missing},
\end{align}
so that \cref{eq:velocity-ps} becomes
\begin{multline}
	\ev{\tilde{\Pi}_{ij}(\vb{k}, \tau) \tilde{\Pi}^*_{ij}(\vb{k'}, \zeta)} = \\
	\delta(\vb{k-k'})
	\int \dd[3]{p} \UETCv(p, \tau, \zeta) \UETCv(q, \tau, \zeta)
	\qty[1 + (\vu{k}\vdot \vu{p} )^2 + (\vu{k}\vdot \vu{q})^2 + (\vu{k}\vdot \vu{p} )^2  (\vu{k}\vdot \vb{q})^2].
\end{multline}
We can now extract the kernel of the anisotropic stress defined in \cref{eq:Pi} and obtain
\begin{equation}
	P_{\tilde{\Pi}}(k, \tau, \zeta) =
	\frac{1}{(2\pi)^3}
	\int \dd[3]{p} \UETCv(p, \tau, \zeta) \UETCv(q, \tau, \zeta)
	\qty[1 + (\vu{k}\vdot \vu{p} )^2] \qty[1 + (\vu{k}\vdot \vu{q})^2].
	\label{eq:unequal-stress-final}
\end{equation}
This result agrees with \Refa{Caprini:2009yp}, but differs from \Refa{Niksa:2018ofa}. The reason is that in \Refa{Niksa:2018ofa}, the tensor indices on the two terms in square brackets on the right-hand side of \cref{eq:stress-full} were assumed identical, but this is not the case as can be seen explicitly in \cref{eq:projection-niksa}
and \cref{eq:projection-missing}.

The following \cref{sec:free-decay-model,sec:PosKer} are devoted to the construction of a model for the velocity UETC $\UETCv(p, \tau, \zeta)$ in \cref{eq:unequal-stress-final}.

\section{Modelling freely decaying turbulence}
\label{sec:free-decay-model}

In this section, we provide definitions and review properties of freely decaying turbulence.
This will be complemented with new theoretical input regarding positive kernels in \cref{sec:PosKer}, leading to the new velocity UETC model which we propose in this work.
Our new model will be validated with relativistic, simulations in \cref{sec:results-vel}.

\subsection{Velocity power spectrum}
\label{sec:free-decay-velps}

The real-space two-point correlation function of a statistically homogeneous, isotropic and divergence-free velocity field reads~\cite{Davidson:2004}
\begin{equation}
	b_{ij}(\vb{r}, \tau)=\frac{\ev{v_i(\vb{x},\tau)v_j(\vb{x}+\vb{r},\tau)}}{\vrms^2(\tau)}
	\equiv\Sigma_t(r,\tau)\bot_{ij}(\vu{r})+\Sigma_\ell(r,\tau)\hat r_i\hat r_j,\label{eq:SigmaGamma}
\end{equation}
where $\vrms^2(\tau)\equiv \ev{v_i(\vb{x},\tau)v_i(\vb{x},\tau)}$ is the rms velocity squared, and the functions $\Sigma_t$ and $\Sigma_\ell$ represent respectively the transverse and longitudinal correlation functions.
Since the velocity field is divergence-free, we have that $\pdv*{b_{ij}}{r_i} = 0$ and the correlation functions are related through
\begin{equation}
	\Sigma_\ell'(r)=\frac{2}{r}[\Sigma_t(r)-\Sigma_\ell(r)].
	\label{eq:gamma-prime}
\end{equation}
The spectral density $\Psdv(k,\tau)$
\begin{equation}
	\ev{v_i(\vb{k}, \tau) v_j^*(\vb{q}, \tau)} \equiv (2\pi)^3 \delta\qty(\vb{k} - \vb{q}) \bot_{ij}(\vu k) \Psdv(k, \tau)
	\label{eq:velocity-sd}
\end{equation}
(comparing with \cref{eq:velocity-ps}, note that we denote quantities at equal time with one time variable only, for conciseness) can be expressed in terms of the correlation functions $\Sigma_\ell$ and $\Sigma_t$ by matching the Fourier transform of \cref{eq:velocity-sd} and the trace of \cref{eq:SigmaGamma} \cite{Caprini:2006jb}:
\begin{equation}
	\Psdv(k,\tau)=\vrms^2(\tau)\int \dd[3]{r}\, \e^{i\vb{k}\cdot\vb{r}}\qty[\Sigma_t(r,\tau)+\frac{1}{2}\Sigma_\ell(r,\tau)].
\end{equation}
In spherical coordinates and using \cref{eq:gamma-prime}, this becomes
\begin{align}
	\Psdv(k,\tau) = &
	\; 2 \pi \,\vrms^2(\tau) \int \dd{r}\qty[ r^3 \Sigma_\ell'(r,\tau) + 3 r^2 \Sigma_\ell(r,\tau)] \frac{\sin kr}{kr}  \label{eq:calP-1d}
	\\
	=               & \; 2 \pi\, \vrms^2(\tau) \int \dd{r}\qty[ r^3 \Sigma_\ell'(r,\tau) + 3 r^2 \Sigma_\ell(r,\tau)] \qty[1 - \frac{(kr)^2}{6} + \order{(kr)^4}].
	\label{eq:spectral-large-scale}
\end{align}
We assume that the correlation function of \cref{eq:SigmaGamma} has compact support because it vanishes outside the causal horizon.
As a consequence, the leading order term in \cref{eq:spectral-large-scale} is $\eval{r^3 \Sigma_\ell}_0^\infty = 0$, and the spectral density on large scales is given by
\begin{equation}
	\Psdv(k\rightarrow 0,\tau) = \frac{2\pi}{3}\vrms^2(\tau) k^2\int_0^\infty \dd{r} r^4\,\Sigma_\ell(r,\tau) + \order{(kr)^4} \,.\label{eq:Pvcausal}
\end{equation}
Therefore, causality in the early Universe points to Batchelor turbulence, at least on super-horizon scales~\cite{Caprini:2006jb}.
The coefficient of the amplitude of the large-scale part of the spectrum in \cref{eq:Pvcausal} is the Loitsyansky integral $\int \dd{r} r^4 \Sigma_\ell(r)$.
This quantity was conjectured to be constant in time in the absence of long-range correlations   (see \eg \Refa{Davidson:2004}), which would then be consistent with the causality requirement.

The power spectrum $\Psv(k)$ encodes how
$\vrms^2$, or equivalently the non-relativistic kinetic energy per unit enthalpy,
is distributed into the different length-scales:
\begin{equation}
	\vrms^2(\tau) \equiv
	\int \frac{\dd k}{k} \Psv(k, \tau).
	\label{eq:kinetic-energy}
\end{equation}
In our notation, it is related to the spectral density through
\begin{equation}
	\Psv(k,\tau) = \frac{k^3}{\pi^2}\Psdv(k,\tau).
	\label{eq:velocity_ps_mathcal}
\end{equation}
Note that the power spectrum as defined here is related to the energy spectrum $E(k,\tau)$ as defined in \Refa{Davidson:2004} through
\begin{equation}
	\Psv(k,\tau) = 2kE(k,\tau).
\end{equation}
To characterize the typical length-scale of the system, we also define the integral scale $\xi$ in terms of the longitudinal correlation function, or equivalently in terms of the power spectrum
\begin{equation}
	\xi(\tau)\equiv \int_0^\infty \dd r\,\Sigma_\ell(r,\tau) = \frac{\pi}{4\bar v^2} \int k^{-1} \Psv(k, \tau) \dd{\ln k}.
	\label{eq:integral-scale}
\end{equation}
We use the integral scale to define  dimensionless,  time-dependent wavenumbers
\begin{equation}
    K(\tau) \equiv \mathcal{A}\, k\, \xi(\tau)\, ,
\end{equation} with $\mathcal{A}$ a normalization constant given in \cref{eq:constants1} below.

Concerning the shape of the velocity power spectrum, we assume in our modelling
that
in the inertial range it is determined by the well-known Kolmogorov $K^{-2/3}$ law \cite{1941DoSSR..30..301K}, and on large scales by the causal $K^5$ slope, motivated by \cref{eq:Pvcausal}.
These properties are captured by the
\vonK\ spectrum \cite{von_karman_progress_1948,Davidson:2004,Caprini:2009yp, Niksa:2018ofa}
\begin{equation}
	\Psv\qty(k, \tau) = \mathcal{B}\, \vrms^2(\tau) \frac{K^5(\tau)}{\qty[1 + K^2(\tau)]^{17/6}}.
	\label{eq:power-spectrum}
\end{equation}
The coefficients $\mathcal{A}$ and $\mathcal{B}$ are chosen to ensure that the definitions of the kinetic energy and the integral scale of \cref{eq:kinetic-energy,eq:integral-scale} are consistent\footnote{Note that in \Refa{Niksa:2018ofa}, the authors introduce a factor of $5/12$ in the denominator of the velocity power spectrum: their motivation was to localize the peak of the spectrum around $K=1$. We do not follow this convention here. Other initial spectral shapes, derived from simulations, are also used for example in \Refs{Brandenburg:2017neh,Pol:2019yex,RoperPol:2022iel}.}:
\begin{align}
	\mathcal{A} &\equiv \frac{55 \Gamma(1/3)}{12 \sqrt{\pi} \Gamma(17/6)} \approx 4.02
	\label{eq:constants1}\\
	\mathcal{B} &\equiv \frac{8  \Gamma(17/6)}{3 \sqrt{\pi} \Gamma(1/3)} \approx 0.97.
	\label{eq:constants2}
\end{align}

\cref{eq:power-spectrum} assumes that the \vonK~spectral shape is preserved during the time evolution of the turbulent field.
This power spectrum will be used for the semi-analytical evaluation of the GW signal (see~\cref{sec:numintegration}): self-similarity in time is therefore an underlying assumption of the resulting GW spectra.
Furthermore, \cref{eq:power-spectrum} will also be used as the initial condition for our hydrodynamical simulations.
The simulations will show, on the other hand, that the \vonK~spectral shape is not fully preserved by time evolution: while for $K \gtrsim 1$, a power law consistent with $K^{-2/3}$ returns after $\order{10}$ eddy turnover times, a shallower one than $K^5$ emerges for $K \ll 1$ (see~\cref{sec:ResETC}).
Nevertheless, the GW spectra output by the simulations, and those resulting from the semi-analytical integration, are in very good agreement in their common wavenumber $K$ region: we therefore anticipate that differences in the time evolution of the large-scale part of
the turbulent spectrum do not significantly affect the final results.
This is due to the fact that the production of gravitational waves is fast, as we shall see.

As a consequence of the fast GW sourcing, though, the particular choice of the initial
velocity power spectrum determines the final GW spectral shape.
Equation~\eqref{eq:power-spectrum} is the only input we use for exploring the SGWB signal from decaying turbulence.
Nevertheless, we are able to draw general lessons from our analysis.

\subsection{Evolution of the velocity field in decaying turbulence}
\label{sec:free-decay-evol}

In this section we provide a  very simplified review of the time evolution of the freely-decaying turbulent power spectrum, and of average quantities such as
the kinetic energy per unit enthalpy $\vrms^2(\tau)$ and the integral scale $\xi(\tau)$.

In line with the model of \cref{sec:free-decay-velps}, we assume that there is only one principal length scale in the flow, $\xi$ (see \eg \Refa{george} for a turbulence description not satisfying this hypothesis).
Since the power spectrum is dimensionless, it must generally take the form (consistently with \cref{eq:power-spectrum})
\begin{equation}
	\Psv(k, \tau) = \qty(\frac{\vrms(\tau)}{\vrmsst})^2\, \varphi[k\xi(\tau)]\,,
	\label{eq:pv_scaled_one_scale}
\end{equation}
where the function $\varphi(k\xi(\tau))$ must satisfy integrability conditions following \cref{eq:kinetic-energy,eq:integral-scale}, and $*$ indicates quantities evaluated at the start of turbulence decay.
At large scales, the absence of relevant dynamical processes implies a simple power law spectrum
\begin{equation}
	\Psv[k\xi(\tau)\ll 1, \tau] \simeq \vrms^2(\tau)\,[k\xi(\tau)]^{\beta+1}\,,
\end{equation}
where, for the moment, the spectral index $\beta$ is left unspecified.

Further assuming that at large scales the power spectrum is constant in time\footnote{This assumption excludes \emph{a priori} the presence of a non-helical inverse cascade.
While observed in simulations of MHD turbulence \cite{Brandenburg:2014mwa,Reppin:2017uud,Christensson:2000sp} (see also \Refs{Campanelli:2015ypt,Olesen:2015dga} for theoretical interpretations), this phenomenon seems to be absent in purely hydrodynamic turbulence: see \Refa{Brandenburg:2016odr} and the results presented in \cref{sec:ResETC}.}, leads to
\begin{equation}
	\vrms^2(\tau)\xi(\tau)^{1+\beta}=\mathrm{const}\,.
	\label{eq:vxigeneral}
\end{equation}
Under this assumption, the self-similarity of the power spectrum takes the form \cite{Christensson:2000sp,Brandenburg:2016odr}
\begin{equation}
	\Psv(k, \tau) = \qty(\frac{\xi_*}{\xi})^{ 1 + \beta}\varphi[k\xi(\tau)]\,.
	\label{eq:Pvxi}
\end{equation}
The individual time evolution of the rms velocity and the integral scale can be derived, if we further assume that there is only one principal timescale in the flow, the eddy turnover time $\tau_\xi=\xi/\vrms$.
With this assumption, one can write
\begin{equation}
	\dv{\vrms^2}{\tau} =  - c_v \frac{\vrms^2}{\tau_\xi},
	\label{eq:kolmog}
\end{equation}
where $c_v$ is a constant.
Together with \cref{eq:vxigeneral}, one can solve \cref{eq:kolmog} to derive
\begin{align}
	\vrms^2(\tau) = \vrms_*^2 \left(1 + \frac{\tau - \tdevel}{\ndecay\tauxist}  \right)^{-p}
	\label{eq:VelEvMod}
	\\
	\xi(\tau) = \xi_* \left(1 + \frac{\tau - \tdevel}{\ndecay\tauxist}  \right)^{q}\,,
	\label{eq:XiEvMod}
\end{align}
where the decay exponents $p$ and $q$ take the values
\begin{align}
	p& =\frac{2(\beta+1)}{\beta+3}
	\label{eq:pchialpha}\\
	& q=\frac{2}{\beta+3},
	\label{eq:qchialpha}
\end{align}
and $\ndecay$ is a constant representing the number of initial eddy turn-over times that the kinetic energy takes to decay, and the integral scale to grow.
In particular $p$ and $q$ satisfy
\begin{equation}
	p = (1 + \beta)q = 2(1-q) \,.
	\label{eq:pVsq}
\end{equation}
Note that this model does not predict the value of $\beta$, although consistency with the arguments of  \cref{sec:free-decay-velps} would indicate $\beta=4$ as would follow from the constancy of the Loitsyansky integral.
In \cref{sec:ResETC} we will report on our numerical investigation of the evolution of the velocity power spectrum in the framework of the model described above.

The validity of the decay laws given in \cref{eq:VelEvMod,eq:XiEvMod}, together with the actual value of the parameter $\beta$ (and consequently of the decay law exponents through \cref{eq:pVsq}), are currently open problems of the theory of turbulence.
Invariance of the Navier-Stokes equations under appropriate rescaling leads to the self-similarity of the velocity power spectrum, and links the decay law exponents $p$ and $q$ to the scaling parameter \cite{Olesen:1996ts,Ditlevsen:2002xs}.
The scaling parameter can then be related to the slope of the initial velocity power spectrum at large scales, if one assumes that the latter also satisfies the self-similarity conditions (see \eg \Refa{Campanelli:2007tc}).
In this context, an initial power spectrum leading to $\beta=4$, as motivated in \cref{sec:free-decay-velps}, provides $p$ and $q$ values consistent with the classical Kolmogorov theory, namely $p=10/7$ and $q=2/7$ \cite{landau2013fluid,Davidson:2004} (see also \cite{2010PhST..142a4003T} for a numerical result).
The relation between the scaling parameter and the slope of the initial power spectrum at large scales has also been derived in the context of renormalization group analyses of the incompressible MHD equations, as holding at the fixed point \cite{camargo_tasso,Shiromizu:1998he}.
However, the initial velocity power spectrum does not need to necessarily satisfy either the self-similarity condition, nor does it need to exhibit power law behaviour.
Relaxing these assumptions changes the prediction on the subsequent free decay model, disconnecting the value of $\beta$ from the initial conditions \cite{Campanelli:2015ypt,Brandenburg:2016odr}.
Recently, new models of the turbulent free-decay have emerged, following which the turbulence decay is governed by the integral scale rather than by the asymptotic behaviour at large scales \cite{george,doi:10.1063/1.4901448,schaefer,meldi:hal-01298925,Zhou:2021vah}.
According to these findings there would be then no universal decay regime, since the details of the spectral shape near the peak are related to the turbulence production mechanism.

Note that in both this Section and \cref{sec:free-decay-velps} we have neglected the dissipative scales.
The turbulent kinetic energy dissipates at the viscous scale $\lambda$, at which the Reynolds number is $\mathrm{Re}(\lambda)=v_\lambda \,\lambda\, a/\nu=1$, where $\nu$ denotes the kinematic viscosity.
The viscous scale $\lambda$ increases with time, and the turbulent velocity on a given scale $2\pi/k$ goes to zero once the dissipation scale has grown to $\lambda(\tau)=2\pi/k$.
Furthermore, $\lambda(\tau)$ grows faster than the integral scale $\xi(\tau)$: the turbulent inertial range is therefore fully dissipated when the two scales become equal, so that $\mathrm{Re}(\xi)=1$.
As previously demonstrated \cite{Caprini:2009pr,Caprini:2009yp}, turbulence is expected to free-decay for several Hubble times, before it is fully dissipated. For example, with the initial set of parameters $T_*= 100~\mathrm{GeV}$, $\vrmsst= 0.03$, $\xi_*\mathcal{H}_*= 0.01$, $q=2/7$, $p=10/7$, one obtains $\mathrm{Re}(\xi)=1$ at $T_\mathrm{fin}\sim 0.3~\mathrm{GeV}$, corresponding to about $1000~\tauxist$.
The kinetic viscosity in the early Universe is indeed extremely small: for the same initial conditions, one finds $\mathrm{Re}(\xi_*)\sim 10^{11}$, and the inertial range spans about $8$ orders of magnitude in wavenumber.
In principle, we should have inserted a cutoff at $k>2\pi/\lambda(\tau)$ in the turbulent power spectrum given in \cref{eq:power-spectrum}, and we should account, in the GW production, for the viscous dissipation of the kinetic energy at each scale $k$ in addition to the overall free decay described by \cref{eq:VelEvMod,eq:XiEvMod}.
However, as we will discuss in \cref{sec:gw-instant}, the bulk of the GW signal is generated faster than the turbulent decay: therefore, the dynamics at the viscous scale plays no role in practice in the present analysis, and it is justified to neglect it altogether.

\subsection{Unequal time correlator of the velocity field}
\label{sec:UETCvelocity}

Together with the overall free-decay, we also need to model the time decorrelation of the turbulent velocity field.
This is encapsulated by the normalized velocity UETC
\begin{equation}
	\label{eq:NorVelUETC}
	\tdec(k, \tau, \zeta)  \equiv \frac{\UETCv(k,\tau,\zeta)}{\sqrt{\Psdv(k,\tau)\,\Psdv(k,\zeta)}}.
\end{equation}
The first model to have been developed was Kraichnan's random sweeping approximation \cite{kraichnan:1964},
in which the decorrelation function in the inertial range is a Gaussian in the variable $k\vrms|\tau - \zeta| $,
for small time differences.
The thinking behind the model is that modes are decorrelated by being
``swept'' by the large-scale flow, whose rms velocity is $\vrms$.
This model has good numerical support in the inertial range for both stationary and decaying turbulence (see for example \Refs{sanada_random_1992,he_computation_2004,Gorbunova:2021cpn}).
We review it in \cref{sec:kraichnan}.

However, for GW production calculations, we need to model the decorrelation around the peak of the velocity power spectrum.
To this end, we study --- for the first time with hydrodynamical simulations --- the decorrelation of the turbulent velocity field outside the inertial range, as far as possible in the low wavenumber part of the spectrum.
The simulation results are presented in \cref{sec:results-uetc}, and show that a good model is obtained by replacing the rms velocity in the Gaussian with a time and scale dependent average decorrelation velocity, based on
the sweeping velocity proposed in Ref.~\cite{kaneda_lagrangian_1993}, which we present in \cref{sec:kaneda}.

Furthermore, we also need to take into account the formal requirement that the UETCs (velocity and anisotropic stress) must be positive definite kernels, which is not manifest in the simple Gaussian model.
This will be presented in \cref{sec:PosKer}.

\subsubsection{Kraichnan random sweeping approximation}
\label{sec:kraichnan}

The random sweeping approximation provides an explanation for the origin of the Gaussian functional form for the decorrelation.
Kraichnan's sweeping hypothesis is based on the assumption that
Fourier modes of the velocity field
in the inertial range are advected without deformation by a locally uniform velocity field $\vb{V}$, which may be time-dependent
\cite{kraichnan:1964}.
This amounts to neglecting the non-linear terms and writing
\begin{equation}
	\pdv{\vb{v}}{\tau} + i \qty[\vb{k} \vdot \vb{V}(\tau)] \vb{v} \simeq 0.
	\label{eq:sweeping-ode}
\end{equation}
Since $\vb{V}$ is locally uniform, the evolution of modes with different wavenumbers \(\vb{k}\) is decoupled.
The turbulent velocity field can then be explicitly integrated: from \cref{eq:sweeping-ode} one finds
\begin{equation}
	\vb{v}(\vb{k}, \tau) \simeq \vb{v}(\vb{k},\zeta) \exp(-i\int_{\zeta}^{\tau}\vb{k}\vdot \vb{V}(s) \dd{s}),
\end{equation}
where $\zeta < \tau$.
Under the assumption that the locally uniform velocity field $\vb{V}$ is statistically independent of the turbulent velocity field $\vb{v}$ at the initial time, the unequal time correlator of $\vb{v}$ can be expressed as
\begin{equation}
	\UETCv(k,\tau,\zeta) \simeq \Psdv(k,\zeta) \ev{\exp(-i\int_{\zeta}^{\tau}\vb{k}\vdot \vb{V}(s) \dd{s})}.
\end{equation}
The average of the exponential can be calculated assuming that its argument is a Gaussian random variable.
It then follows that the velocity field decorrelates with a characteristic Gaussian law\footnote{We recall that for a Gaussian random variable, \begin{equation}
		\ev{\exp(X)} = \ev{\sum_{n=0}^{\infty} \frac{X^{n}}{n!}} = \sum_{k=0}^{\infty}\frac{\ev{X^{2}}^{k}}{2^{k} k!} = \exp(\frac{\ev{X^{2}}}{2}).
	\end{equation}},
\begin{equation}
	\UETCv(k,\tau,\zeta) \simeq \Psdv(k,\zeta) \exp\qty(- \frac{1}{2} \ev{X^2}),
\end{equation}where
\begin{equation}
	X^{2} \equiv \iint_{\zeta}^{\tau} (\vb{k} \vdot \vb{V}(s))(\vb{k} \vdot \vb{V}(s')) \dd{s} \dd{s'}.
\end{equation}
This expression can be further simplified assuming that on average  the locally uniform velocity has the same amplitude in the three spatial directions \cite{kraichnan:1964}, leading to
\begin{equation}
	\UETCv(k,\tau,\zeta) \simeq \Psdv(k,\tau) \exp\qty[-\frac{1}{2} k^2 \vsweep^2(\tau, \zeta)(\tau - \zeta)^2],
	\label{eq:expvsweep}
\end{equation}
where $\vsweep(\tau,\zeta)$ is the average sweeping velocity which takes the form \cite{he_computation_2004,dong_study_2008}
\begin{equation}
	\vsweep^{2}(\tau, \zeta) = \frac{1}{3 (\tau-\zeta)^{2}}\iint_{\zeta}^{{\tau}} \ev{\vb{V}(s)\cdot \vb{V}(s')} \dd{s} \dd{s'}.
	\label{eq:sweep-equal-time}
\end{equation}
Kraichnan's model assumes stationary turbulence
and sets the sweeping velocity to the rms velocity divided by $\sqrt{3}$, under the hypothesis of statistical isotropy \cite{kraichnan:1964,wilczek_wave-numberfrequency_2012}.
Since this model has good numerical support in the inertial range \cite{sanada_random_1992,he_computation_2004,Gorbunova:2021cpn},
The sweeping velocity should reduce to
\begin{equation}
	\vsweep^{2} (\tau, \tau) = C_v^2\,\frac{ \vrms^2(\tau)}{3}
	\label{eq:Cv2}
\end{equation}
at equal time and in the inertial range, with $C_v^2$ a numerical factor whose value will be inferred from  simulations.
As presented in \cref{sec:results-uetc},  simulations indicate $C_v^2 \simeq 1$,mserving as a verification of the random sweeping approximation and of statistical isotropy.

\subsubsection{Scale dependent sweeping velocity}
\label{sec:kaneda}

\begin{figure}
	\centering
	\includegraphics{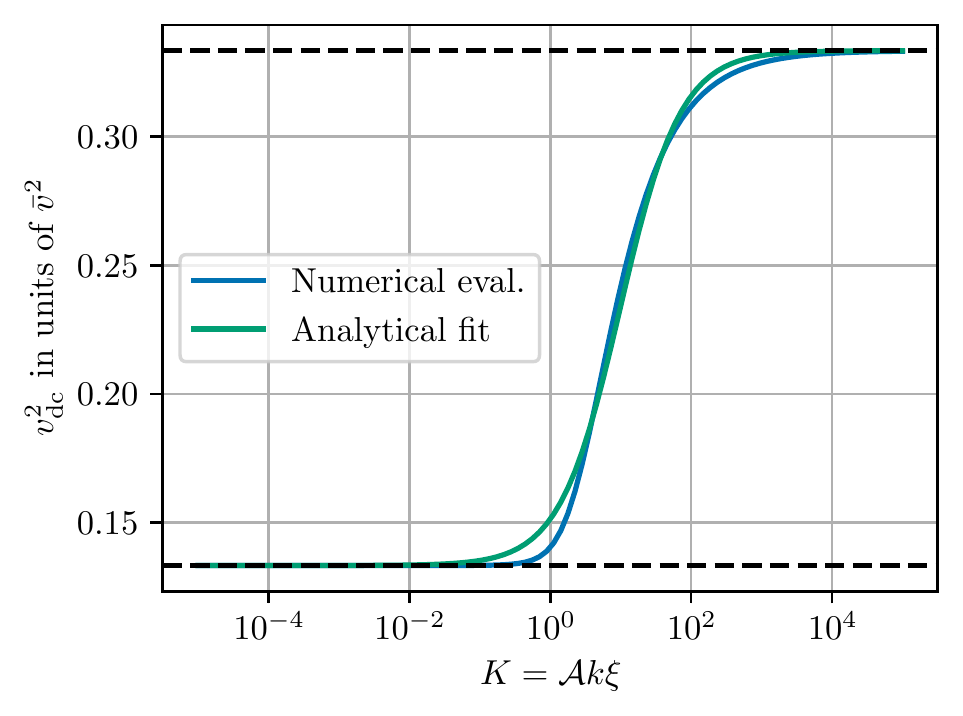}
	\caption{Extension of the decorrelation velocity at scales larger than the integral scale, following the model given in \cref{eq:Vlarge}. It interpolates smoothly between $\vrms^2/3$ in the inertial range as in \cref{eq:Cv2} and $2\vrms^2/15$ on large scales. These limits are shown with the black dashed lines.}
	\label{fig:IRsweep}
\end{figure}

The random sweeping approximation reviewed in \cref{sec:kraichnan}
applies in the inertial range.
However, in order to evaluate the GW generation, we need to model the time decorrelation of the velocity field on all scales (see~for example \cref{eq:unequal-stress-final}).
The decorrelation dynamics at scales larger than the integral scale in freely decaying turbulence have received little attention: numerical studies such as those performed in \Refs{he_computation_2004,dong_study_2008}, for example, study decorrelation only in the inertial range.
The largest scale analysed in  \Refa{sanada_random_1992} is $k=1$, corresponding to the scale of the forcing, namely the peak of the power spectrum: it appears from this analysis that even this scale decorrelates more slowly than those of the inertial range.

In the context of the literature dedicated to GW production by turbulence, various decorrelation models have been proposed, see  \Refs{Gogoberidze:2007an,Caprini:2009yp,Niksa:2018ofa}.
\Refa{Caprini:2009yp} assumed that the large scales do not decorrelate. The only time dependence of the velocity spectrum outside the inertial range was therefore due to the free decay, and an exponential decorrelation was inserted for wavenumbers in the inertial range by means of a step function. This introduced a non-physical discontinuity (see~Eq.~(57) of \cite{Caprini:2009yp}).
Furthermore, as pointed out in \Refa{Niksa:2018ofa}, \Refs{Gogoberidze:2007an,Caprini:2009yp} used the Lagrangian eddy turnover time as typical decorrelation time,
which is scale-dependent in the inertial range, and therefore
in conflict with the numerical results of \Refs{sanada_random_1992,he_computation_2004}.

\Refa{Niksa:2018ofa} recommended the use of the
model of  Ref.~\cite{kaneda_lagrangian_1993},
which improves Kraichnan's random sweeping approximation by taking interactions between modes into account.
The result is an Eulerian eddy turnover time which is weakly scale-dependent for wavenumbers up to $\order{10^2}$ times that corresponding to the integral scale.
We adopt the same form for the Eulerian eddy turnover time
to construct our model of the velocity UETC, but
correct it to ensure the mathematical consistency of
the correlation, as discussed in \cref{sec:PosKer}.

Following \Refs{kaneda_lagrangian_1991,kaneda_lagrangian_1993},
we define an instantaneous scale-dependent decorrelation velocity
\begin{equation}
	\vdc^2(k, \tau) \equiv
	\int_0^\infty h\qty(\frac{q}{k}) \Psv(q, \tau) \dd{\ln q},
	\label{eq:Vlarge}
\end{equation}
where the function $h(x)$ is given by\footnote{Note that the function $h(x)$ of \Refs{kaneda_lagrangian_1991,kaneda_lagrangian_1993} has a factor of $2$ difference due to our definition of the power spectrum in \cref{eq:kinetic-energy}. }
\begin{equation}
	h(x) = \frac{1}{48}(13-8x^2+3x^4) + \frac{1}{32 x} (1-x^2)^3 \ln \frac{1+x}{\abs{1-x}}.
	\label{eq:weird-h}
\end{equation}
Note that the calculations performed in \Refa{kaneda_lagrangian_1993} in principle apply only to the inertial range.  We will see that extending it to lower wavenumbers is justified by the numerical data.

\cref{eq:Vlarge} reduces to $\vdc^2(\tau)\simeq \vrms(\tau)^2 / 3$ far in the inertial range, consistent with the result of the random sweeping hypothesis, and gives $C_v^2=1$ (see~\cref{eq:Cv2}).
However, it also has another term, relevant at scales closer to the integral scale $\xi$, which is proportional to $- v_k^2\sim - \Psv(k)$.
This negative contribution comes from the analysis at small time intervals and appears to slow down the process of decorrelation.
In order to account for decorrelation at all scales, one can therefore attempt to use \cref{eq:Vlarge} as an extension of the sweeping velocity \cite{Niksa:2018ofa}.
Pushing \cref{eq:Vlarge} to large scales, one can appreciate that $\vdc(k,\tau)$ provides a continuous interpolation from $\vrms^2 / 3$ in the inertial range to $2\vrms^2 / 15$ at large scales, as shown in \cref{fig:IRsweep}.
In the semi-analytical GW evaluation, we will solve the integral of \cref{eq:Vlarge} numerically, although a good analytical fit to $\vdc(k,\tau)$ is (we recall that $K=\mathcal{A}k\xi(\tau)$)
\begin{equation}
	\vdc^2(k, \tau) \simeq \frac{\vrms^2}{3} \qty(\frac{1+0.2 K}{\sqrt{5/2} + 0.2 K})^2,
	\label{eq:vdcapprox}
\end{equation}
that we compare to the numerical result in \cref{fig:IRsweep} (note that this fit differs slightly from what was given in \Refa{Niksa:2018ofa}).

In order to describe how the relevant scales in the velocity spectrum decorrelate, one might substitute, in the exponential of \cref{eq:expvsweep}, the decorrelation velocity given in \cref{eq:Vlarge}.
This is the approach adopted in \Refa{Niksa:2018ofa}.
However, this breaks the time symmetry of the UETC $\UETCv(k,\tau,\zeta)$ (see~\cref{eq:velocity-ps}). To restore this symmetry,
the decorrelation velocity entering the exponential in \cref{eq:expvsweep} must depend on both times.  
The next section is devoted to ensuring the symmetry and positivity requirements that are necessary to model a valid UETC.

\section{Unequal time correlators as positive kernels}
\label{sec:PosKer}

The two-point correlation function of any random variable must be a positive kernel \cite{Genton:2002}, in other words it must satisfy certain properties --- most notably Mercer's theorem.
This applies to the turbulent velocity field UETC \cref{eq:velocity-ps}, as well as to  $P_{\tilde{\Pi}}(k, \tau,\zeta)$ \cref{eq:Pi},  describing the random anisotropic stresses arising from the turbulent field.
The model of the velocity field --- and consequently of the anisotropic stress --- UETCs that we construct in this section ensures that they are positive kernels.
According to Mercer's theorem, this implies that the GW energy density power spectrum is indeed also positive, see~\cref{eq:sbgw}.
This issue had already been raised in \Refa{Caprini:2009yp}, but it is analysed here in more depth directly on the velocity field UETC.

\subsection{The unequal time correlation function as a non-stationary Gaussian kernel}
\label{sec:mercer}

As we have discussed in \cref{sec:free-decay-model}, the turbulent rms velocity and  integral scale are evolving in time, while the turbulent source decorrelates.
GWs are therefore generated by a non-stationary, random process, whose UETC we need to model.
To do so, we make use of some general properties of two-point correlators of stochastic processes, which we review in the following.

The two-point correlator of a complex valued
stochastic process $\phi(\tau)$ is defined by
\begin{equation}
	K(\tau_1, \tau_2) \equiv \ev{\phi^*(\tau_1) \phi(\tau_2)}.
\end{equation}
We assume that the real and imaginary part of the stochastic process are uncorrelated, in which case $K(\tau_1, \tau_2)$ is real.
$K(\tau_1, \tau_2)$ is a positive semi-definite kernel which satisfies several properties: symmetry $K(\tau_1, \tau_2)=K(\tau_2, \tau_1)$; the Cauchy-Schwartz inequality; and the Mercer condition (see \eg~\Refa{Genton:2002})
\begin{equation}
	\iint_{I\times I} \dd{\tau_1} \dd{\tau_2} f^*(\tau_1) f(\tau_2) K(\tau_1,\tau_2)
	\geq 0\,,
	\label{eq:mercer-condition}
\end{equation}
for any interval $I$.
This follows from the fact that, for any well-behaved function $f$, the stochastic variable
$X = \int \dd{\tau} f(\tau) \phi(\tau)$ must have positive variance $\langle \abs{X}^2 \rangle \geq 0$.
One can show that the linear combination and
multiplication of two kernels yield another kernel~\cite{1909RSPTA.209..415M}.

The velocity UETC $\UETCv(k,\tau,\zeta)$ is the two-point correlation function of the stochastic process $v_i(\vb{k}, \tau)$ (see~\cref{eq:velocity-ps}), and is therefore a positive semi-definite kernel.
In \cref{sec:UETCvelocity} we analysed the time evolution and decorrelation properties of the velocity field,
based on our understanding of the turbulence physical process.
In this section, we explain how the fact that a UETC must be a positive semi-definite kernel leads to our choice of the decorrelation velocity  at unequal times $\vdc(k,\tau,\zeta)$, and how to symmetrize $\UETCv(k,\tau,\zeta)$.

The simplest example of a positive semi-definite kernel is the separable kernel $K(\tau_1, \tau_2) = g^*(\tau_1) g(\tau_2)$, which
factorizes Mercer's condition, \cref{eq:mercer-condition}, into a modulus.
This type of kernel was already proposed in the context of GW generation by turbulence and primordial magnetic fields in Refs~\cite{Caprini:2009yp,Caprini:2009fx,Caprini:2009pr}, where it is referred to as the coherent assumption.
Another common form is the stationary kernel $K(\tau_1, \tau_2) =K(\tau_1-\tau_2)$.
Stationary kernels are positive if and only if there exist a positive finite function $F$ such that~$K(\tau_1-\tau_2) = \int \cos[\omega (\tau_1-\tau_2)] F(\omega) \dd{\omega}$.
From this result one can easily see that a kernel of the form $K(\tau_1, \tau_2) \propto \delta(\tau_1-\tau_2)$ is positive definite: it is referred to as incoherent in Refs.~\cite{Caprini:2009yp,Caprini:2009pr}.

Most importantly for us, stationary kernels also include the Gaussian kernel $$K(\tau_1, \tau_2) = \exp[-\alpha(\tau_1-\tau_2)^2],$$ with $\alpha>0$, which is the form provided by the Kraichnan sweeping scenario \cref{eq:expvsweep}.
However, here we would like to model the power spectrum of freely decaying turbulence, which is, by definition, non-stationary.
A simple but well-defined departure from stationary kernels is provided by locally stationary kernels, of the form $K(\tau_1, \tau_2) = K_1\qty({\tau_1+\tau_2}/{2}) K_2\qty(\tau_1-\tau_2)$,
where $K_1$ is a non-negative function and $K_2$ a positive stationary kernel \cite{Silverman:1957}.
The UETC of a freely-decaying turbulent velocity field, though, cannot assume this form.
This is because the sweeping velocity depends explicitly on time, as can be seen from \cref{eq:expvsweep}.
For the problem at hand, we therefore also need to go beyond local stationarity.

We do so by introducing process-convolution kernels~\cite{higdon:1999}.
This specific class of non-stationary kernels is defined through a two-valued
complex function $g(\tau,u)$ as follows:
\begin{equation}
	K(\tau_1, \tau_2) = \int g^*(\tau_1, u) g(\tau_2, u) \dd{u}.
	\label{eq:pckernel}
\end{equation}
It is easy to show that process-convolution kernels satisfy Mercer's condition \cref{eq:mercer-condition}.
The process-convolution kernel adapted to our case is the non-stationary Gaussian kernel, defined with the function
\begin{equation}
	g(\tau,u) = \qty(\frac{2}{\pi J(\tau)})^ {1/4}\exp[-\frac{(\tau-u)^2}{J(\tau)}],
	\label{eq:gtwovalue}
\end{equation}
where $J(\tau)$ is a free function.

\subsection{Symmetrized unequal time  correlator of the velocity field}
\label{sec:unequal-time-velocity}

To define the kernel representing the velocity unequal time correlator, we choose for a given $k$
\begin{equation}
	J_k(\tau) =  \frac{1}{[k\,\vdc(k,\tau)]^2},
\end{equation}
where $\vdc(k,\tau)$
is given in \cref{eq:Vlarge}.
Inserting this definition into \cref{eq:gtwovalue} and further into \cref{eq:pckernel}, one finds a family of non-stationary Gaussian kernels
\begin{equation}
	K_k(\tau,\zeta)  = \sqrt{\frac{2\, \vdc(k, \tau) \vdc(k, \zeta) }{\vdc^2(k, \tau) + \vdc^2(k, \zeta)}}
	\exp[-k^2 (\tau-\zeta)^2 \frac{\vdc^2(k,\tau) \vdc^2(k,\zeta)}{\vdc^2(k,\tau)+\vdc^2(k,\zeta)}].
	\label{eq:kernelvdc}
\end{equation}
Note that these kernels satisfy $K_k(\tau, \tau) = 1$, and are therefore of a suitable form to be used as a normalized UETC $R(k,\tau,\zeta)$.

We can now use this result to construct a UETC $\UETCv(k,\tau,\zeta)$ which satisfies the Mercer condition.
First we note that the UETC evaluated at equal times is the power spectrum,
$\UETCv(k,\tau,\tau) = \Psdv(k, \tau)$.
Hence, multiplying the kernel \cref{eq:kernelvdc} by
$\sqrt{\Psdv(k, \tau)\Psdv(k, \zeta)}$
will give a function of the two times which correctly
reduces to the power spectrum at equal times.
This function is also a kernel,
as $\sqrt{\Psdv(k, \tau)\Psdv(k, \zeta)}$ is a separable kernel, and the product of two kernels is a kernel.
Our model is then
\begin{equation}
	\UETCv\qty(k, \tau, \zeta) =
	\sqrt{\Psdv(k, \tau)\Psdv(k, \zeta)} \sqrt{\frac{2 \vdc(k, \tau) \vdc(k, \zeta) }{\vdc^2(k, \tau) + \vdc^2(k, \zeta)}}
	\exp[- \frac{1}{2}k^2 (\tau-\zeta)^2 \vdc^2(k,\tau, \zeta)],
	\label{eq:unequal-velocity-ps}
\end{equation}
where the average decorrelation velocity $\vdc(k,\tau,\zeta)$ is given by the harmonic average of the equal-time decorrelation velocity:
\begin{equation}
	\vdc^{2}(k,\tau,\zeta)=2\,\frac{\vdc^{2}(k,\tau)\vdc^{2}(k,\zeta)}{\vdc^{2}(k,\tau)+\vdc^{2}(k,\zeta)}. \label{eq:vsweepcomplete}
\end{equation}
Note that this is symmetric in the two times $\tau$ and  $\zeta$.

To summarize, we have shown that the velocity UETC of \cref{eq:unequal-velocity-ps} is a positive kernel, \ie a well-defined two point correlator.
This guarantees that the anisotropic stress UETC $P_{\tilde{\Pi}}(k,\tau,\zeta)$ of \cref{eq:unequal-stress-final} is also a positive kernel, being the product of two positive kernels.
Finally, the integrand of \cref{eq:sbgw} is also a positive kernel. Indeed, $\mathcal{G}$ is a separable kernel since it results from the square of \cref{eq:hdot-short}.
As a consequence, the GW energy density power spectrum defined in \cref{eq:sbgw} is always positive, as it should be.

Having managed to model the decorrelation of the freely-decaying turbulence with \cref{eq:unequal-velocity-ps}, we can go beyond the approach of \Refa{Caprini:2009yp}.
In that paper the shear stress decorrelation was modelled directly in the anisotropic stress
UETC as a top-hat function of the kind $\Theta[k\abs{\tau-\zeta} - x_c]$.
We remark that such a function is not a positive kernel, since it is the Fourier transform of a sinc function:
in order to derive a positive definite GW power spectrum, one must indeed take $x_c <\pi$
\cite{Caprini:2009yp,Caprini:2009pr}.


\begin{figure}
	\centering
	\includegraphics[width=0.49\textwidth]{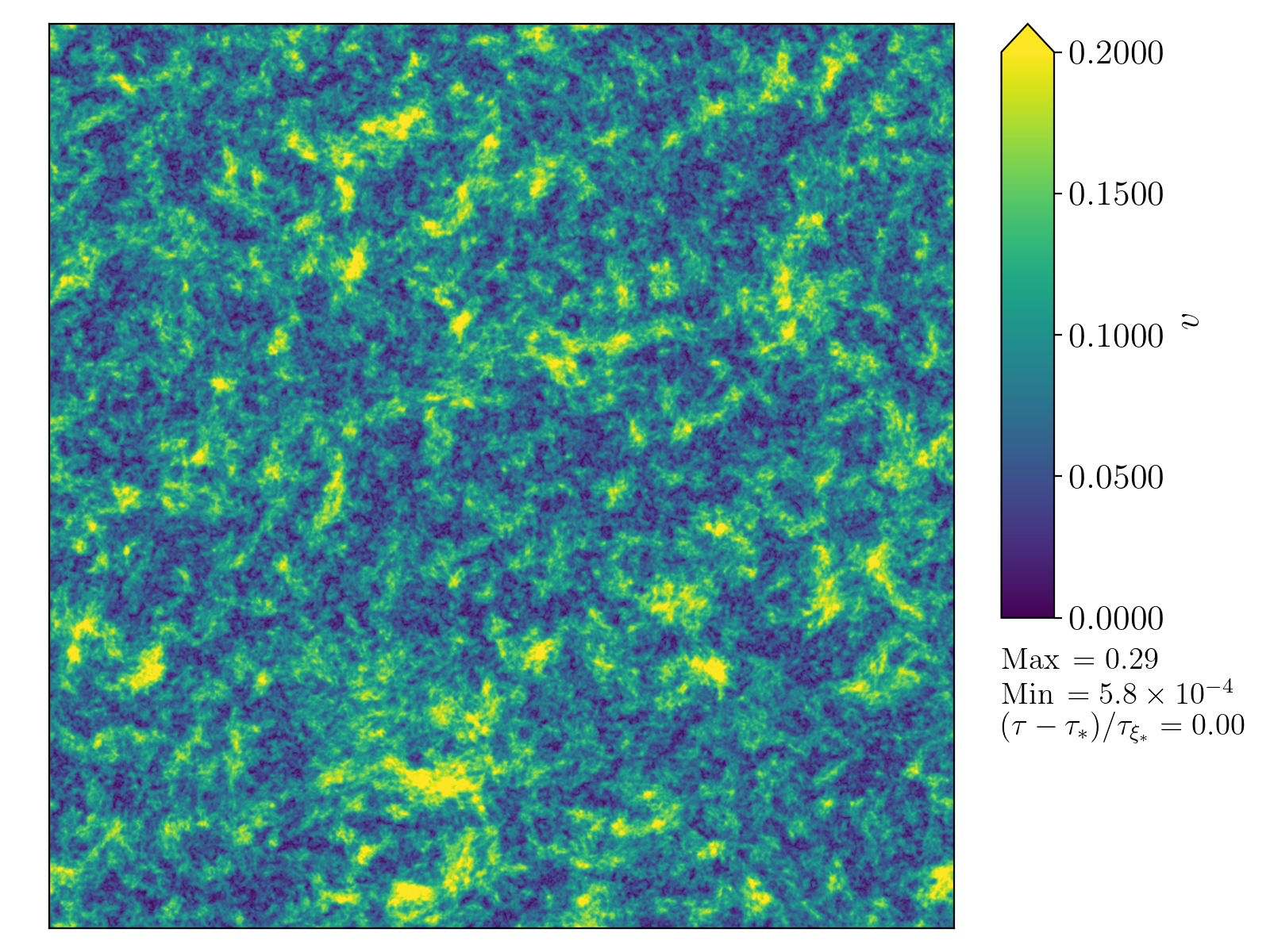}
	\includegraphics[width=0.49\textwidth]{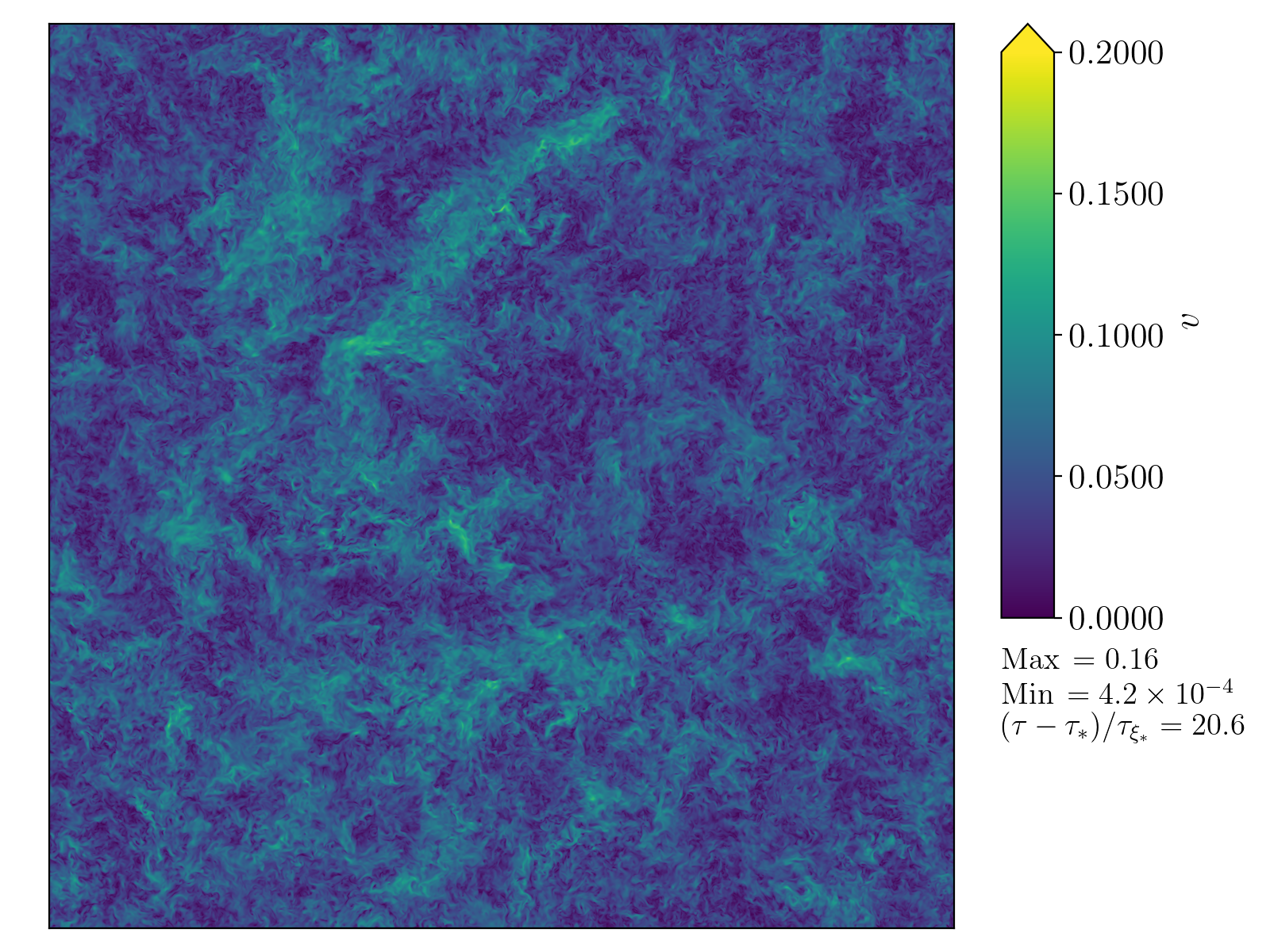}
	\caption{\emph{Left panel}: Slice through simulation ($\mathrm{A}'$) showing the velocity initial conditions in real space. \emph{Right panel}: Same slice as the left panel but
    after a time $\Delta \tau=20.6 \tauxist$ has elapsed.}
	\label{fig:simslices}
\end{figure}

\section{Simulations}
\label{sec:numerical}

We carry out a series of direct hydrodynamical simulations in order to
study the freely-decaying turbulent flow, in particular the UETCs of
the velocity field and its overall time evolution, as well as the SGWB signal produced.  For our
simulations, we use a modified version of the relativistic
hydrodynamics code SCOTTS, previously developed to study the coupled
evolution of the scalar field and the relativistic fluid, and the
production of GWs during a thermal phase
transition~\cite{Hindmarsh:2013xza,Hindmarsh:2015qta,Hindmarsh:2017gnf,Cutting:2019zws}.
Here we are mainly interested in the dynamics of the fluid after the
phase transition has completed. We therefore
initialize the scalar field to lie in the broken phase at the start of
the simulation. Given the form of the effective potential and equation
of state in Ref.~\cite{Cutting:2019zws}, this leaves us
solely with the evolution of the
relativistic fluid.

Ideally, one would want to simulate the coupled
field-fluid model~\cite{Enqvist:1991xw}, in order to observe the
generation of turbulence at all stages from the bubble collisions onwards~\cite{Cutting:2019zws}.
In this work, however, we do not analyse the
complete system up to fully developed turbulence.  We rather
concentrate on the
decay of turbulence,
inserting the turbulence
parameters (\eg kinetic energy and integral scale) as input values,
disconnected from the rest of the phase transition dynamics.
Here, we will initialize only the vortical component of the
fluid velocity.
This permits a cleaner analysis of the results and comparison to the
semi-analytical modelling of \cref{sec:free-decay-model,sec:PosKer}.
We defer studying the compressional component and
simulating the turbulence generation to future work. See Fig.~\ref{fig:simslices} for a visualisation of the velocity field in a slice through one of our simulations at initialisation and after many eddy turnover times have passed.

In the rest of this section, we specify the equations of motion of the fluid that are solved by the code, and then explain how we fix the initial conditions and evaluate the UETCs. Finally, we discuss how we compute GW power spectra from the simulations.

\subsection{Hydrodynamic evolution}
\label{sec:hydroeq}

Given that we are discarding the scalar field dynamics, the energy momentum tensor of the system becomes that of a perfect fluid,
\begin{equation}
	T^{\mu\nu}= (\epsilon + p) U^\mu U^\nu + g^{\mu\nu}p\text,
\end{equation}
with $\epsilon$ the internal energy density in the fluid, $p$ the
pressure, $U^\mu=\gamma(1,\vb{v})$ the fluid four-velocity and with
Lorentz factor $\gamma = 1/\sqrt{1-v^2}$.
The equation of state is
that of an ultrarelativistic gas with $p=\epsilon/3$.  It has been
shown that the hydrodynamic equations of motion of an
ultrarelativistic fluid in an expanding universe expressed in comoving
coordinates and conformal time are the same as the hydrodynamic
equations in Minkowski space-time, provided the dynamical quantities
are replaced by appropriately scaled
variables~\cite{Brandenburg:1996fc}.  However, the source term for the
GWs does not scale in the same way. Therefore, the Minkowski
space-time code SCOTTS
can be used to study hydrodynamic turbulence in an expanding background, but the output GW power spectrum can be matched to the expanding background result only for sources which are  short-lived compared with the Hubble time \cite{Hindmarsh:2015qta}.

The dynamical quantities that the code evolves are the fluid energy density $E=\gamma \epsilon$ with equation of motion~\cite{Hindmarsh:2015qta}
\begin{equation}
	\dot{E} + \partial_{i}(Ev^i) + p\left[\dot{\gamma} + \partial_i(\gamma v^i)\right] = 0\text,
\end{equation}
and the fluid momentum density $Z_i = \gamma^2(\epsilon + p)v_i$, the components of which evolve according to
\begin{equation}
	\dot{Z_i} + \partial_j(Z_i v^j) + \partial_i p = 0\text.
\end{equation}
The evolution algorithm follows the approach taken in
Ref.~\cite{WilsonMatthews} with a leapfrog and operator splitting
approach to updating the dynamical quantities.
For this work we
use a van Leer scheme for the advection
update~\cite{VANLEER1977276,Anninos:2002gz}, whereas earlier uses of
the SCOTTS code used an upwind donor cell scheme. For all of our simulations we pick the lattice
spacing and timestep such that their ratio is $\dd x/\dd \tau = 2.5$\footnote{In the code $\dd{x}$ takes the numeric value of $1$ for all the runs in this paper.}.  
We denote
the starting time of our simulations as $\tau = \tdevel$.
Note that while our simulations do not contain a physical viscosity,
the introduction of a lattice spacing $\dd x$ effectively sets a
dissipation scale. This is due to the numerical viscosity introduced
by the discretization of the fluid equations. Therefore,
simulations with larger $\xi_*/\dd x$ have a larger effective
Reynolds number.

\subsection{Initial conditions of the  simulation}
\label{sec:initial}
In our simulations the velocity field $\vb{v}$ is not necessarily divergence-free. However, as outlined in the rest of this section, we initialize the velocity field to be divergence-free in our simulations. In \cref{sec:appendix-sim-vrms-comp} we show that in this case the longitudinal component of the velocity field remains negligible throughout our simulations, and so we treat $\vb{v}$ as divergence-free in subsequent sections.

At the start of our simulations, we initialize the fluid energy density $E$ everywhere. For our simulations we choose to set $E\, \dd x^4 = 1/16$. Next we initialize the velocity field of the simulations in Fourier space. We do this indirectly by initializing the spatial components of the four-velocity $\vb{U}=\gamma \vb{v}$, such that
\begin{equation}
	\vb{U}_{\vb{x}} = \frac{1}{L^3}\sum_{\vb{k}} \vb{U}_{\vb{k}} \e^{i \vb{k} \cdot \vb{x}},
\end{equation}
where $\vb{x}$ and $\vb{k}$ refer to the discrete real and momentum space coordinates on the lattice, and $L^3$ is the total number of lattice sites.
Each mode $\vb{U}_{\vb{k}}$ is randomly distributed and follows Gaussian statistics with mean zero and variance determined by the desired power spectrum serving as initial condition.
Initializing $\vb{U}_{\vb{k}}$ rather than $\vb{v}_{\vb{k}}$ prevents individual modes of $\vb{v}_{\vb{k}}$ being drawn from the tails of the Gaussian with unphysical velocities larger than one.
The $\mathbf{U}_{\vb{k}}$ field is then projected onto its vortical component with the projector $\bot_{ij}(\vu{k})$, see \cref{eq:PSsim}.
Since $\vb{U}$ is real, we impose that $\vb{U}_{\vb{k}} = \vb{U}^* _{-\vb{k}}$.
Once the $\vb{U}_{\vb{k}}$ field is correctly initialized we perform a Fourier transform to find the real space configuration of the $\vb{U}$ field, and from this we can calculate the velocity field $\vb{v}$.

This method allows us to initialize the simulation with arbitrary choices for the $\vb{U}$ field power spectrum $\mathcal{P}_U(k,\tau) = {k^3} P_U(k,\tau)/{\pi^2}$.
Here $P_U(k,\tau)$ corresponds to the spectral density of the $\vb{U}$ field, defined via the equal time correlator (see~also \cref{eq:velocity-sd})
\begin{equation}
	\langle U_{i}(\vb{k},\tau) U_{j}^*(\vb{q},\tau) \rangle = (2\pi)^3 \bot_{ij}(\vu{k}) \delta(\vb{k}-\vb{q})P_{U}(k,\tau)\text,
\end{equation}
where in our initial conditions we have set the longitudinal component of $\mathbf{U}$ to zero.
We initialize the simulations with a relativistic version of the \vonK~spectrum \cref{eq:power-spectrum}, supplemented with a high-wavenumber cutoff:
\begin{equation} \label{eq:sim-input}
	\mathcal{P}_U(k,\tdevel) =  C \frac{(k/\tilde{k})^5}{[1 + (k/\tilde{k})^2]^{17/6}} \exp\left[-\left(\frac{k}{k_\mathrm{max}}\right)^2\right]\text.
\end{equation}
The two constants $\tilde{k}$ and $C$ determine respectively the peak location and amplitude.
The values of $C$ and $\tilde{k}$ are chosen so that the initial rms velocity $\vrmsst=\sqrt{ \vrms^2(\tdevel)}$ (defined in \cref{eq:kinetic-energy}) and integral scale $\xi_* = \xi(\tdevel)$ (defined in \cref{eq:integral-scale}) extracted from the simulation will be close to a desired value.
A list of parameters for the simulations we perform is given in \cref{tab:list}.
In \cref{eq:sim-input} we insert an exponential cutoff to ensure that the power does not extend to the lattice scale, with $k_\mathrm{max}\, \dd{x}=\pi/4$ used in our simulations.

\begin{table}
	\centering
	\resizebox{\textwidth}{!}{%
		\begin{tabular}{c D{.}{.}{-1} D{.}{.}{-1} D{.}{.}{-1} D{.}{.}{-1} D{.}{.}{-1} c c c c c c c}
			$L^3$                                           &
			\multicolumn{1}{c}{$\vrmsst$}                   &
			\multicolumn{1}{c}{$\xi_* / \dd{x}$}            &
			\multicolumn{1}{c}{$L \dd{x}/\xi_*$}            &
			\multicolumn{1}{c}{$(\tend-\tdevel)/\tauxist$}  &
			\multicolumn{1}{c}{$(\tuetc-\tdevel)/\tauxist$} &
			\multicolumn{1}{c}{$\deltuetc/\tauxist$}        &
			C                                               &
			$\tilde{k} \dd{x}$
			                                                & Label                                                                                            \\
			\hline
			$4096^3$                                        & 0.0988 & 29.8 & 138  & 99.6 & 5.97  & $1.08\,\times\, 10^{-2}$ & 1.99  & $9.19\times10^{-3}$   &
			(A)                                                                                                                                                \\
			\hline
			$2048^3$                                        & 0.0298 & 8.08 & 253  & 99.5 & 4.97  & $1.18\,\times\, 10^{-3}$ & 0.995 & $4.06\times10^{-2}$   &
			(B)                                                                                                                                                \\
			$2048^3$                                        & 0.0274 & 57.2 & 35.8 & 12.9 & 6.47  & $9.18\,\times\, 10^{-4}$ & 0.129 & $3.28\times10^{-3}$   &
			(C)                                                                                                                                                \\
			\cline{2-9}
			$2048^3$                                        & 0.0984 & 8.08 & 253  & 97.4 & 4.87  & $1.32\,\times\, 10^{-2}$ & 0.974 & $4.06\times10^{-2}$   &
			(D)                                                                                                                                                \\
			$2048^3$                                        & 0.0955 & 64.4 & 31.8 & 11.9 & 0.594 & $1.02\,\times\, 10^{-2}$ & 0.119 & $3.28\times10^{-3}$   &
			(E)                                                                                                                                                \\
			\cline{2-9}
			$2048^3$                                        & 0.279  & 8.11 & 253  & 103  & 4.13  & $1.18\,\times\, 10^{-1}$ & 1.03  & $4.06\times10^{-2}$   &
			(F)                                                                                                                                                \\
			$2048^3$                                        & 0.289  & 70.9 & 28.9 & 11.7 & 0.466 & $9.18\,\times\, 10^{-2}$ & 0.117 & $3.28\times10^{-3}$   &
			(G)                                                                                                                                                \\
			\hline
			$2048^3$                                        & 0.0974 & 28.4 & 72.0 & 103  & 6.17  & $1.08\,\times\, 10^{-2}$ & 2.06  & $9.19 \times 10^{-3}$ &
			($\mathrm{A}'$)                                                                                                                                    \\
			$1024^3$                                        & 0.0942 & 25.4 & 40.4 & 111  & 6.69  & $1.08\,\times\, 10^{-2}$ & 2.23  & $9.19 \times 10^{-3}$ &
			($\mathrm{A}''$)                                                                                                                                   \\
		\end{tabular}}
	\caption{List of the simulations used in this work. We list
		the number of sites in the lattice for each simulation
		$L^3$, the initial rms velocity $\vrmsst=\sqrt{
				\vrms^2(\tdevel)}$ (see \cref{eq:kinetic-energy}) and
		integral scale $\xi_*=\xi(\tdevel)$ (see
		\cref{eq:integral-scale}) extracted from the simulation
		on the initial timestep. The relative size of each
		simulation to the integral scale $L \dd{x}/\xi_*$ is
		given. We also list the final time of each simulation
		$\tend$, the UETC reference time $\tuetc$ and the UETC output interval $\deltuetc$, all given in
		units of the eddy turnover time at the integral scale
		$\tauxist = \xi_* / \vrmsst$. Finally we list the parameters
		for the simulation input power spectrum as given in
		\cref{eq:sim-input}. Simulations ($\mathrm{A}'$) and ($\mathrm{A}''$) are
		used to test the effects of finite volume and are discussed
		in \cref{sec:appendix-finite-vol}.}
	\label{tab:list}
\end{table}

\subsection{Velocity power spectra in the simulations}
\label{sec:sim-velps}
During the simulation we regularly output the rotational velocity power spectra $\Psv(k,\tau)$.
We validate that the longitudinal piece of the velocity correlator $\Psdv(k,\tau)$ is negligible over the full duration of all our simulations, providing it is initialized to be zero, see \cref{sec:appendix-sim-vrms-comp}.
To find the velocity power spectrum within our simulations, we first perform a discrete Fourier transform on the real space field to obtain the Fourier modes
\begin{equation}
	\vb{v}_{\vb{k}} = \sum_{\vb{x}} \vb{v}_{\vb{x}} \e^{- i \vb{x} \cdot \vb{k}}\text.
\end{equation}
The rotational power spectrum is then approximated by projecting and binning these modes in momentum space.
The power spectrum for the $r$-th bin is constructed via
\begin{equation}
	\Psv(k_r,\tau) = \frac{1}{L^6} \frac{k_r}{\Delta k} \sum_{r \leq \tfrac{|\vb{k}|}{\Delta k}<r+1} \bot_{ij}(\mathbf{\hat k}) v^i_{\vb{k}}(\tau) {v}^{j*}_{\vb{k}}(\tau)\text,
\end{equation}
where $k_r = 2\pi(r+1/2)/L \dd x$ is the central value of the bin in $k$, and $\Delta k = 2\pi/L \dd x$ is the bin width. Note that the power spectra from the simulation are formed from an average over a spherical shell in $\mathbf{k}$ of Fourier modes.

\subsection{Unequal time correlators in the simulations}
\label{sec:sim-uetc}

We are also interested in studying the UETCs of the velocity field in these simulations as these enable us to model the GW source, see \cref{eq:sbgw,eq:unequal-stress-final}.
However, evaluating $\UETCv(k,\tau, \zeta)$ for many values of $\tau$ and $\zeta$ is very costly within a  simulation, in part because each pair of $\tau$ and $\zeta$ represents snapshots of the field $\vb{v}$ that must be stored concurrently in memory.
Instead, we define a reference time $\tuetc$ at which we store in memory the field $\vb{v}(\vb{k},\tuetc)$, and then compute the UETC at regular intervals with the field on the current timestep, \ie
\begin{equation}
	\ev{ v_{i}(\vb{k},\tau) v_j^*(\vb{q},\tuetc) }= (2\pi)^3 \bot_{ij}(\vu{k}) \delta(\vb{k}-\vb{q})\UETCv(k,\tau, \tuetc).
\end{equation}
We evaluate the unequal-time correlator at times $\tau = \tuetc + n\deltuetc$, where $\deltuetc$ is the interval between UETC outputs (see \cref{tab:list}) and $n$ is some positive integer.
Using a single value of $\tuetc$ per simulation is in principle a limitation of our analysis.
Nonetheless, we are confident that the latter still captures the general behaviour of the time decorrelation of the turbulent field, especially given the good agreement we find between the simulations results and the theoretical model, see~\cref{sec:results-vel}.

\subsection{Gravitational waves}
\label{sec:gw-numerical}

We also output GW power spectra calculated from our simulations.
As SCOTTS is a Minkowski space code, we neglect cosmological expansion in the following.
To find the metric perturbations $h_{ij}$, we evolve an auxiliary tensor $u_{ij}$ according to \Refa{Garcia-Bellido:2007fiu} (see~also \cref{eq:gw-propagation})
\begin{equation} \label{eq:eom-metric}
	\ddot{u}_{ij} - \nabla^2 u_{ij} = 8 \pi T_{ij}\text.
\end{equation}
The metric perturbations can then be recovered from our auxiliary tensor by performing a projection in momentum space,
\begin{equation}
	h_{ij}(\mathbf{k},\tau) = G \Lambda_{ij\ell m}(\vu{k}) u_{\ell m}(\mathbf{k},\tau)
\end{equation}
where $\Lambda_{ij,\ell m}(\mathbf{k})$ is the projector onto transverse-traceless rank two tensors, \cref{eq:lambda_proj}.
We can then define the power spectral density of the metric perturbations, $P_{\dot{h}} (k,\tau)$ via
\begin{equation}
	\langle\dot{h}_{ij}(\mathbf{k},\tau) \dot{h}_{ij}(\mathbf{q},\tau) \rangle = G^2 \langle \Lambda_{ijk\ell }(\vu{k}) \Lambda_{ijmn}(\vu{q}) \dot{u}_{k\ell}(\mathbf{k}) \dot{u}_{mn}(\mathbf{q})\rangle = (2\pi)^3\delta(\mathbf{k}-\mathbf{q}) P_{\dot{h}}(k,\tau)\text,
\end{equation}
and from this construct the power spectrum of the metric perturbations,
\begin{equation}
	\mathcal{P}_{\dot{h}}(k,\tau) = \frac{k^3}{2\pi^2} P_{\dot{h}} (k,\tau)\text.
\end{equation}
From $\mathcal{P}_{\dot{h}}(k,\tau)$ and a suitable scaling, one can then reconstruct the gravitational wave power spectrum (see~\cref{eq:OmGW})
\begin{equation}
	\left.\frac{\dd \Omega_\mathrm{gw}}{\dd \mathrm{ln}\, k}\right|_\tau =  \frac{\mathcal{P}_{\dot{h}}(k,\tau)}{8\pi G \rho_*}\text,
\end{equation}
where $\rho_*$ is the conserved average total energy density (note that we insert a $*$ subscript to make the connection with the expanding Universe quantity, see~\cref{eq:Hstarsim}).

Dimensionful quantities in our simulations are always output normalized according to a corresponding length scale, which we have chosen to be the initial integral scale $\xi_*$ (defined later in \cref{eq:integral-scale}).
Noting that the gravitational constant $G$ has been scaled out of our equation of motion in \cref{eq:eom-metric}, the output from our code is
\begin{equation}
	O_\mathrm{code}(k,\tau) = \frac{\xi_*^2}{8\pi \rho_*}  \frac{\mathcal{P}_{\dot{h}}(k,\tau)}{G^2}
	=\frac{\xi_*^2}{G} \left.\frac{\dd \Omega_\mathrm{gw}}{\dd \mathrm{ln} \,k}\right|_\tau
	\text.
\end{equation}
In this expression, the power spectra of the metric perturbations in the simulations are binned and constructed analogously to the velocity power spectra,
\begin{equation}
	\frac{\mathcal{P}_{\dot{h}}(k_r,\tau)}{G^2} = \frac{1}{L^6} \frac{k_r}{\Delta k} \sum_{r \leq \tfrac{|\vb{k}|}{\Delta k}<r+1} \Lambda_{ij\ell m}(\mathbf{\hat k}) \dot{u}^{ij}_{\vb{k}}(\tau) \dot{u}^{\ell m*}_{\vb{k}}(\tau)\text. \label{eq:metric-pspec-dns}
\end{equation}
The GW power spectrum is thus recoverable from the simulations by fixing a value for $\xi_*^2/G$.

While the GW evaluation is carried out in flat space-time,
we can still infer a value for the Hubble rate via the Friedmann equation
\begin{equation}
	\mathcal{H}_*^2 = \frac{8 \pi G \rho_*}{3}\text.\label{eq:Hstarsim}
\end{equation}
This allows us to rewrite the scaling for the gravitational wave power spectrum as (note that $\rho_*$ is a conserved quantity throughout the simulation)
\begin{equation}\label{eq:OmGWsimul}
	\frac{1}{(\mathcal{H}_*\xi_*)^2} \left.\frac{\dd \Omega_\mathrm{gw}}{\dd \mathrm{ln} \,k}\right|_\tau = \frac{3}{8\pi \rho_* \xi_*^4}O_\mathrm{code}(k,\tau)\text.
\end{equation}
The SGWB output of the simulations will be compared with the semi-analytic results in \cref{sec:compamethods}.
Since the simulations neglect the expansion of the Universe, one can only compare the spectral region $k>\mathcal{H}_*$.
The wavenumbers accessible in the simulations satisfy $k\xi_*>2\pi\xi_*/(L \dd{x}) > 0.03$ (see \cref{tab:list}), and
we will therefore compare with semi-analytical results calculated setting  $\mathcal{H}_*\xi_*=0.001$ (see \cref{sec:numintegration}).


\section{Results: evolution of the velocity field}
\label{sec:results-vel}

In this section, we verify the model of decaying turbulence presented in \cref{sec:free-decay-model} with the hydrodynamic simulations.
We compare the time evolution laws of the turbulent kinetic energy and correlation scale  with the simulation results in \cref{sec:ResETC}.
Then in \cref{sec:results-uetc}, we validate the  symmetrized velocity field decorrelation function constructed in  \cref{sec:UETCvelocity,sec:PosKer}, and thereby show that it can also describe decorrelation at large scales, outside the inertial range.
Note that the turbulence model we adopted has been developed in the context of non-relativistic turbulence.
The highest value of the initial rms velocity that we have tested in the simulations is $\vrmsst \simeq 0.3$ (see~\cref{tab:list}), finding good agreement with the predictions of the non-relativistic turbulence modelling, as far as the UETC, free decay evolution and SGWB spectra are concerned.

\subsection{Velocity power spectrum and averaged quantities}
\label{sec:ResETC}

\begin{figure}
	\centering \subfloat[Simulation (A), unrescaled velocity power
		spectrum.\label{fig:unrescaled}]{
		\includegraphics[width=0.45\textwidth]{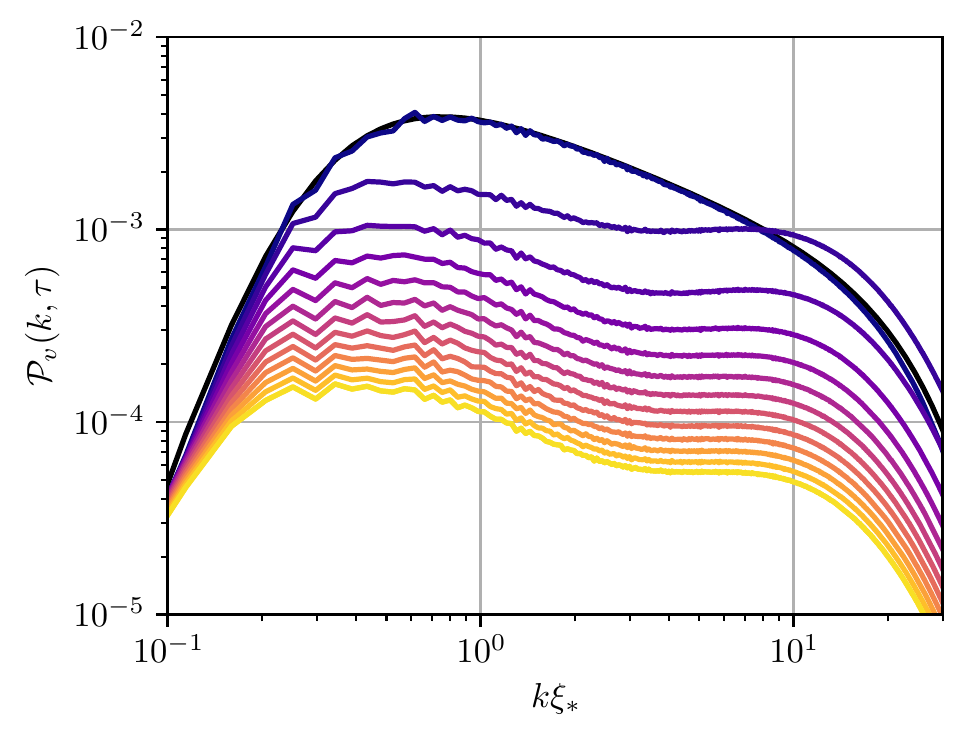}}
	\subfloat[Power spectrum rescaled with \cref{eq:pv_scaled_one_scale}.\label{fig:rescaledonescale}]{
		\includegraphics[width=0.465\textwidth]{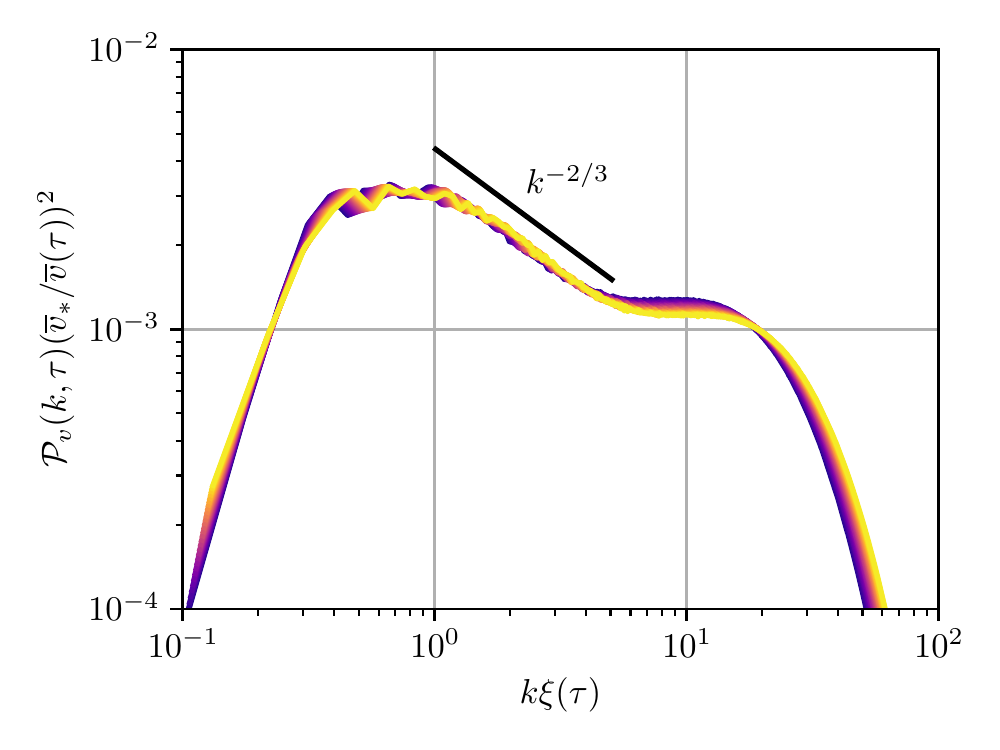}}
	\caption{Evolution of the velocity power spectrum from simulation
		(A). \emph{Left panel}: unrescaled power
		spectrum. The solid black line is the
		initial condition, \cref{eq:power-spectrum} and coloured lines
		show the time evolution from $\tau=\tdevel$ up to $\tau = \tend$
		with interval $\Delta\tau = 9.96\, \tauxist$. Lighter colours refer to later times.
		\emph{Right panel}:
                power spectrum rescaled according to \cref{eq:pv_scaled_one_scale}. The curves are plotted starting from $\tau-\tdevel\simeq 40\,\tauxist$, at fixed intervals $\Delta \tau = 1.99 \tauxist$.
		\label{fig:Pv_and_first_rescale}}
\end{figure}

Here we use the results from  simulation (A) (\cref{tab:list}) to investigate the time evolution of the velocity power spectrum, and the values of the parameters $\beta$, $p$ and $q$, introduced in \cref{sec:free-decay-evol}.

In \cref{fig:Pv_and_first_rescale} we show the time evolution of the velocity power spectrum.  \cref{fig:unrescaled} shows the velocity power spectrum in its raw form, whilst \cref{fig:rescaledonescale} shows the power spectrum rescaled with \cref{eq:pv_scaled_one_scale}, indicating that there is indeed only one principal length scale in the flow, given our initial condition.
The self-similarity is established starting from about $40 \tauxist$, the time to which the first power spectrum line in \cref{fig:rescaledonescale} corresponds.
Note that the large-scale slope of the late-time power spectrum is significantly less steep than the $k^5$ corresponding to the Batchelor power spectrum imposed in the initial condition  \cref{eq:power-spectrum}: therefore, the large-scale power spectrum is not constant in time in the simulation, before about $40 \tauxist$.
While the large-scale slope could be subject to finite volume effects, its constancy in time seems to be well established at late times.

Turning to the form of the  velocity power spectrum at wavenumbers higher than the peak,  we see from \cref{fig:Pv_and_first_rescale}
evidence for a power law in the range $1 \lesssim k\xi_* \lesssim 5 $.
This power law is approximately consistent with a Kolmogorov power spectrum ($k^{-2/3}$) at late times.
Although this was also the power spectrum in the initial conditions, one can see that this power law is lost in the early evolution,
but later regained.
We do not have the inertial range
to give a firm value for the late-time power law index.
For wavenumbers above $k\xi_* \gtrsim 5 $, we see a small build up in power followed by a sharp fall-off for $k\xi_* \gtrsim 10$.
The fall-off is set by the viscous damping scale of the numerical scheme, while the build up seems to be an associated feature of the van Leer advection scheme\footnote{The upwind donor cell advection scheme previously used was associated with a fall-off in power at smaller $k\xi_*$ consistent with it being a lower order advection scheme.}.

\begin{figure}
	\centering
	\subfloat[Evolution of $\vrms^2 \xi^{1+\beta}$. \label{fig:vrms-xi-beta}]{\includegraphics[width=0.45\textwidth]{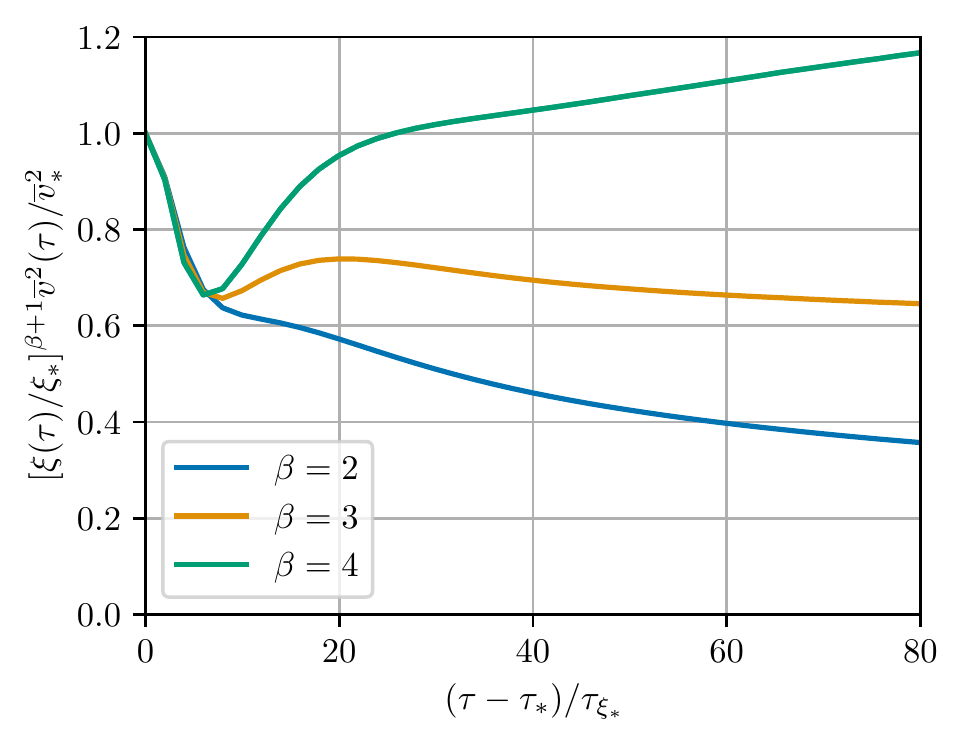}}
	\subfloat[Power spectrum rescaled with \cref{eq:Pvxi}, assuming
		$\beta=3$.\label{fig:rescaledbeta3}]{
		\includegraphics[width=0.465\textwidth]{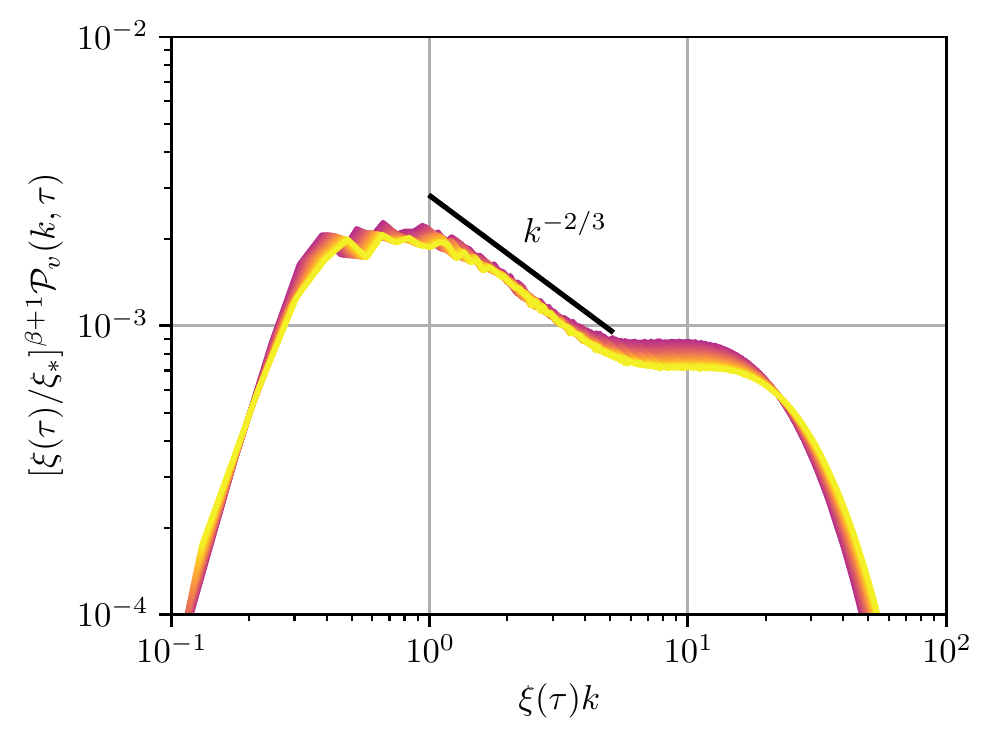}}
	\caption{
		\emph{Left panel}: evolution of $\vrms^2 \xi^{1+\beta}$ for different values of $\beta$ in simulation (A). As shown in \cref{eq:vxigeneral}, this quantity is expected to remain constant, thus indicating that $\beta \simeq 3$.
		\emph{Right panel}:
		Power spectrum rescaled with
		\cref{eq:Pvxi}, using $\beta=3$.
		The power spectra are plotted from $\tau-\tdevel\simeq 40\,\tauxist$ onward, at fixed interval $\Delta \tau = 1.99 \tauxist$.
		\label{fig:collapse}}
	\label{fig:beta3}
\end{figure}

In \cref{fig:vrms-xi-beta} we plot the
combination $\vrms^2 \xi^{1+\beta}$, predicted to be constant when the large-scale power spectrum is constant as well, see \cref{eq:vxigeneral}.
It can be seen that this combination indeed
remains approximately constant after about $40\tauxist$, provided $\beta=3$.
Consequently, in \cref{fig:rescaledbeta3} we analyse again the self similarity of the power spectrum, this time rescaled according to  \cref{eq:Pvxi} setting $\beta=3$. Again, we only plot the rescaled spectra for $\tau-\tdevel\gtrsim 40\,\tauxist$.
It appears that the self-similarity condition  \cref{eq:Pvxi} is satisfied by the simulation with $\beta=3$.
This result is consistent with the findings of \Refa{Brandenburg:2016odr}.
However, the large-scale power spectrum is less steep than $k^4$, which is the power law which would be inferred setting $\beta=3$.

Simulation (A), therefore, seems broadly consistent with the self-similar behaviour outlined in \cref{eq:Pvxi}, at late times.
We have seen that $\beta$ is not set by the slope of the power spectrum at low wavenumbers in the initial condition; rather, it seems to be part of the dynamics of self-similarity. We would need even larger lattices
to study in detail the evolution of the power spectrum at low wavenumbers,
where there appear to be correlations established at scales greater than the integral scale.
Limitations of the size of the computational domain are known to possibly affect the interpretation of the turbulent decay features \cite{doi:10.1063/1.4901448}.

\begin{figure}
	\centering
	\includegraphics[width=0.47\textwidth]{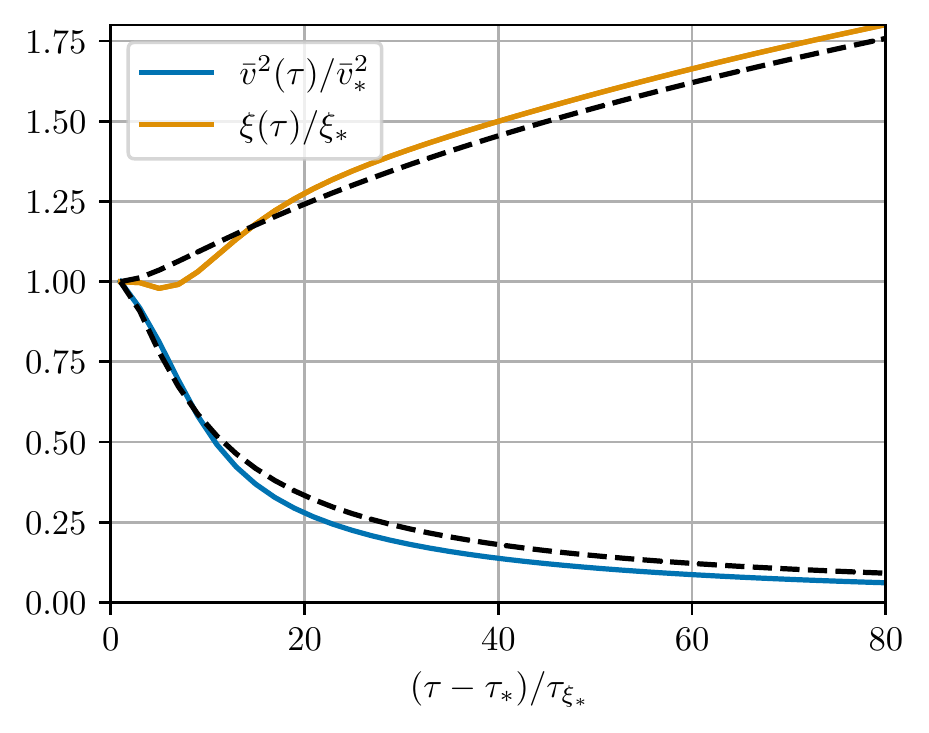}
	\caption{Evolution of the velocity and integral scale in simulation (A).
	  The black-dashed line showcases \cref{eq:VelEvMod,eq:XiEvMod} for $\ndecay=5$, $p=4/3$ and $q=1/3$, where the values of $p$ and $q$ correspond to $\beta=3$; see the relationships in \cref{eq:pchialpha,eq:qchialpha}.
		\label{fig:xi-vrms-evol}}
\end{figure}

In \cref{fig:xi-vrms-evol} we show the ratios
$\vrms^2(\tau)/\vrms_*^2$ and $\xi(\tau)/\xi_*$  extracted from simulation (A), where $\vrms^2(\tau)$ and $\xi(\tau)$ are calculated by integrating the velocity power spectra of \cref{fig:unrescaled}, according to \cref{eq:kinetic-energy,eq:integral-scale}.
We also show the predicted evolution according to \cref{eq:VelEvMod,eq:XiEvMod} with $\ndecay = 5$ (black, dashed lines), where this specific value has been chosen to match the simulation result.
We fix the decay exponents in \cref{eq:VelEvMod,eq:XiEvMod} to $p=4/3$ for the kinetic energy and $q=1/3$ for the integral scale. These exponent values are provided by the condition $\beta=3$ from \cref{eq:pchialpha,eq:qchialpha} and provide a good fit to the simulation outcome.
We shall see later in \cref{sec:numintegration} that the precise values of the decay exponents do not have a significant effect on the final SGWB spectrum.
Instead, we will see in \cref{sec:GWcontinuous} that the initial phase preceding the generation of the turbulence plays a much larger role in determining the spectral shape.

\begin{figure}
	\centering
	\subfloat[Instantaneous exponents $(p,q)$.\label{fig:pq-diagram}]{\includegraphics[width=0.49\textwidth]{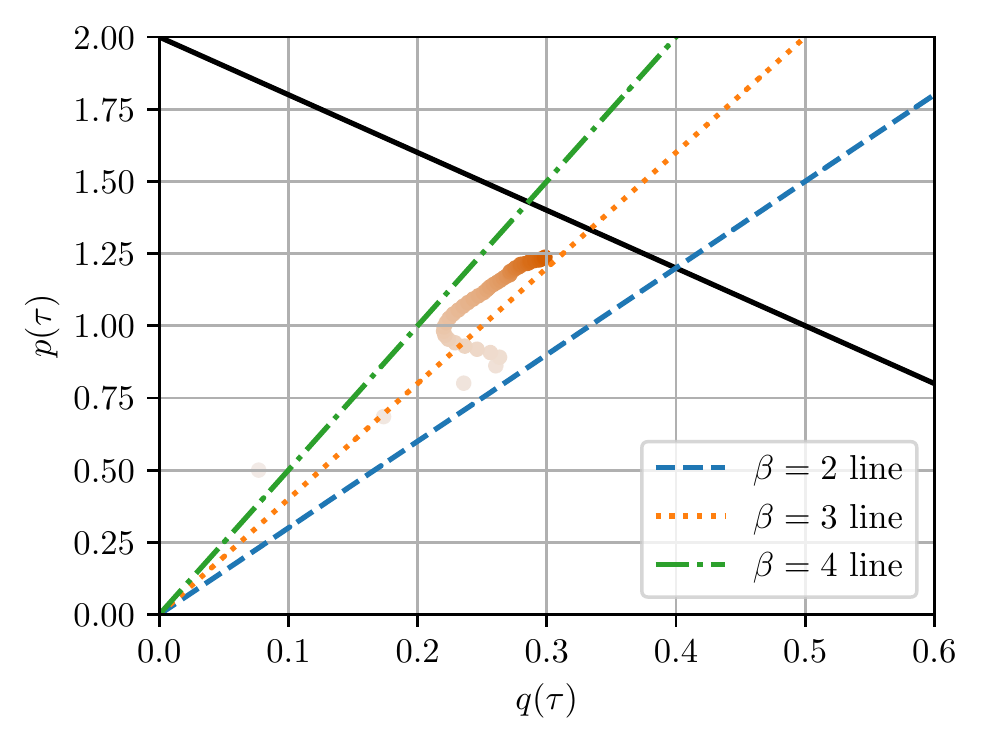}}
	\subfloat[Time evolution $p(\tau)$, $q(\tau)$.\label{fig:p-q-evol}]{\includegraphics[width=0.46\textwidth]{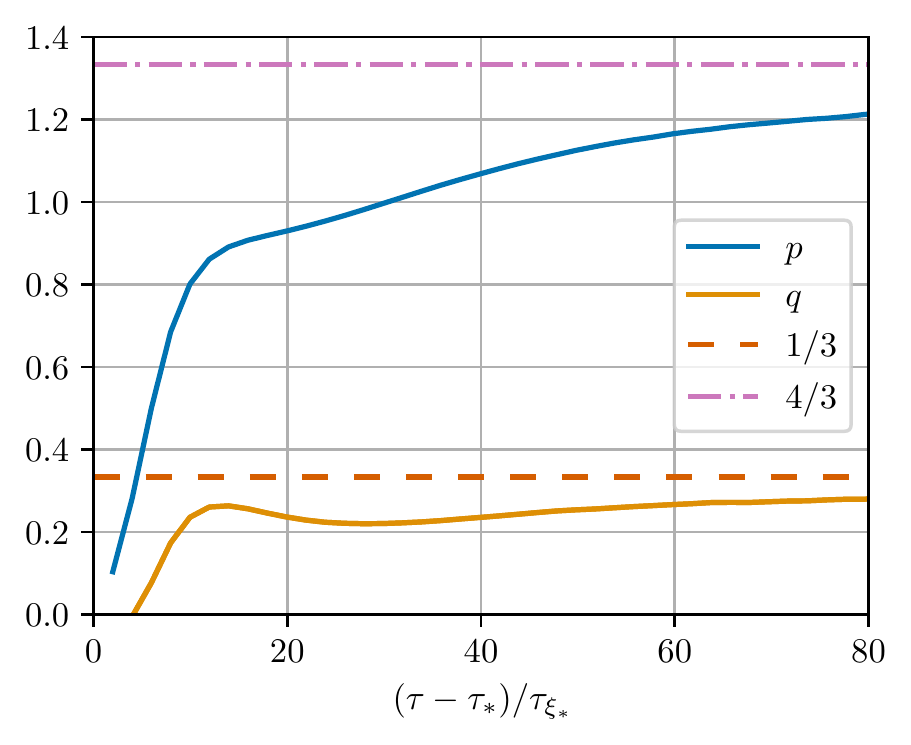}}
	\caption{\emph{Left panel}: Trajectory of the instantaneous exponents $(p,q)$ in simulation (A). Time is represented by the colour scheme: early times are shown with lighter shades and late times with darker shades, starting at $\tau=\tdevel$ and with interval $\Delta \tau \simeq 2 \tauxist$ between markers.
		The dark solid line represents the scale-invariance line $p=2(1-q)$, \cref{eq:pVsq}. The coloured dashed, dotted and dash-dotted lines show the self-similarity relations $p=(1+\beta)q$ for various choices of $\beta$. \emph{Right panel}: Evolution of the instantaneous kinetic energy and integral scale exponents $(p, q)$ as a function of time in simulation (A). The dash-dotted line shows the values expected for $p$ and long dashed line for $q$, with $\beta = 3$.}
	\label{fig:pq-explore}
\end{figure}

We further analyse the behaviour of the power law exponents $p$ and $q$ of the kinetic energy and integral scale in \cref{fig:pq-explore}.
In \cref{fig:pq-diagram}, we adopt the same approach as \Refa{Brandenburg:2016odr} and plot the state of the system in $(p,q)$ space
for a range of times during the simulation.
The values of $p$ and $q$ are found by solving \cref{eq:VelEvMod,eq:XiEvMod} for $p$ and $q$ in terms of the measured values of $\vrms^2(\tau)$, \eqref{eq:kinetic-energy}, and $\xi(\tau)$, \eqref{eq:integral-scale}.
The scale-invariance line $p = 2(1 - q)$, derived in \cref{eq:pVsq}, is represented by the black line in \cref{fig:pq-diagram}.
The self-similarity lines $p=(1+\beta)q$ are shown for various values of $\beta$ as dashed lines.
As in clear from the figure, the point $(p,q)$ is still evolving at the end of the simulation.
However, the evolution is moving in the direction of the intersection of the self-similarity line $p = (1+\beta)q$ with $\beta=3$, and the scale invariance line, $p =2(1 -q)$.
In \cref{fig:p-q-evol}, we display the evolution of $p$ and $q$
against time, and also show the values that can be predicted from requiring scale-invariance and self-similarity for $\beta = 3$, namely $p = 4/3$ and $q =1/3$.

\subsection{Unequal time velocity correlations}
\label{sec:results-uetc}

In this subsection we report on our results for the decorrelation of the velocity field at different times. We introduce a new model for the unequal time correlator and test it against the numerical simulations.

Our model for the normalized UETC is
obtained by combining \cref{eq:expvsweep,eq:Vlarge,eq:vsweepcomplete}
giving
\begin{equation}
	\tdec_\mathrm{A+}(k,\tau,\zeta) =
	\frac{\vdc(k,\tau,\zeta)}{\sqrt{\vdc(k, \tau) \vdc(k, \zeta)}}
	\exp\qty[-\frac{1}{2} k^2
		\vdc^{2}(k,\tau,\zeta)(\tau-\zeta)^2].
\end{equation}
The factor in front of the exponential guarantees that the shear stress UETC is a positive definite kernel and hence that the resulting gravitational wave power spectrum is positive definite, as explained in \cref{sec:mercer}.
This is to be compared with the model of \Refa{Niksa:2018ofa},
\begin{equation}
	\label{eq:RNSSdef}
	\tdec_\mathrm{NSS}(k,\tau,\zeta) =
	\exp\qty[-\frac{1}{2} k^2
		\vdc^{2}(k,\zeta)(\tau-\zeta)^2], \quad \tau > \zeta \, .
\end{equation}

\begin{figure}
	\centering
	\subfloat[Model of \Refa{kaneda_lagrangian_1993}. ]{\includegraphics[width=0.48\textwidth]{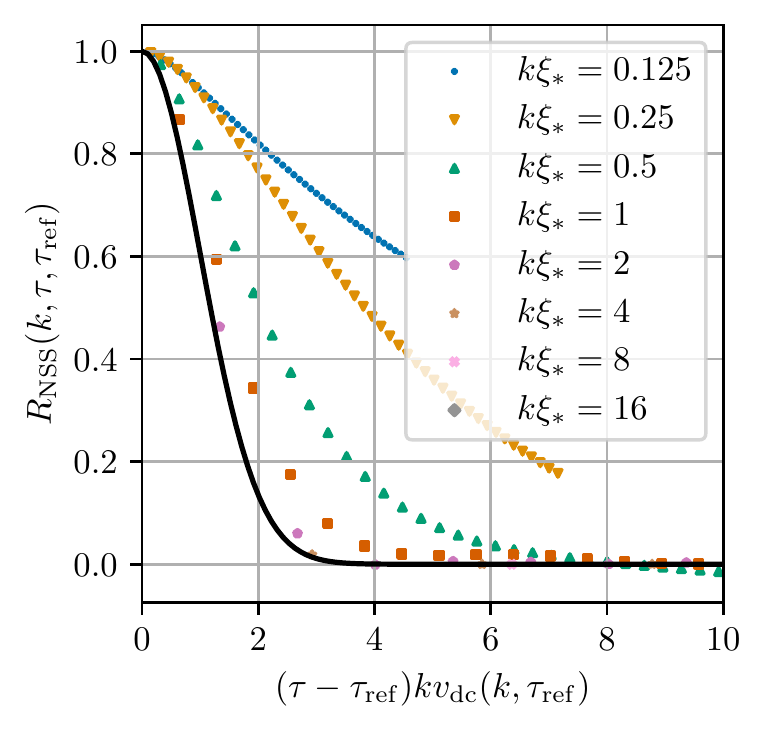}}
	\subfloat[Our model. ]{\includegraphics[width=0.48\textwidth]{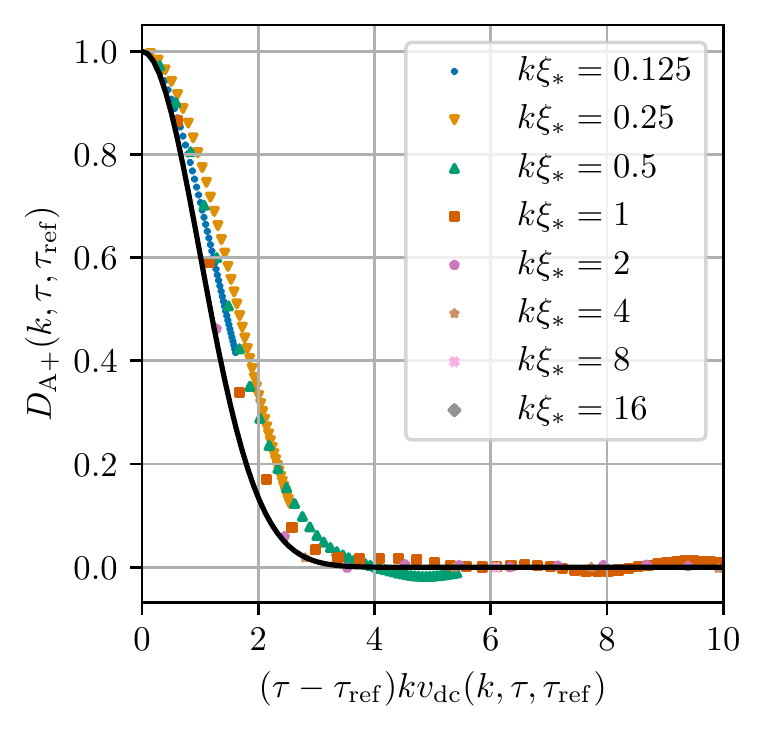}}
	\caption{Comparison of normalized unequal time correlator models with data from simulation (A): on the left, the model of \Refa{kaneda_lagrangian_1993} as used in \Refa{Niksa:2018ofa}, on the right our model. The $y$-axis displays the combination of data and model functions which should produce a Gaussian
	curve in the argument of the $x$-axis;
	this is discussed in further detail in connection with \cref{eq:Rfigure2b}.
	The solid dark line is the prediction of the model in each case.}
	\label{fig:decorr}
\end{figure}

In \cref{fig:decorr},  we compare the two models to data from simulation (A), by
plotting combinations of data and model functions which should be of Gaussian form.
Results from the other simulations are given in \cref{sec:appD}.
The construction of the numerical UETC, $\UETCv(k,\tau,\zeta)$
is described in \cref{sec:sim-uetc}; in our simulations we measure the correlation of the velocity field at times $\zeta = \tuetc$ and $\tau$.

For our model, the data is plotted as the quantity
\begin{equation}
	\tgauss(k, \tau, \tuetc) \equiv \frac{\UETCv(k,\tau,\tuetc)}{\sqrt{\Psdv(k,\tau)\,\Psdv(k,\tuetc)}}  \sqrt{\frac{\vdc^2(k, \tau) + \vdc^2(k, \tuetc)}{2 \vdc(k, \tau) \vdc(k, \tuetc) }},
	\label{eq:Rfigure2b}
\end{equation}
against the combination $k\,\vdc(k,\tau,\tuetc)(\tau-\tuetc)$,
while for the model of \Refa{Niksa:2018ofa}, we plot the normalized velocity UETC \eqref{eq:NorVelUETC}
as a function of $k\,\vdc(k,\tuetc)(\tau-\tuetc)$. We choose a range of values for the wavenumber $k$, both larger and smaller than the inverse integral scale $\xi_*$.
The reference time is taken to be approximately one initial eddy turnover time, $\tuetc \simeq \tauxist = \xi_*/\vrmsst$ (see \cref{tab:list}).
The success of the models can be judged by how well all the data from simulations collapses onto a Gaussian in the $x$-axis argument,
indicated by the black lines.

From \cref{fig:decorr}
it appears that both models work well in the inertial range $k\xi_* \gg 1$.
This demonstrates that
the sweeping model, with sweeping velocity $\vrms/\sqrt{3}$,
characterizes very well the decorrelation of the system at high wavenumbers, and that we may take $C_v^2 = 1$ in \cref{eq:Cv2}.
This also indicates that models using the scale-dependent Lagrangian eddy turn-over time
$v^2_\mathrm{dc}(k,\tuetc) \propto \Psv(k, \tuetc)$
in the inertial range \cite{Kamionkowski:1993fg,Gogoberidze:2007an}
give a poor fit in the inertial range.

On the other hand, the decorrelation model based on \cref{eq:RNSSdef} degrades somewhat for $ 0.125\leq k\xi_*\leq 1$.
Our decorrelation model is a significant improvement on these scales, which we ascribe partly to the fact that the average decorrelation velocity takes into account
the slowing of the decorrelation with the decay of the kinetic energy.

Because of limitations in the dynamical range of the simulations, it is not possible to analyse very small values of $k\xi_*$.
However, we will see that the GW power spectrum decreases at small $k\xi_*$, and so it is less important to model accurately the UETC at wavenumbers well below the peak.

Note that according to recent work in \Refa{Gorbunova:2021cpn}, the Gaussian form of the random sweeping hypothesis should change to an exponential form
at large time differences.
Again, this effect should be most influential at low $k$, below the peak of the power spectrum. We leave further study of UETC models for future work.

\section{Results: the gravitational wave spectrum}
\label{sec:sgwbresults}

This section is organized as follows.
In \cref{sec:stationary} we study a stationary source, meaning that we assume that $\UETCv(p,\tau,\zeta)$ only depends on the time difference $\tau-\zeta$.
This is not the correct model for freely decaying turbulence, but we consider it nonetheless since it is useful to illustrate some properties of the GW power spectrum of a freely decaying source.

We present the case of freely decaying turbulence in the subsequent \cref{sec:gw-instant,sec:GWcontinuous}.
\cref{sec:gw-instant} contains the main results of our work, and is dedicated to the case of instantaneously generated turbulence, meaning that we insert as initial conditions a fully developed turbulent spectrum.
Finally, in \cref{sec:GWcontinuous}, we go beyond the  instantaneous generation scenario, and include a growth phase for the turbulence kinetic energy.

Before continuing, we first set the basis for the calculation of the stochastic GW background that will follow.
Combining \cref{eq:sbgw,eq:unequal-stress-final}, one obtains (recall that $\vb{q} = \vb{k} - \vb{p}$)
\begin{equation}
	\eval{\dv{{\Omega}_\mathrm{gw}}{\ln k}}_{\tau} =
	\frac{k^3}{3 \pi^5}
	\int \dd[3]{p}
	\qty[1 + (\vu{k}\vdot \vu{p} )^2] \qty[1 + (\vu{k}\vdot \vu{q})^2]
	\iint_{\tini}^{\tau} \mathcal{G}(k,\tau,\eta,\zeta)
	\UETCv(p, \eta, \zeta) \UETCv(q, \eta, \zeta)
	\frac{\dd{\eta}}{\eta}\frac{\dd{\zeta}}{\zeta}.
	\label{eq:OmPv}
\end{equation}
From this expression, it appears that the computation of the GW background involves performing a five-dimensional integration for each mode $k$.
Integrating over the azimuthal angle of $\vb{p}$ immediately gives a factor of $2\pi$ and one is left with a four-dimensional integration.
Technical details on how to tackle this four-dimensional integration are presented in \cref{sec:4dinteg}.
In particular, we advocate the change of variables $\vb{p}=\vb{k}/2 + \vb{h}$ (and hence $\vb{q} = \vb{k}/2 - \vb{h}$): this makes the symmetry around the zero of the cosine of the declination angle $\alpha = \hat{h}\cdot\hat{k}$ explicit.

After the complete decay of turbulence, occurring at a time that we denote $\tfin$, the GW power spectrum in the radiation era evolves as
\begin{equation}
	\eval{\dv{\Omega_\mathrm{gw}}{\ln k}}_{\tau\geq \tfin} = \frac{8}{3 \pi^2}
	k^3
	\iint_{\tini}^{\tfin} \mathcal{G}(k,\tau, \eta,\zeta) P_{\tilde{\Pi}}(k, \eta,\zeta)\frac{\dd{\eta}}{\eta} \frac{\dd{\zeta}}{\zeta}\label{eq:gwradera}.
\end{equation}
This can be obtained by matching the sourced and free solutions of \cref{eq:gw-propagation} at time $\tfin$.
Note that the function $\mathcal{G}$ of \cref{eq:green_with_all_terms} depends explicitly on $\tau$, meaning that the GW spectrum in principle still depends on time after the source has decayed.
We are interested in the GW spectrum at a time $\tau$ in the radiation era long after the source has decayed, and such that all relevant wavenumbers are inside the horizon, $k\tau>1$.
This way we can neglect the second and third terms in \cref{eq:green_with_all_terms}, proportional to $1/(k\tau)$ and $1/(k\tau)^2$, since they become subdominant with respect to the first one.
Furthermore, we can expand the first term of \cref{eq:green_with_all_terms} to obtain
\begin{align}
	\eval{\dv{\Omega_\mathrm{gw}}{\ln k}}_{\tau\geq \tfin}
	=  \, & \frac{4}{3 \pi^2} k^3 \iint_{\tini}^{\tfin} \cos k(\eta - \zeta) P_{\tilde{\Pi}}(k, \eta,\zeta)\frac{\dd{\eta}}{\eta} \frac{\dd{\zeta}}{\zeta}  \nonumber                                                     \\
	      & + \frac{4}{3 \pi^2} k^3 \cos(2k\tau) \iint_{\tini}^{\tfin} \sin k(\zeta + \eta) P_{\tilde{\Pi}}(k, \eta,\zeta)\frac{\dd{\eta}}{\eta} \frac{\dd{\zeta}}{\zeta} \nonumber                                       \\
	      & + \frac{4}{3 \pi^2} k^3 \sin(2k\tau) \iint_{\tini}^{\tfin} \cos k(\zeta + \eta ) P_{\tilde{\Pi}}(k, \eta,\zeta)\frac{\dd{\eta}}{\eta} \frac{\dd{\zeta}}{\zeta} + \order{(k\tau)^{-1}} \label{eq:green-approx}
\end{align}
The residual time dependence consists of irrelevant, rapid oscillations in time.
We can therefore perform a time average of the free SGWB
solution over a long time interval around $\tau$, satisfying $\Delta
\tau \gg 1/k$ for every interesting $k$, to obtain
\begin{equation}
	\dv{\Omega_\mathrm{gw}}{\ln k}
	\approx \frac{4}{3 \pi^2} k^3 \iint_{\tini}^{\tfin} \cos k(\eta - \zeta) P_{\tilde{\Pi}}(k, \eta,\zeta)\frac{\dd{\eta}}{\eta} \frac{\dd{\zeta}}{\zeta}\,.
	\label{eq:OmPvAVG}
\end{equation}
This is the effectively time-independent GW power spectrum that we aim to calculate.
Recall that
$\Omega_\mathrm{gw}$ is normalized to the radiation energy density, see \cref{eq:sbgw}.
Therefore, long after the source has decayed,
the averaged energy density of the GW signal would be expected to decrease as $a(\tau)^{-4}$.

\subsection{The stationary assumption}
\label{sec:stationary}

In this section, we simplify the evaluation of the GW spectrum integral of \cref{eq:OmPvAVG} by assuming that the turbulence is stationary.
Under this assumption, the rms velocity $\vrms$ and the integral scale $\xi$ remain constant in time.
We further set the decorrelation velocity to $\vdc\equiv \vrmsst / \sqrt{3}$ on all scales.
In other words, we neglect the $k$-dependence in \cref{eq:vdcapprox}.
Motivated by what we see in the simulations, we also set $C_v^2=1$ in \cref{eq:Cv2}.
Consequently, the only time dependence left in the UETC in \cref{eq:unequal-velocity-ps} is the time difference in the exponential.
In order to evaluate \cref{eq:OmPvAVG} under the hypothesis that the source is stationary, it therefore becomes natural to change the time integration variables to  $\tmid= (\tau+\zeta)/2$ and $\tdiff=\tau-\zeta$, where the integration over the latter is symmetric around zero.
We give the form that \cref{eq:OmPvAVG} takes under this stationary assumption as \cref{eq:Omstationary}.

In order to proceed analytically, we rely on the exponential decorrelation, and extend the integration limits of
$\tdiff$ to the interval $[0,\infty[$.
Furthermore, one can also assume that the time difference $\tdiff$ is small compared to the absolute time $\tmid$.
Under these two further assumptions, which are not strictly speaking a direct consequence of stationarity, one can integrate \cref{eq:Omstationary} analytically.

A stationary source in principle lasts forever.
However, this is not very realistic in the early Universe setting.
We will therefore fix the source duration to a finite value that we express in terms of $\ncut$, the number of eddy turnover times for which the turbulence lasts.
In other words, we integrate from $\tdevel$ until $\tdevel+\ncut \tauxist$.
The result then depends explicitly on $\ncut$.

In \cref{sec:analytical} we provide the details of the integration procedure.
The asymptotic behaviour of the resulting SGWB spectrum, and its scaling with the turbulence parameters are given by
\begin{align}
	{\dv{{\Omega}_\mathrm{gw}}{\ln k}}
	\qty(K \ll 1) & =
	c_1\,\frac{\ncut\,(\mathcal{H}_*\xi_*)^2}{\vrmsst+\ncut\mathcal{H}_*\xi_*}\, \vrmsst^6\, \qty(\frac{K}{\vrmsst})^3; \label{eq:stat-ir} \\
	{\dv{{\Omega}_\mathrm{gw}}{\ln k}}
	\qty(K \gg 1) & = \frac{\ncut\,(\mathcal{H}_*\xi_*)^2}{\vrmsst+\ncut\mathcal{H}_*\xi_*}\,
	\qty[c_2\,\vrmsst^6\,
	\qty(\frac{K}{\vrmsst})^{-\frac{7}{3}}+
	c_3\,
	\vrmsst^{4/3}\,
	\exp(-\frac{3}{2 \vrmsst^2})
	\qty(\frac{K}{\vrmsst})^{-\frac{5}{3}} ] \label{eq:stat-uv}
\end{align}
(note that $K=\mathcal{A}\xi k$, and the value of the numerical coefficients $c_i$ are given in \cref{sec:analytical}).

\cref{eq:stat-ir,eq:stat-uv} show that the scaling of the GW signal with the parameter $\mathcal{H}_*\xi_*$ depends on the source duration.
If the sourcing lasts less than one Hubble time,
meaning $\ncut \tauxist=\ncut \xi_*/\vrmsst< \mathcal{H}_*^{-1}$, one can neglect the second term in the denominator of \cref{eq:stat-ir,eq:stat-uv}, and the SGWB scaling becomes quadratic in $\mathcal{H}_*\xi_*$.
On the other hand, if the source lasts longer than one Hubble time, the scaling is linear in  $\mathcal{H}_*\xi_*$
(see for example \Refs{Hindmarsh:2015qta,Hindmarsh:2017gnf,Caprini:2009yp,Caprini:2006jb,Caprini:2019egz}).
If we had set the duration of the source to infinity, as required in principle by the stationary assumption, we would have consistently
found a linear dependence on $\mathcal{H}_*\xi_*$.

Gravitational wave generation from sound waves also presents the same duration-dependent scaling with $\mathcal{H}_*\xi_*$.
This was previously derived within the sound shell model  \cite{Hindmarsh:2015qta,Hindmarsh:2019phv,Hindmarsh:2020hop,Guo:2020grp}.
In the sound shell model one can also assume stationarity and set $\ncut \sim 1$ as in \Refa{Guo:2020grp}.
Note that, in the acoustic case, the decorrelation time depends on the sound speed, instead of on the rms fluid velocity.
The typical lifetime of the source is connected to the kinetic energy.
For the turbulent case, low velocities correspond to longer eddy turnover times $\tauxist=\xi_*/\vrmsst$, and therefore sources that last longer.
The kinetic energy is in turn expected to be connected to the PT strength (see for example \Refa{Ellis:2020awk}).
The latter then ultimately influences the scaling of the SGWB amplitude \cite{Caprini:2019egz}.

The turbulent kinetic energy also naturally enters the SGWB amplitude, through $\vrmsst$.
In particular, the stationary assumption leads to a stronger scaling with $\vrmsst$ than both the case of instantaneous generation (see \cref{sec:gw-instant}) and the SGWB signal arising from sound waves \cite{Hindmarsh:2019phv,Hindmarsh:2020hop}: at large scales and/or for low initial turbulent velocity, \cref{eq:stat-ir,eq:stat-uv} show that the SGWB amplitude from turbulence within the stationary assumption scales as $\vrmsst^6$, as opposed to $\vrmsst^4$ for both sound waves and an instantaneous turbulence generation (as we shall see in \cref{sec:gw-instant}).

We remark that the value of $\vrmsst$ also influences the high-frequency slope of the SGWB, see~\cref{eq:stat-uv}. This is a consequence of the exponential decorrelation in the velocity power spectrum \cref{eq:unequal-velocity-ps}. For large $\vrmsst > 0.5$, the second term in \cref{eq:stat-uv} dominates: consequently, the SGWB spectrum becomes shallower, and its amplitude scales as $\vrmsst^{4/3}$.
The slope of the SGWB at low wavenumber, on the other hand, is due to the steep (causal) increase of the velocity power spectrum \cref{eq:power-spectrum}, causing the convolution in \cref{eq:unequal-stress-final} to be flat for $k< 1/\xi_*$ (uncorrelated in space for scales larger than $\xi_*$) \cite{Caprini:2006jb}.
The complete spectral shape can be obtained by integrating numerically \cref{eq:stationary}: it smoothly interpolates between the slopes given by \cref{eq:stat-ir,eq:stat-uv}, peaking at the wavenumber $k \simeq 2 \vrmsst / (\mathcal{A}\xi_*)$.
Examples of spectra are given in \cref{sec:analytical}.

\subsection{Instantaneous turbulence generation}
\label{sec:gw-instant}

In this section, we consider turbulence generated instantaneously, and freely decaying afterwards.
We consider a situation where gravitational wave production begins when the fluid has
the velocity spectrum of fully developed turbulence, at time $\tdevel$.
We use three different methods to compute the SGWB spectrum.
First, we measure the SGWB in the simulations.
Second, we numerically integrate \cref{eq:OmPvAVG} inserting the model for freely-decaying turbulence derived in the first part of this paper, summarized in \cref{eq:unequal-velocity-ps} (see also section \cref{sec:results-vel}).
Third, we assume that the GW source is constant in time, and analytically integrate  \cref{eq:green-approx} under this assumption\footnote{The reason why we do not perform the time average in this case will be clarified in \cref{sec:const}.}.
We show that there is good agreement between the three methods for the resulting SGWB.
In particular, the analytical approximation provides an easy-to-use formula, \cref{eq:constant_approx}, in which the scaling with the turbulence parameters (initial rms velocity, integral scale, duration) is apparent.

\subsubsection{Simulations}
\label{sec:results-gw-sim}

\begin{figure}
	\centering
	\subfloat[Simulation (A), $\Delta\tau/\tauxist = 3.98$.]{\includegraphics[width=0.45\textwidth]{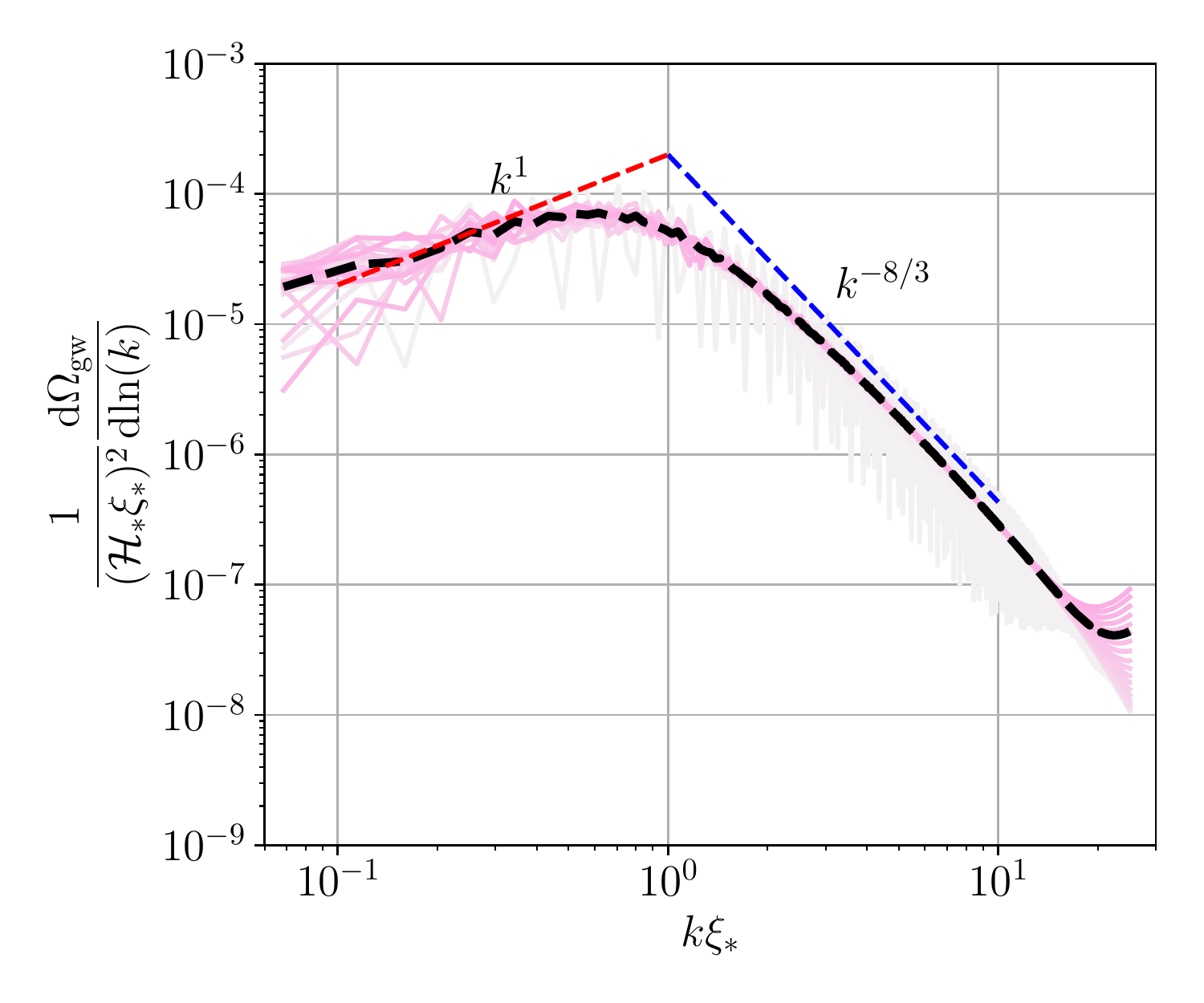}} \\
	\subfloat[Simulation (D), $\Delta\tau/\tauxist = 4.87$.]{\includegraphics[width=0.45\textwidth]{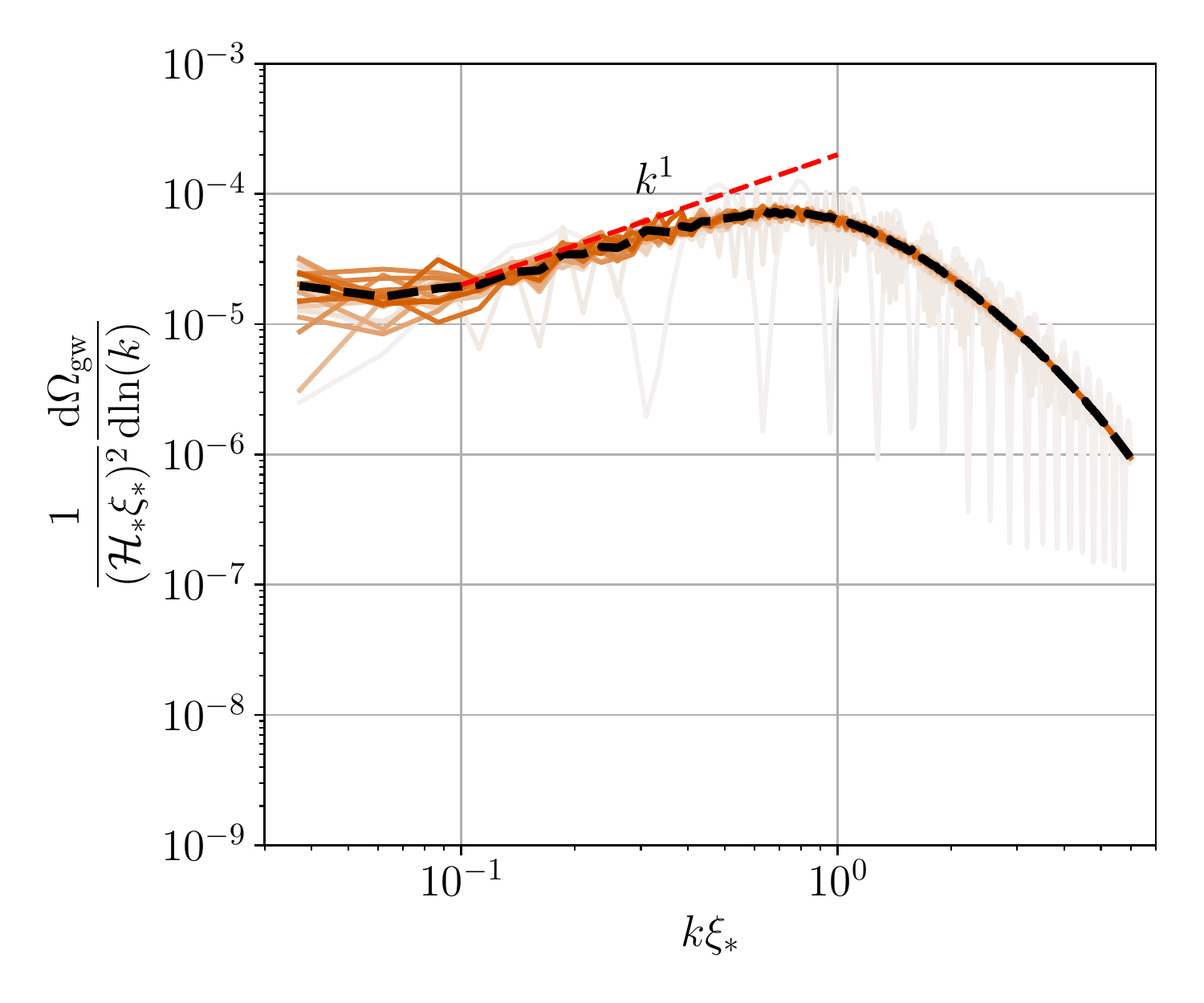}}
	\subfloat[Simulation (E), $\Delta\tau/\tauxist = 0.594$.]{\includegraphics[width=0.45\textwidth]{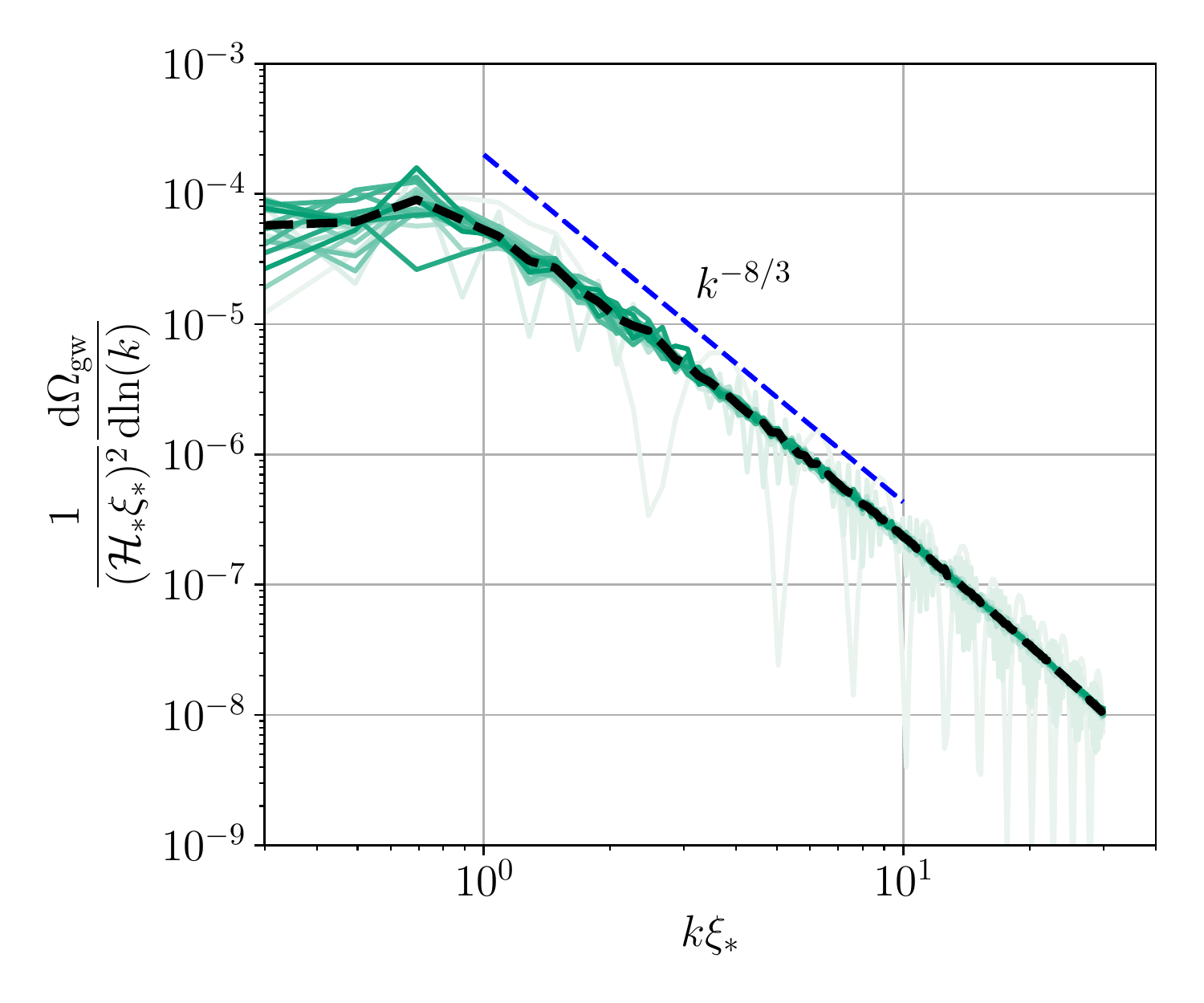}}
	\caption{Gravitational wave power spectrum from our simulations with
		$\vrmsst \approx 0.1$.
		In the y-axis, we divide the by $(\mathcal{H}_*\xi_*)^2$ since the simulations are in flat space-time, see~\cref{eq:OmGWsimul} and the following discussion.
		The coloured lines show the gravitational wave power spectrum
		at intervals $\Delta \tau$ starting from $\tau - \tdevel = \Delta \tau$ and finishing at $\tau = \tend$ .
		Darker shades correspond to later times. The black dashed line
		shows an average over the gravitational wave power spectrum in the
		last
		half of the elapsed simulation time.
		The red dashed line shows a $k^1$ power-law, while the blue dashed line shows a $k^{-8/3}$.
		The sampling of relatively few $k$-space modes into discrete bins leads to some noise in the very lowest wavenumbers. It is also possible that, for these lowest wavenumbers, finite volume effects lead to a flattening of the power spectrum as this is not seen in our results using numerical integration.
		We have cut off the
		spectrum at high wavenumbers as it progressively gets polluted due to
		numerical precision errors in projecting $\dot{u}_{ij}(\vb{k})$ to
		$\dot{h}_{ij}(\vb{k})$.\label{fig:gwps-sim-vrms-0.1} }
\end{figure}

In \cref{fig:gwps-sim-vrms-0.1} we present GW power spectra from three simulations with $\vrmsst\approx 0.1$, corresponding to
simulations (A), (D) and (E) in \cref{tab:list}. Equivalent
plots for the other simulations are given in
\cref{sec:appendix-sim-gws}. We explore different regions of the power spectrum with each simulation.
Simulation (D) has a large value
of $L \dd{x}/\xi_*$ and is therefore able
to probe the GW power spectrum around the peak and towards smaller
$k$, whereas simulation (E) is better able to probe the inertial range due to the smaller value of $\dd
x/\xi_*$.
Simulation (A) has more lattice sites, allowing for a larger dynamic range. For this simulation we chose $L \dd{x}/\xi_*$ and $\dd x/\xi_*$ such that they lie in between the values of simulations (D) and (E). Simulation (A) then probes the peak of the spectrum and intermediate wavenumbers, confirming the power laws at low and high wavenumbers in these regions.

The SGWB power spectrum builds up very
rapidly - this is further discussed in \cref{sec:const}; its shape reflects the initial power spectrum, $\mathcal{P}_U$ of \cref{eq:sim-input}, representing fully developed turbulence.

By averaging over the gravitational wave power spectrum in the latter
half of each simulation, we are able to smooth out time-dependent
oscillations in the GW power spectrum. This averaged power spectrum is shown
by the black dashed line in \cref{fig:gwps-sim-vrms-0.1}. A clear $k^{-8/3}$ power
law is found for the inertial range in simulations (A) and (E), and an approximate $k^1$
power law at small wavenumbers is seen in (A) and (D).
Taking a time average over the second half of the simulation is justified since the simulation wavenumbers satisfy $k>2/(\tend-\tdevel)$ for all simulations.
This can be verified by comparing the k-range in
\cref{fig:gwps-sim-vrms-0.1} (figures for the other simulations are in \cref{sec:appendix-sim-gws})
with the values given in \cref{tab:list} (see also \cref{tab:notation}).

When we compute the power spectrum from our simulations, we bin modes which have a wavenumber within a given shell of $k$,
as given by \cref{eq:metric-pspec-dns}.
At late times, the frequency of the oscillations of each mode increases and the modes in each bin begin to oscillate out of phase. Consequently, our simulations average over these oscillations and do not capture them, even when we do not perform an average over time. This process can clearly be seen for high wavenumbers in \cref{fig:gwps-sim-vrms-0.1}.

\begin{figure}
	\centering
	\includegraphics[width=0.7\textwidth]{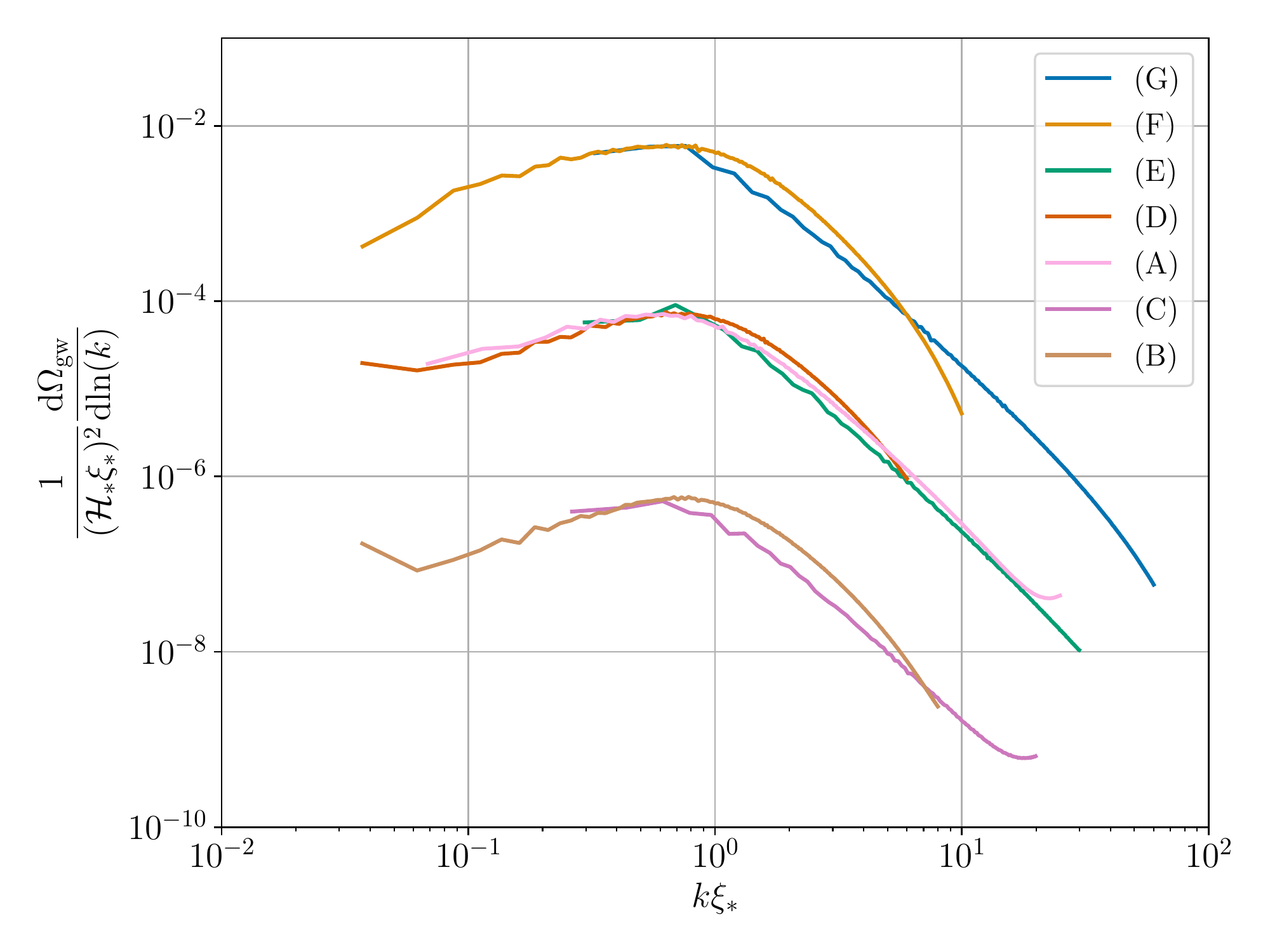}
	\caption{Averaged GW power spectra for simulations (A)-(G) from \cref{tab:list}. The coloured lines shown here correspond to  averaging the GW power spectra over the last half of the simulations. Simulations (B) and (C) have $\vrmsst\simeq0.03$, (A), (D) and (E) have $\vrmsst\simeq0.1$, and (F) and (G) have $\vrmsst\simeq0.3$. A cut off has been applied to the spectrum at high wavenumbers due to numerical precision noise.
	}
	\label{fig:gwps-sim-multi}
\end{figure}

To show the variation of the GW power spectrum for different $\vrmsst$
and for different choices of $\xi_*/\dd x$, we plot the averaged GW
power spectrum from all simulations in \cref{fig:gwps-sim-multi}. As can be seen,
simulations with different choices of $\xi_*/\dd x$ explore different
parts of the spectrum, with small $\xi_*/\dd x$ mapping out low
wavenumbers and the peak of the spectrum, and larger $\xi_*/\dd x$
exploring the inertial range.
Indeed, for $\vrmsst\simeq0.1$ we find good agreement between simulations (A) and (D) from low wavenumbers up to the vicinity of the peak, and with (E) at high wavenumbers.
This gives us confidence
that the results
obtained for other $\vrmsst$ are also valid.

In subsequent sections we will show how these simulations can be compared with the results of the four dimensional numerical integration, and with the analytical result obtained under the assumption of a constant source in time.
We plot all three approaches in \cref{fig:sim-analytic-comp} and find good agreement.

\subsubsection[GW spectrum via numerical integration of the
	anisotropic stress UETC]{Gravitational wave spectrum via numerical integration of the anisotropic stress UETC}
\label{sec:numintegration}

We have developed a numerical method allowing exact integration of the time-averaged SGWB spectrum, \cref{eq:OmPvAVG}.
This includes integration of the full angular dependence as well, improving on previous attempts~\cite{Caprini:2009yp,Niksa:2018ofa}.
We insert \cref{eq:unequal-velocity-ps} into \cref{eq:unequal-stress-final}, and in turn into \cref{eq:OmPvAVG}.
Making use of \cref{eq:velocity_ps_mathcal}, the initial spectral density $\Psdv(k,\tini)$ is set to the
\vonK\ spectrum  \cref{eq:power-spectrum},
representing fully developed incompressible turbulence.

We also need to insert the time evolution laws for $\vrms^2(\tau)$ and $\xi(\tau)$.
In accordance with the initial conditions adopted in the simulations (see \cref{sec:initial}), we assume instantaneous generation of turbulence.
Specifically, prior to $\tdevel$ the velocity field is zero everywhere, then at time $\tdevel$, turbulence appears fully developed with initial rms velocity $\vrmsst$ and initial correlation scale $\xi_*$.
Turbulence then starts decaying on a timescale of order $\tauxist$, following \cref{eq:VelEvMod,eq:XiEvMod}.
To summarize, for the instantaneous generation scenario:
\begin{align}
	 & \vrms^2(\tau)=
	\begin{cases}
		0                                                                              & \text{if } \tau < \tdevel \\
		\vrmsst ^2 \qty(\dfrac{\tau - \tdevel +\ndecay\tauxist}{\ndecay\tauxist})^{-p} & \text{if } \tau > \tdevel
	\end{cases}
	\quad\text{and} \quad
	\xi(\tau) =
	\begin{cases}
		\text{Not defined}                                                      & \text{if } \tau < \tdevel \\
		\xi_*\qty(\dfrac{\tau - \tdevel +\ndecay\tauxist}{\ndecay\tauxist})^{q} & \text{if } \tau > \tdevel
	\end{cases}.
	\label{eq:disc_evol}
\end{align}

In the following we set $\ndecay=5$, the value that best fits the simulations, see \cref{fig:xi-vrms-evol}. For the decay exponents $p,~q$, we tested both sets of values that one infers from \cref{eq:pchialpha,eq:qchialpha}, setting $\beta=3$ and $\beta=4$.
The unequal time decorrelation velocity appearing in the exponential of \cref{eq:unequal-velocity-ps} is expressed in terms of the equal time one as in \cref{eq:vsweepcomplete}, with $\vdc(k,\tau)$ given in \cref{eq:Vlarge}.
The four-dimensional integral of \cref{eq:OmPvAVG} is calculated with Monte Carlo integration with importance sampling, using the VEGAS algorithm~\cite{1978JCoPh..27..192L, Lepage:2020tgj}, an iterative and adaptive Monte Carlo scheme.
More details on the implementation are given in \cref{sec:4dinteg}.

\begin{figure}
	\centering
	\includegraphics[width=\textwidth]{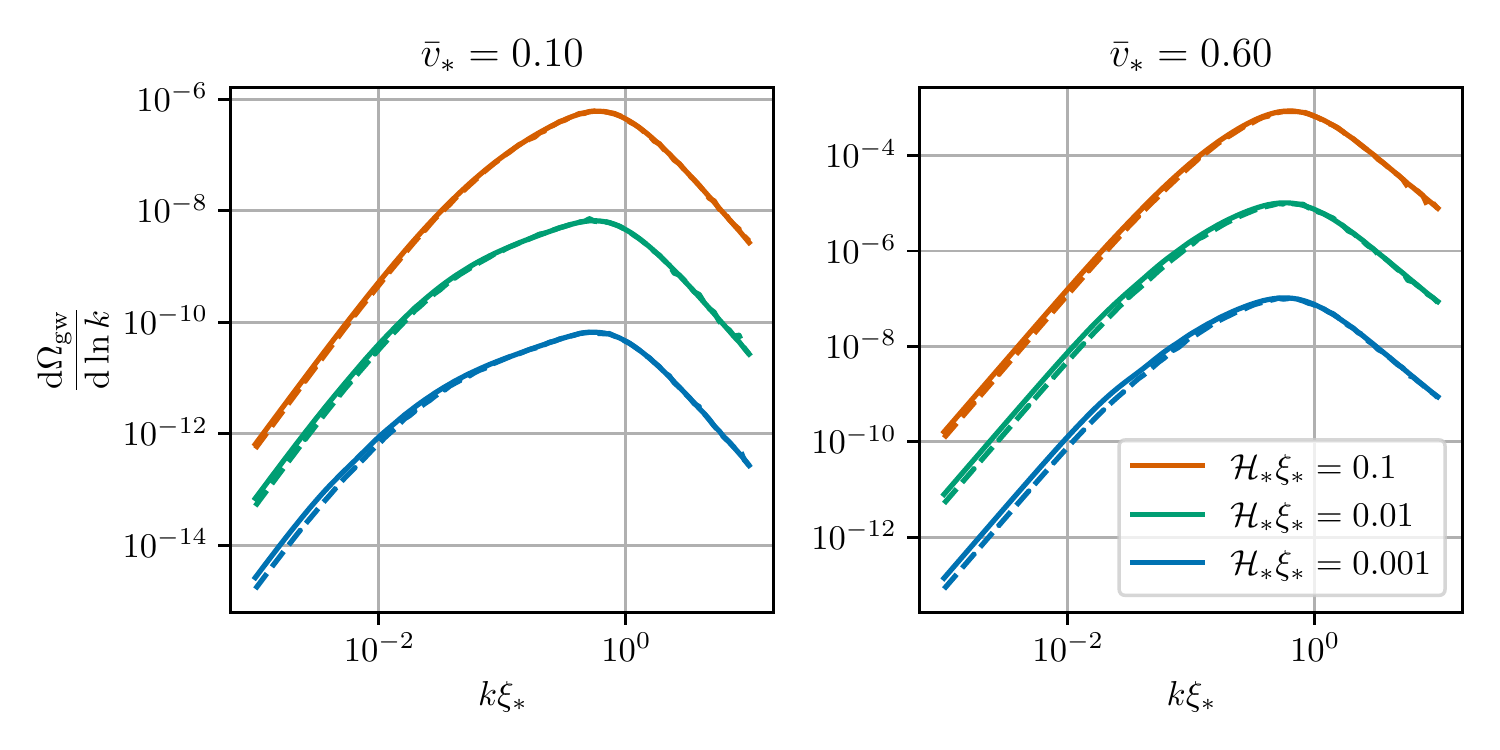}
	\caption{
		Examples of GW power spectra computed via numerical integration of \cref{eq:OmPvAVG} under the assumption of instantaneous turbulence generation (see \eqref{eq:disc_evol}).
		The spectra are computed using different values of the initial rms velocity, $\vrmsst$, the initial integral scale relative to the Hubble scale, $\mathcal{H}_*\xi_*$.
		Each panel corresponds to a different value of $\vrmsst$, as specified by the title, and each colour indicates a different value of $\mathcal{H}_*\xi_*$, as specified by the legend in the bottom right panel.
		The solid lines are computed setting $\beta = 3$, whereas the dashed lines setting $\beta = 4$.
		We recall that the evolution of $\vrms(\tau)$ and $\xi(\tau)$ are determined by $\beta$ through \cref{eq:pchialpha,eq:qchialpha}.}
	\label{fig:varying-beta}
\end{figure}

As discussed at the end of \cref{sec:free-decay-evol},  turbulence is fully dissipated after about $1000\tauxist$ (the inertial range has disappeared), for reasonable choices of the initial parameters $T_*,~\vrmsst,~ \xi_*\mathcal{H}_*$.
However, there is no need to time-integrate \cref{eq:OmPvAVG} for so long: the bulk of the GW power spectrum is sourced on a much shorter timescale.
This is due to two reasons.
First, the generation of GWs is very localized in time: GWs with wavenumber $k$ are produced by the source on a timescale  $\order{1/k}$, since the Green's function in \cref{eq:OmPv}, given in \cref{eq:green_with_all_terms}, has period $2\pi/k$.
Second, the decay of turbulence occurs over a timescale of the order of the initial eddy turnover time $\tauxist$, which effectively cuts off the source (more precisely, about five $\tauxist$, since $\ndecay\simeq 5$ is the value that best fits the simulations -  see \cref{sec:free-decay-evol}).
All GW wavenumbers with $k\gtrsim 1/(\ndecay\tauxist)$ are therefore sourced before the turbulent kinetic energy has decayed appreciably. This includes the peak region of the SGWB spectrum.
On the other hand, GW wavenumbers with $k\lesssim 1/(\ndecay\tauxist)$ are excited when the decay of the turbulent kinetic energy has already occurred: GW production on these wavenumbers is therefore negligible. This is the large scale spectral region, proportional to $k^3$.
We therefore expect the GW power spectrum to stop growing after a timescale of the order of $1/(\ndecay\tauxist)$.
We have numerically checked that the integration in \cref{eq:OmPv} indeed converges to a fixed result after a few tens of $\tauxist$.

In the code, we actually integrate for $100 ~ \tauxist$: therefore, in practice we output the SGWB spectra long after the source has decayed, when the spectral amplitude does not grow in time any longer (the integral 
in \cref{eq:OmPvAVG} has converged).
Furthermore, the $k$-range of interest is such that $k>1/(100\,\tauxist)$ for every chosen value of $\vrmsst$ and $\mathcal{H}_*\xi_*$, as can be appreciated from  \cref{fig:varying-beta}.
Therefore, we are allowed to integrate directly the time-averaged GW spectrum \cref{eq:OmPvAVG}.
We have indeed checked that numerically integrating the full expression \cref{eq:green-approx} provides a GW spectrum which is oscillating around the amplitude of the integral of the time-averaged GW spectrum, \cref{eq:OmPvAVG}, shown in \cref{fig:varying-beta}.
We do not plot these spectra, as they do not convey any additional information.

\cref{fig:varying-beta} shows the GW power spectrum computed from the numerical integration code, for different values of the parameters $\vrmsst$ and $\mathcal{H}_*\xi_*$ and for both $\beta=3$ and $\beta=4$.
It can be appreciated that the peak amplitude scales like $(\mathcal{H}_*\xi_*)^2$.
The spectral peak is independent of the value of $\vrmsst$, and corresponds to the peak of the velocity power spectrum \cref{eq:power-spectrum}, occurring at $k\xi_*\simeq 2.7/\mathcal{A}$.
This is consistent with the finding of Ref.~\cite{Caprini:2009fx} for a source which is discontinuous in time.

The spectra which have $\mathcal{H}_*\xi_* = 10^{-3}$ in \cref{fig:varying-beta} are
minimally affected by the expansion of the Universe: the chosen values of $\vrmsst$ and $\mathcal{H}_*\xi_*$ satisfy $100\,\tauxist \leq \mathcal{H}_*^{-1}$, and so the dynamics take place fully at  sub-horizon scales.
It is therefore appropriate to compare these spectra to the time-averaged spectra resulting from the simulations, in the region $k>\mathcal{H}_*$ (see \cref{eq:OmGWsimul} and discussion thereafter). We will do so in~sub-\cref{sec:compamethods}.
On the other hand, for larger values of $\mathcal{H}_*\xi_*$ in \cref{fig:varying-beta}, we integrate for longer than one Hubble time.

We find that the actual values of the exponents $p$ and $q$ do not substantially impact the SGWB spectrum, indicating that the precise details of the turbulent decay are unimportant in this respect.
This is due to the nature of the free decay of turbulence in combination with the typically short timescale associated with GW production.
\cref{fig:varying-beta}  shows the SGWB spectrum from the numerical integration for both $\beta=3$ and $\beta=4$.
For all values of $\vrmsst$ and $\mathcal{H}_*\xi_*$ the spectra only differ in the region $k< 1/(\ndecay\tauxist)$, as expected for reasons we have outlined above.
For the same reasons, we can safely ignore the role of the viscous scale $\lambda$ in all our evaluations.

\subsubsection{Constant source approximation}
\label{sec:const}

In this section, we show that the GW source can be approximated as almost constant in time while the majority of the GW signal is being generated.
This has already been pointed out in \Refa{RoperPol:2022iel}, in the case of MHD turbulence.
Since the Green's function in \cref{eq:OmPv} has period $2\pi/k$, the GW production for each wavenumber $k$ occurs on a timescale of the order of $1/k$.
Therefore, over the typical decay time for the turbulence source, namely $\ndecay\tauxist$ -- as we assumed for \cref{eq:disc_evol} --
there is sourcing of gravitational waves for wavenumbers $k>1/(\ndecay\tauxist) \equiv \vrmsst/(\ndecay\xi_*)$.
This constitutes a large fraction of the GW spectrum, including the region around the peak $k\,\xi_*\simeq 2.7/\mathcal{A}$.
Consequently, the SGWB signal on the scales of interest -- specifically, around the peak -- is fully established before the turbulence has decayed appreciably.
On scales larger than the peak, on the other hand, the GW production is less efficient due to the decay of the source.
On even larger scales, one expects the $k^3$ increase typical of uncorrelated sources in both time and space \cite{Caprini:2009fx}.

We assume in what follows that the turbulent GW source is instantaneously generated, remains constant for a duration $\ncut \tauxist$, and sharply turns off at a time $\tdevel+\ncut \tauxist$.
The number $\ncut$ is of the same order as $\ndecay$, and we set $\ncut$ to $7$ as it gives the best fit to the result of the numerical integration.
We neglect both the overall time dependence of the turbulent fluid velocity power spectrum and its time decorrelation in \cref{eq:unequal-velocity-ps}
This means that the anisotropic stress spectral density \cref{eq:unequal-stress-final} becomes time-independent, and the time and momentum integrals in \cref{eq:OmPv} decouple.
We can then rewrite \cref{eq:gwradera} as
\begin{equation}
	\eval{\dv{{\Omega}_\mathrm{gw}}{\ln k}}_{ \tdevel+\ncut \tauxist}
	=\frac{8}{3\pi^2}\,k^3\,\mathcal{T}_\mathrm{gw}(k,\tdevel,\ncut \tauxist)\, P_{\tilde\Pi}(k,\tdevel) \,.\label{eq:SGWBfinconst}
\end{equation}
Let us first analyse  the part due to the double time integral $\mathcal{T}_\mathrm{gw}$.
We evaluate it at the source turn-off:
\begin{equation}
	\mathcal{T}_\mathrm{gw}(k,\tdevel,\ncut \tauxist) \equiv \iint_{\tdevel}^{\tdevel+\ncut\tauxist} \mathcal{G}(k, \tdevel+\ncut\tauxist, \eta, \zeta)\,\frac{\dd{\eta}}{\eta}\frac{\dd{\zeta}}{\zeta}.
\end{equation}
Restricting our consideration of this integral to wavenumbers satisfying $k>1/(\tdevel+\ncut\tauxist)$, we are again entitled to drop the second and third term from the Green's function \cref{eq:green_with_all_terms}, to obtain~\cite{RoperPol:2022iel}:
\begin{align}
	\mathcal{T}_\mathrm{gw}(k,\tdevel,\ncut \tauxist) & =\iint_{\tdevel}^{\tdevel+\ncut\tauxist} \cos k(\tdevel+\ncut\tauxist - \eta)\cos k(\tdevel+\ncut\tauxist - \zeta) \frac{\dd{\eta}}{\eta}\frac{\dd{\zeta}}{\zeta} \label{eq:mathT} \\
	                                                  & = \left\{\cos k(\tdevel+\ncut\tauxist) \qty[\cosi (k (\tdevel+\ncut\tauxist)) -\cosi(k \tdevel)] \right. \nonumber                                                                 \\
	                                                  & \qquad + \left.\sin k(\tdevel+\ncut\tauxist) \qty[\sini (k (\tdevel+\ncut\tauxist ))-\sini(k \tdevel)]\right\}^2.
\end{align}
The upper limits of both time integrals must be set to $\tfin=\tdevel+\ncut\tauxist$ in order to match the results of both the simulations and the numerical integration (see~\cref{sec:compamethods}).
As pointed out in \Refa{RoperPol:2022iel}, the abrupt switching off of the source means that the evolution after $\tfin$ would give an enhancement of the SGWB power at high frequency, which would break this agreement.
Since we evaluate the time integrals at $\tfin$, we cannot perform a time average; this is why we do not use \cref{eq:OmPvAVG} in this Section.

In \Refa{RoperPol:2022iel}, a simplified expression for \cref{eq:mathT} has been found, tuned to interpolate the envelope of the $k$-oscillations.
In this work, we want to compare with the results of the simulations (\cref{sec:results-gw-sim}) and of the numerical integration (\cref{sec:numintegration}), which both evaluate the time-averaged SGWB.
We therefore adapt the result of \Refa{RoperPol:2022iel} to account for time averaging, as follows:
\begin{equation}
	\mathcal{T}_\mathrm{gw}(k,\tdevel,\ncut \tauxist)\approx
	\begin{cases}
		\ln^2\qty(1 + \ncut \tauxist / \tdevel) & \text{ for}~\sqrt{2}\, k<1/(\ncut\tauxist)\,,  \\
		\ln^2\qty[1 + (\sqrt{2}k\tdevel)^{-1}]  & \text{ for}~\sqrt{2}\, k>1/(\ncut\tauxist) \,.
	\end{cases}
	\label{eq:tgw-approx}
\end{equation}
The transition occurs at the wavenumber corresponding to the source lifetime, namely $\ncut\tauxist$ (see \Refa{Caprini:2009fx}).

\begin{figure}
	\centering
	\includegraphics{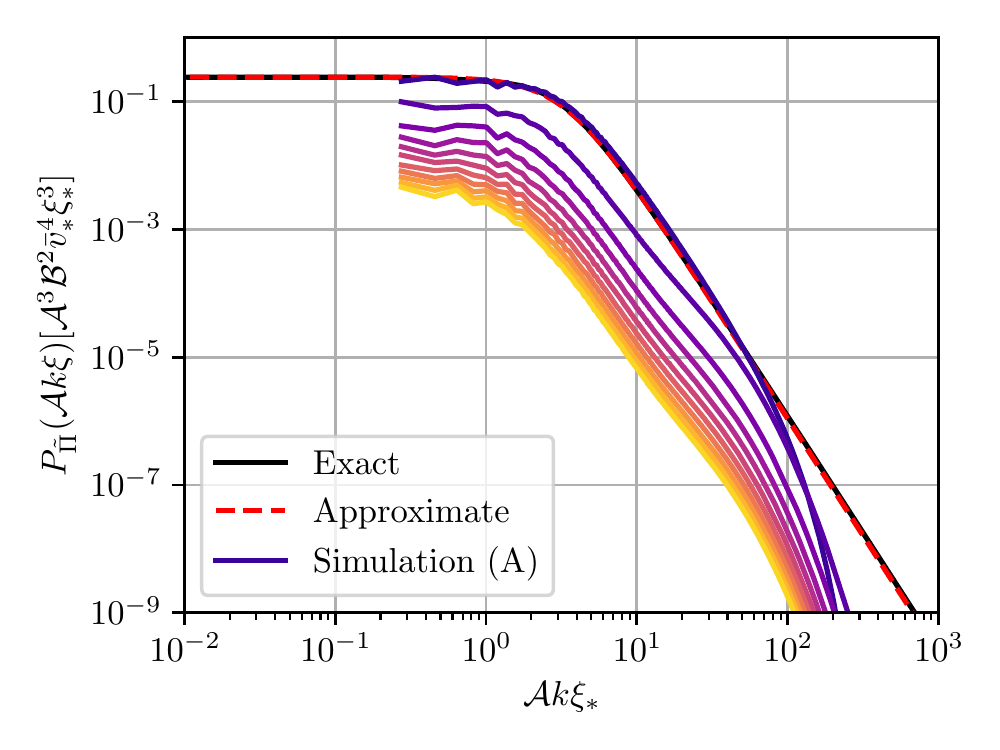}
	\caption{Anisotropic stress spectral density $P_{\tilde{\Pi}}(k,\tdevel)$ at the initial time in the constant source approximation, from the exact integration of \cref{eq:Piintegral} (black solid line) and the analytical approximation \cref{eq:Piapprox} (red dashed line). We also show $P_{\tilde{\Pi}}(k,\tau)$ extracted from simulation (A) (colored solid lines) between $\tau=\tdevel$ and $\tau=\tend$ with interval $\Delta \tau = 9.96 \tauxist$. Lighter colours indicate later times.}
	\label{fig:plotPi}
\end{figure}

We now turn to the description of the anisotropic stress spectral density at initial time $P_{\tilde\Pi}(k,\tdevel)$, also entering \cref{eq:SGWBfinconst}.
This is given by
the convolution in momentum of \cref{eq:unequal-stress-final} (without residual time dependence).
It takes the form (we use the notation of \cref{sec:analytical}):
\begin{align}
	P_{\tilde\Pi}(k,\tdevel) & =\frac{\pi^2}{2}\mathcal{A}^3{\xi_*^3}
	\int_0^\infty \dd H\,H^2
	\int_0^1 \dd\alpha  \,
	\frac{\mathrm{Proj}(H,\alpha)}{\side^3(H,\alpha)\side^3(H,-\alpha)}\Psv[\side(H,\alpha), \tdevel]\Psv[\side(H,-\alpha), \tdevel]\label{eq:Piintegral}
	\\
	                         & \simeq
	\frac{P_{\tilde\Pi}(0)}{1+\qty(\frac{\mathcal{A} k\xi_* }{3.0})^2+\qty(\frac{\mathcal{A} k\xi_* }{3.5})^{11/3}}\,,\label{eq:Piapprox}
\end{align}
where the factors of $3.0$ and $3.5$ come from an analytical fit, tuned to match  the exact numerical integration of \cref{eq:Piintegral}, and $P_{\tilde\Pi}(0)$ is the value of the anisotropic stress spectral density at $k=0$: %
\begin{equation}
	P_{\tilde\Pi}(0)= \mathcal{B}^2\mathcal{A}^3\,\vrmsst^4\,{\xi_*^3} \int_0^\infty \dd{H}  \frac{1024\,\sqrt[3]{2}\,\pi^2 H^6}{(4 H^2+4)^{17/3}} \int_0^1 \dd{\alpha}  (\alpha^2+1)^2= \mathcal{B}^2\mathcal{A}^3\,\vrmsst^4\,{\xi_*^3}\,\, \frac{7 \pi^{5/2} \,\Gamma \left(\frac{13}{6}\right)}{8\, \Gamma \left(\frac{17}{3}\right)}\,.
\end{equation}
In \cref{fig:plotPi}, we  compare the anisotropic stress spectral density from simulation (A) with both the numerical result \cref{eq:Piintegral}, and the analytical approximation
\cref{eq:Piapprox}.
As can be appreciated from the figure, at the initial time $\tdevel$ there is excellent agreement between the simulation result and the anisotropic stress spectral density derived in the context of our turbulent model.
For illustration purposes, in \cref{fig:plotPi} we also display the time evolution of the anisotropic stress spectral density from simulation (A): it is decaying, due in turn to the decay of the turbulent source itself.

\begin{figure}
	\centering
	\includegraphics[width=\textwidth]{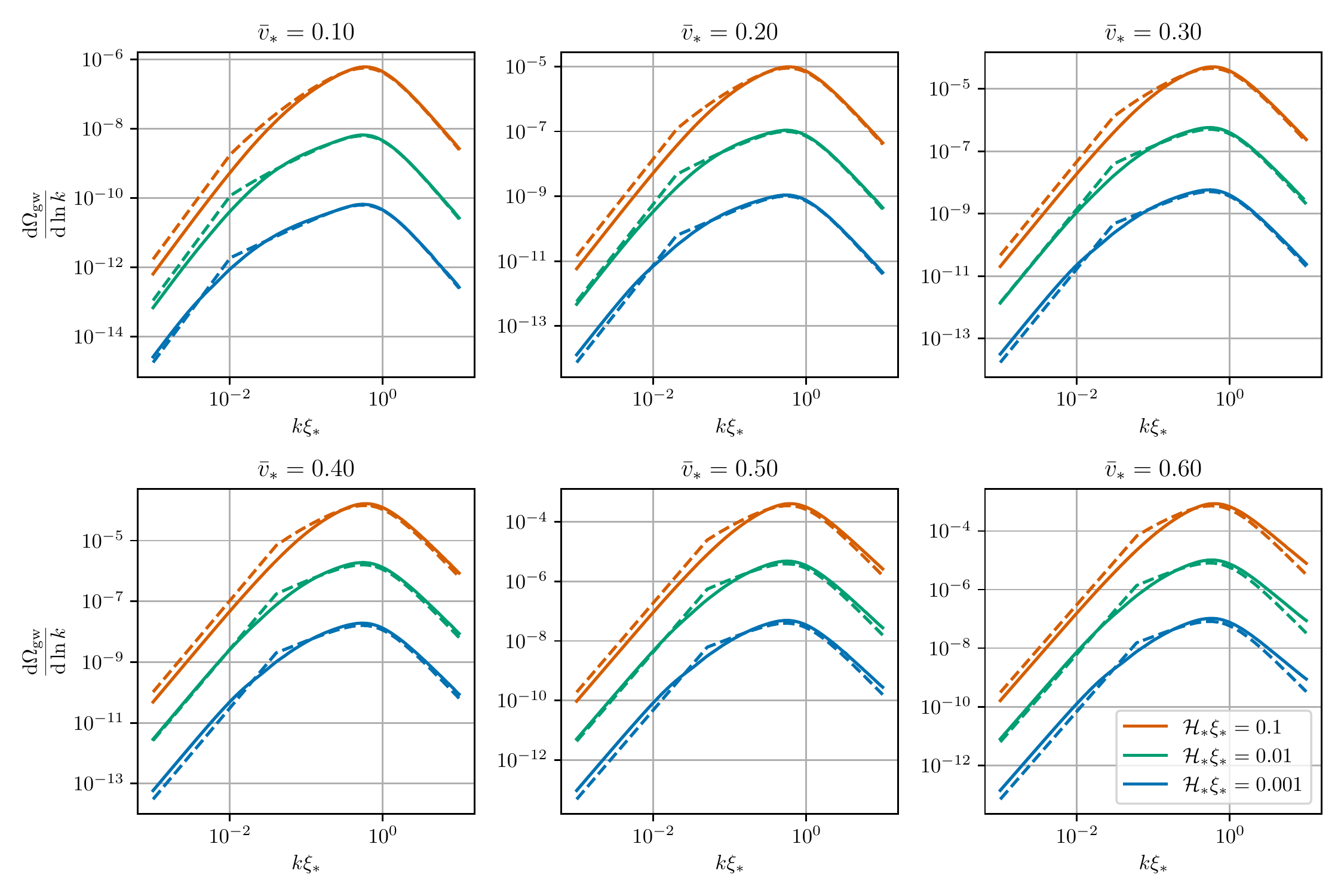}
	\caption{Reproduction of \cref{fig:varying-beta}, where here we also show the constant source approximation given in \cref{eq:constant_approx} for different values of $\vrmsst$ and $\mathcal{H}_*\xi_*$.  The solid lines match those of \cref{fig:varying-beta} (with $\beta=3$), whereas the dashed lines give the constant source approximation for an equivalent value of $\vrmsst$ and $\mathcal{H}_*\xi_*$. We fix $\ncut=7$ in the constant source approximation \cref{eq:constant_approx} for all values of $\vrmsst$ and $\mathcal{H}_*\xi_*$.
	}
	\label{fig:SGWB_num_int}
\end{figure}

The spectral shape of the SGWB from a constant source, \cref{eq:SGWBfinconst}, is given by the combination of the $k$-dependence coming from both  $\mathcal{T}_\mathrm{gw}(k,\tdevel,\ncut \tauxist)$ and  $P_{\tilde{\Pi}}(k,\tdevel)$.
Recalling that the values of $\mathcal{A}$ and $\mathcal{B}$ can be found in \cref{eq:constants1,eq:constants2}, we find
\begin{multline}
	\dv{{\Omega}_\mathrm{gw}}{\ln k}
	= \frac{7 \sqrt{\pi} \,\Gamma(13/6)}{3 \Gamma(17/3)} \frac{(\mathcal{A}\, k\, \xi_*)^3 \mathcal{B}^2\,\vrmsst^4}{1+\qty(\mathcal{A} \,k\,\xi_* / 3.0)^2+\qty(\mathcal{A}\, k\,\xi_* / 3.5)^{11/3}} \\
	\times \begin{cases}
		\ln^2\qty(1 + \ncut \tauxist \mathcal{H}_*) & \text{ for}~\sqrt{2} k<1/(\ncut\tauxist)   \\
		\ln^2\qty[1 +\mathcal{H}_*/ (\sqrt{2}\,k)]  & \text{ for}~\sqrt{2}\, k>1/(\ncut\tauxist)
	\end{cases}\text.
	\label{eq:constant_approx}
\end{multline}
The $k$-dependence of $\mathcal{T}_\mathrm{gw}(k,\tdevel,\ncut \tauxist)$ has been described in Ref.~\cite{Caprini:2009fx} in the context of coherent sources,
meaning those for which the UETC can be factorized.
\cref{eq:mathT} is effectively a time-domain Fourier cosine transform, therefore the decay properties of $\mathcal{T}_\mathrm{gw}(k,\tdevel,\ncut \tauxist)$ at large $k$ depend on its time differentiability -- and in the case being studied here, it is discontinuous in time.
This gets combined with the behaviour of $P_{\tilde{\Pi}}(k,\tdevel)$, which is uncorrelated for wavenumbers $k< 1/(\mathcal{A}\xi_*)$, while at high $k$ it shares the same power-law decrease as $\Psdv(k,\tdevel)$ (hence, for an initially von K\'arm\'an spectrum, it decreases as $k^{-11/3}$ at high $k$) \cite{Caprini:2007xq}.
Altogether, for instantaneous turbulence generation of a von K\'arm\'an spectrum, the expected slopes for low $\vrmsst$ are (with notation $K=\mathcal{A}\xi_* k$):
\begin{align}
	\text{if}~~\ncut\mathcal{H}_*\xi_*<\vrmsst \,, \qquad &
	\dv{{\Omega}_\mathrm{gw}}{\ln k} \propto \label{eq:Omslopesprediction}
	\begin{cases}
		\ncut^2\, \vrmsst^2\, (\mathcal{H}_*\xi_*)^2\, K^3 & \text{ for}~\ncut K\ll\vrmsst \,,          \\
		\vrmsst^4\, (\mathcal{H}_*\xi_*)^2\, K^{1}         & \text{ for}~ \vrmsst \ll \ncut K \text{ and } K \ll 1 \,, \\
		\vrmsst^4\, (\mathcal{H}_*\xi_*)^2\, K^{-8/3}      & \text{ for}~ 1 \ll K \,,
	\end{cases} \\
	\text{if}~~\ncut\mathcal{H}_*\xi_*>\vrmsst\,, \qquad  &
	\dv{{\Omega}_\mathrm{gw}}{\ln k} \propto \label{eq:OmslopespredictionLONG}
	\begin{cases}
		\vrmsst^4 K^3 \ln^2\qty(1 + \ncut \mathcal{H}_*\xi_* / \vrmsst) & \text{ for}~\ncut K\ll\vrmsst \,,                                    \\
		\vrmsst^4 K^3 \ln^2\qty[1 + \mathcal{A}\mathcal{H}_*\xi_*/ (\sqrt{2}\,K)]
		                                                                & \text{ for}~\vrmsst/\ncut \ll K\ll \mathcal{A}\mathcal{H}_*\xi_* \,, \\
		\vrmsst^4\, (\mathcal{H}_*\xi_*)^2\, K^{1}                      & \text{ for}~ \mathcal{A}\mathcal{H}_*\xi_* \ll \ncut K \text{ and } K \ll 1 \,,     \\
		\vrmsst^4\, (\mathcal{H}_*\xi_*)^2\, K^{-8/3}                   & \text{ for}~ 1 \ll K \,.
	\end{cases}
\end{align}
This behaviour is in accordance with that found in \Refs{Caprini:2009fx,RoperPol:2022iel}.
Note that the $P_{\tilde\Pi}/k^2$ slope observed in \Refa{Brandenburg:2019uzj} at high $k$ can be interpreted in light of what has been discussed above as well as the results of \Refa{Caprini:2009fx}.

In \cref{fig:SGWB_num_int}, we plot the constant source approximation,  \cref{eq:constant_approx}, and compare it to the result of the full numerical integration presented in \cref{sec:numerical}.
As discussed above, we set $\ncut=7$ to match the SGWB spectra obtained from the numerical integration.
The spectral shapes of the SGWBs evaluated under the constant source approximation are in good agreement with those produced with the numerical integration,
over the tested range of values for the input parameters $\vrmsst$ and $\mathcal{H}_*\xi_*$. 
The discrepancy around the transition between regimes at $k = 1/(\sqrt{2} \ncut \tauxist)$, see~\cref{eq:constant_approx}, 
is most notable for sources that last longer than one Hubble time (meaning $\ncut\tauxist\mathcal{H}_*>1$), where the intermediate power law for wavenumbers smaller than the peak is less pronounced.
Note that in \Refa{RoperPol:2022iel} it was possible to adjust the source duration (i.e.~the value of $\ncut$) in the constant approximation, as a function of $\vrmsst$ and $\mathcal{H}_*\xi_*$. This was done by comparison with the SGWBs output by the simulations, which covered also the large scale part of the spectrum.
Here, on the other hand, we fix  $\ncut=7$ for all initial $\vrmsst$ and $\mathcal{H}_*\xi_*$.

The UETC of decaying turbulence \cref{eq:unequal-velocity-ps} has both a factorizable (coherent) component and a decorrelating one.
The coherent component shapes the SGWB spectrum at large scales and around the peak.
On the other hand, the decorrelation influences the spectral shape at small scales, as discussed in \cref{sec:stationary}.
The discrepancy that can be observed in \cref{fig:SGWB_num_int} at high wavenumber for large $\vrmsst$
can therefore be interpreted in light of  what found under the assumption of a purely stationary source, see~the asymptotic behaviour at high $k$ of \cref{eq:stat-uv}.
The constant source approximation, in which the decorrelation is absent, predicts a steeper power law -- given in \cref{eq:Omslopesprediction} --  than that found by the numerical integration, which contains the full time evolution of the source.

All in all, it emerges that both the overall decay of turbulence as well as its decorrelation play a relatively minor role in shaping the SGWB signal: the first influences the spectral shape only at large scales $k\xi_*< \vrmsst/\ndecay$, while the second does so only for large initial velocity $\vrmsst$ and at small scales $k\xi_*\gg 1$.

\subsubsection{Comparison of the SGWB spectra}
\label{sec:compamethods}

\begin{figure}
	\centering
	{\includegraphics[width=0.8\textwidth]{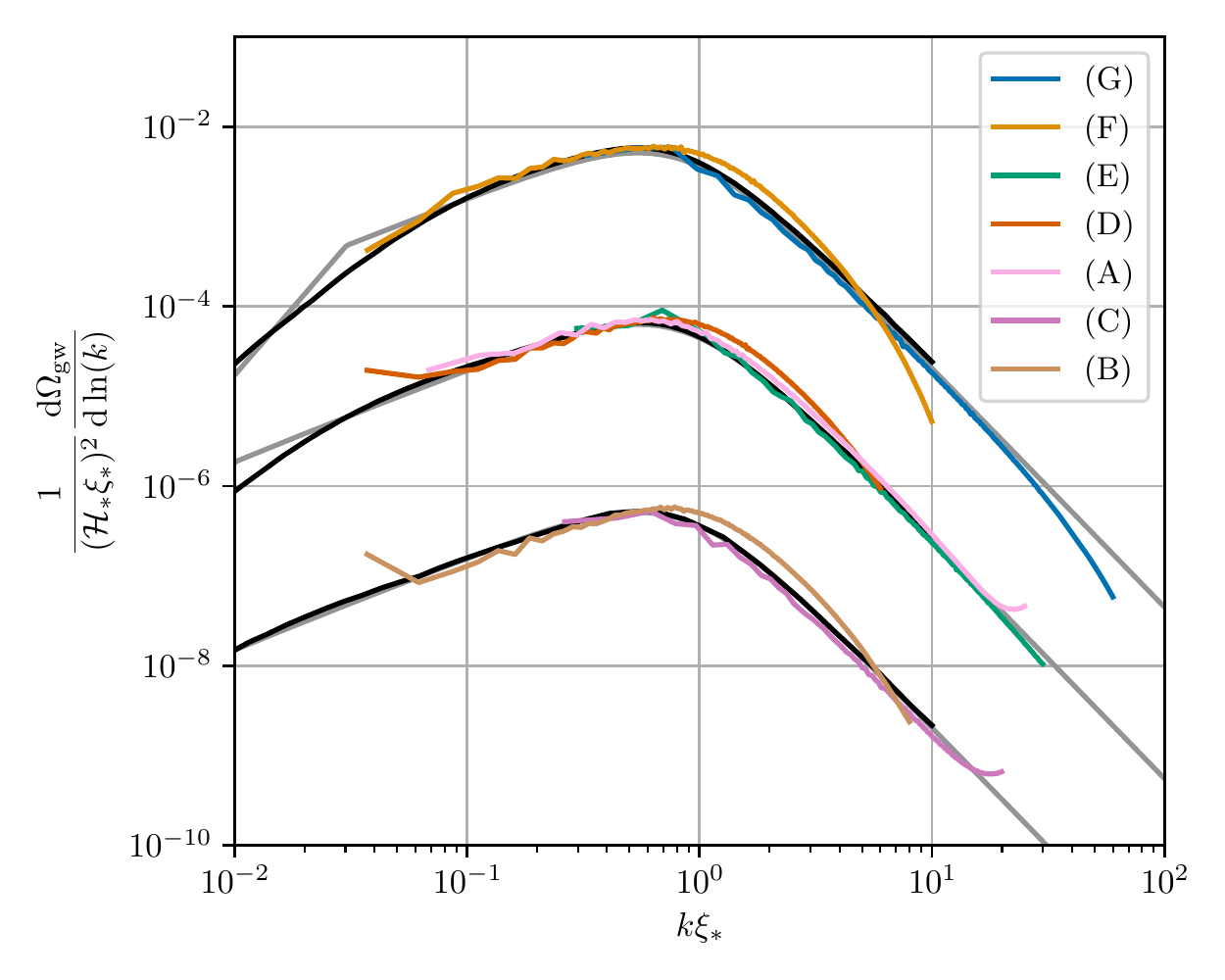}}
	\caption{GW power spectrum for instantaneous turbulence generation. The gray lines show the analytical approximation of \cref{eq:constant_approx} based on a constant source lasting for $\ncut = 7$ eddy turnover times. The black lines show the result of the 4d numerical integration of \cref{sec:numintegration}. From top to bottom, these lines correspond to $\vrmsst=0.3$, $\vrmsst=0.1$ and $\vrmsst=0.03$ respectively. In all cases we fix $\mathcal{H}_* \xi_* = 0.001$.
		We also show the averaged GW power spectra for simulations (A)-(G) from \cref{tab:list}, which are plotted using colored lines as specified in the legend.
		The GW power spectra from simulations has been cut off at high wavenumbers due to numerical precision noise.
	}
	\label{fig:sim-analytic-comp}
\end{figure}

In \cref{sec:results-gw-sim,sec:numintegration,sec:const}, we have presented three different methods of estimating the GW power spectrum when turbulence is generated instantaneously.
We first presented the GW spectra from simulations, then we performed an exact numerical integration and finally computed an analytical approximation assuming a constant source.

In this section, we compare the GW spectra obtained from the three different methods.
The comparison is performed (i) long after the source has dissipated and (ii) for the lowest value of $\mathcal{H}_*\xi_*$
for which we performed the numerical integration, namely $\mathcal{H}_*\xi_* = 10^{-3}$. Condition (i) is necessary to perform the time average, while
condition (ii) is necessary since the simulations neglect the effect of expansion. For more detail, we refer the reader to the discussions at the end of \cref{sec:gw-numerical,sec:numintegration}.

We plot the late-time averaged GW power spectra from our simulations in \cref{fig:sim-analytic-comp}, and compare them to those computed in the full numerical integration and in the constant source approximation.
We show the GW power spectra from the numerical integration and the constant source approximation for $\vrmsst=0.03\text,~0.1$ and $0.3$.
There is excellent agreement between the SGWBs obtained with the three methods, for all the values of $\vrmsst$ we simulated.
They are all consistent with respect to the GW power spectrum amplitude, the breadth of the peak, the peak location at $k\xi_* \simeq 2.7/\mathcal{A}$ and the power laws at low and high wavenumbers.
In particular, they all agree in the regions we expect the simulations to be accurate, namely at low wavenumbers for simulations (B), (D) and (F), and at high wavenumbers for simulations (C), (E) and (G). Simulation (A), which is our largest simulation, has close agreement with the other methods for the full range of wavenumbers depicted.
We have therefore validated that the numerical integration and the constant approximation are
accurate in the limit of $\mathcal{H}_*\xi_*\ll1$. Furthermore, as they include the effect of expansion, we expect them to also be accurate outside this limit.
As discussed in the previous section and depicted in \cref{fig:SGWB_num_int}, there is a discrepancy between the constant source approximation and the full numerical integration for wavenumbers close to the kink at $k = 1/(\sqrt{2} \ncut \tauxist)$. This can also be seen in \cref{fig:sim-analytic-comp}.

As the constant source approximation displays  excellent agreement around the peak and obtains the correct asymptotic power laws, \cref{eq:constant_approx} can be used as a simple  approximation to the GW power spectrum which does not require simulations.

\subsection{Phase of turbulence growth: results from the numerical integration}
\label{sec:GWcontinuous}

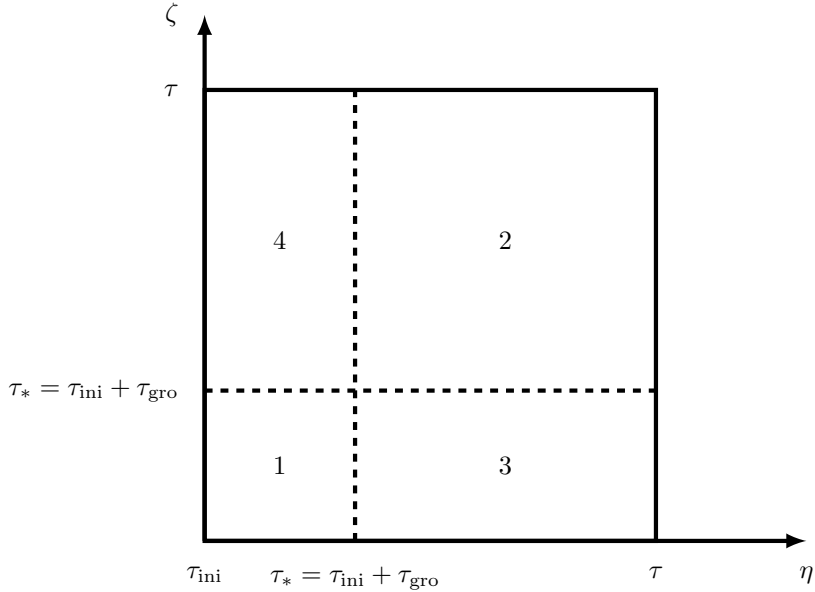
\begin{figure}
	\centering
	\begin{tikzpicture}[scale=2,ultra thick, node distance=2mm]

		\coordinate(begin) at (0,0);
		\coordinate(onset) at (1,0);
		\coordinate(onset2) at (0,1);
		\coordinate(end) at (3,3);
		\coordinate(tau) at (4,0);
		\coordinate(zeta) at (0,3.5);
		\coordinate(fin) at (3,0);
		\coordinate(fin2) at (0,3);
		\node (zone1) at (0.5, 0.5) {$1$};
		\node (zone2) at (2, 2) {$2$};
		\node (zone3) at (2, 0.5) {$3$};
		\node (zone4) at (0.5, 2) {$4$};

		\draw (begin) rectangle (end);
		\draw[dashed] (onset) -- +(0, 3);
		\draw[dashed] (onset2) -- +(3, 0);
		\draw[-latex] (begin) -- +(tau);
		\draw[-latex] (begin) -- +(zeta);

		\node [below=of begin] {$\tini$};
		\node [below=of onset] {$\tdevel=\tini + \tgro$};
		\node [left=of onset2] {$\tdevel=\tini + \tgro$};
		\node [below=of tau] {$\eta$};
		\node [left=of zeta] {$\zeta$};
		\node [below=of fin] {$\tau$};
		\node [left=of fin2] {$\tau$};

	\end{tikzpicture}

	\caption{Diagram for the evolution of turbulence in terms of $\eta$ and $\zeta$. The injection of kinetic energy starts at $\tini$ and turbulence develops  in region $1$ on a timescale $\tgro$, the value of which depends on the model: it might correspond to the PT duration $\beta^{-1}$, or to the eddy turnover time $\tauxist$. In region $2$, the turbulence is freely decaying. The growth and free decay phases are correlated: regions $3$ and $4$ also contribute to the production of GWs.}
	\label{fig:two-time}
\end{figure}

In this section we go beyond the scenario of instantaneous turbulence generation, considered so far, and include a growth phase for the turbulence kinetic energy.
Since the Reynolds number in the early Universe is very large (of the order of $10^{13}$ at the  electroweak scale, see \cref{sec:numintegration}), turbulence is expected to arise from vorticity generated during the PT \cite{Cutting:2020nla},  or due to the interaction of shocks \cite{Pen:2015qta,Dahl:2021wyk}.
In order to model the GW source properly, one would have to model the onset of turbulence by simulating the complete system of scalar field and fluid.  We leave this complicated problem  for future work.

In the present analysis we model the (purely vortical) turbulence growth phase heuristically.
This allows us to  gauge in a simple way the importance of this phase, as far as the SGWB spectral shape is concerned.
\Refs{Caprini:2009fx,Caprini:2009yp,Caprini:2009pr} have demonstrated analytically that time continuity of the GW sourcing process must be ensured, since it plays a role in shaping the SGWB signal, and Refs.~\cite{Pol:2019yex,Brandenburg:2021aln,RoperPol:2021xnd} also obtained  different SGWB spectral shapes in the forced case.
As we shall see, we confirm this result in the present work, via the numerical integration of the SGWB source (see \cref{sec:numintegration}).

Starting from the initial time $\tini$, we assume that turbulence is sourced on a timescale $\tgro$.  At time $\tdevel=\tini + \tgro$, turbulence is fully developed, and then starts decaying (see~\cref{fig:two-time}).
The duration of the growth phase would depend on the particular mechanism sourcing the turbulence. If the sourcing process is related to the PT dynamics, one would expect the duration of the growth phase to be of the order of the duration of the PT.
Alternatively, one can assume that it takes approximately one initial eddy turnover time $\tauxist = \xi_* / \vrmsst$ to build the turbulent cascade.
This is the assumption we make in the following, namely that $\tgro = \tauxist$.

Concerning the correlations of the velocity field, the Gaussian decorrelation scenario of \cref{sec:kraichnan,sec:unequal-time-velocity} holds in principle only in the free-decay phase.
However, in the absence of a better model, we assume that \cref{eq:unequal-velocity-ps} is valid also during the growth phase, and in the mixed regions.
Therefore, in the UETC of \cref{eq:unequal-velocity-ps}, the times $\tau$ and $\zeta$ can pertain either both to the growth phase, or both to the free decay phase, or to the two phases mixed (see~\cref{fig:two-time}).

We consider two scenarios for the generation of turbulence: a $\mathcal{C}^0$ {growth phase} and a $\mathcal{C}^{1}$ {growth phase}.
First we adopt the model of Ref.~\cite{Caprini:2009pr} and suppose that the vortical kinetic energy grows linearly in time before turbulence enters the free-decay phase. Furthermore, we assume that the integral scale remains constant during the growth phase, and starts growing during the phase of free decay in the $\mathcal{C}^0$ scenario:
\begin{equation}
	\begin{split}
		\vrms^2(\tau) & = \vrmsst ^2 \begin{cases}
			\dfrac{\tau - \tini}{\tgro}                                         & \text{if } \tau < \tdevel \\
			\qty(\dfrac{\tau - \tdevel +\ndecay\tauxist}{\ndecay\tauxist})^{-p} & \text{if } \tau > \tdevel
		\end{cases} \\
		\text{and} \quad
		\xi(\tau) & = \xi_*\begin{cases}
			1                                                                  & \text{if } \tau < \tdevel  \\
			\qty(\dfrac{\tau - \tdevel +\ndecay\tauxist}{\ndecay\tauxist})^{q} & \text{if } \tau > \tdevel.
		\end{cases}
	\end{split}
	\label{eq:C0}
\end{equation}

In the second scenario the kinetic energy and the integral scale evolve in such a way that they are both  continuous and differentiable at time $\tdevel$. The kinetic energy is also continuous and differentiable at initial time $\tini$. This scenario requires defining two extra functions.
For the growth phase of the kinetic energy we use a smooth step function with the properties $\smoothstep(1) = 1$,  $\smoothstep'(1) = 0$ (and also, $\smoothstep(0)=\smoothstep'(0) = 0$):
\begin{equation}
	\smoothstep(x) = \begin{cases}
		0         & x< 0  \\
		3x^2-2x^3 & 0<x<1 \\
		1         & x>1.
	\end{cases}
\end{equation}
To connect with the decay phase, we use a smooth power law with the property $\smoothpl(1, p) = 1$ and $\smoothpl'(1, p) = 0$:
\begin{equation}
	\smoothpl(x, p) = (1 - p) x^p + p x^{p-1}\, ,
\end{equation}
where $p<1$.
Using the above functions, we set for the $\mathcal{C}^1$ scenario:
\begin{equation}
	\begin{split}
		\vrms^2(\tau) & = \vrmsst^2 \begin{cases}
			\smoothstep\qty(\dfrac{\tau - \tini}{\tgro})                                 & \text{if } \tau < \tdevel \\
			\smoothpl\qty(\dfrac{\tau - \tdevel+\ndecay \tauxist}{\ndecay \tauxist}, -p) & \text{if } \tau > \tdevel
		\end{cases} \\
		\text{and} \quad \xi(\tau) & = \xi_* \begin{cases}
			1                                                                           & \text{if } \tau < \tdevel  \\
			\smoothpl\qty(\dfrac{\tau - \tdevel+\ndecay \tauxist}{\ndecay \tauxist}, q) & \text{if } \tau > \tdevel.
		\end{cases}
	\end{split}
	\label{eq:C1}
\end{equation}

\begin{figure}
	\centering
	\includegraphics[width=\textwidth]{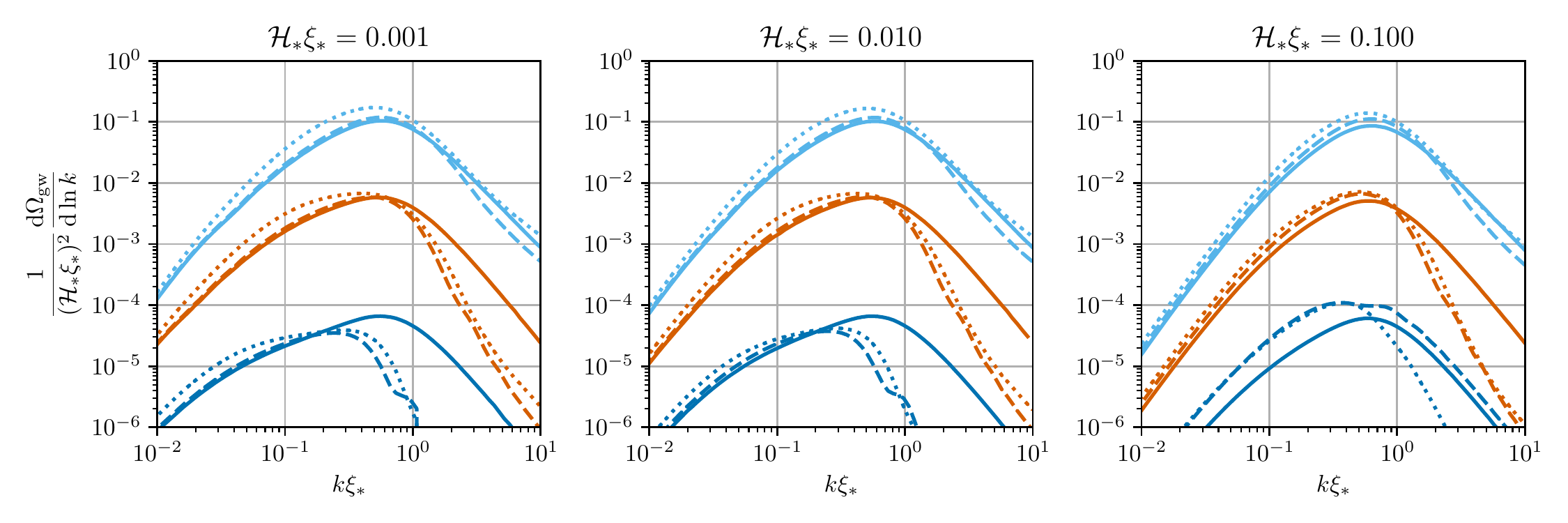}
	\caption{Gravitational wave power spectrum in the  instantaneous generation scenario (solid lines), with a $\mathcal{C}^0$ growth phase (dashed lines) and with a $\mathcal{C}^1$ growth phase (dotted lines).
		From bottom to top, $\vrmsst = 0.1, 0.3$ and $0.6$.
		The left panel shows $\mathcal{H}_* \xi_* = 10^{-3}$, the middle panel $\mathcal{H}_* \xi_* = 10^{-2}$ and the right panel $\mathcal{H}_* \xi_* = 10^{-1}$.}
	\label{fig:c0vsc1}
\end{figure}

\begin{figure}
	\centering
	\includegraphics[width=\textwidth]{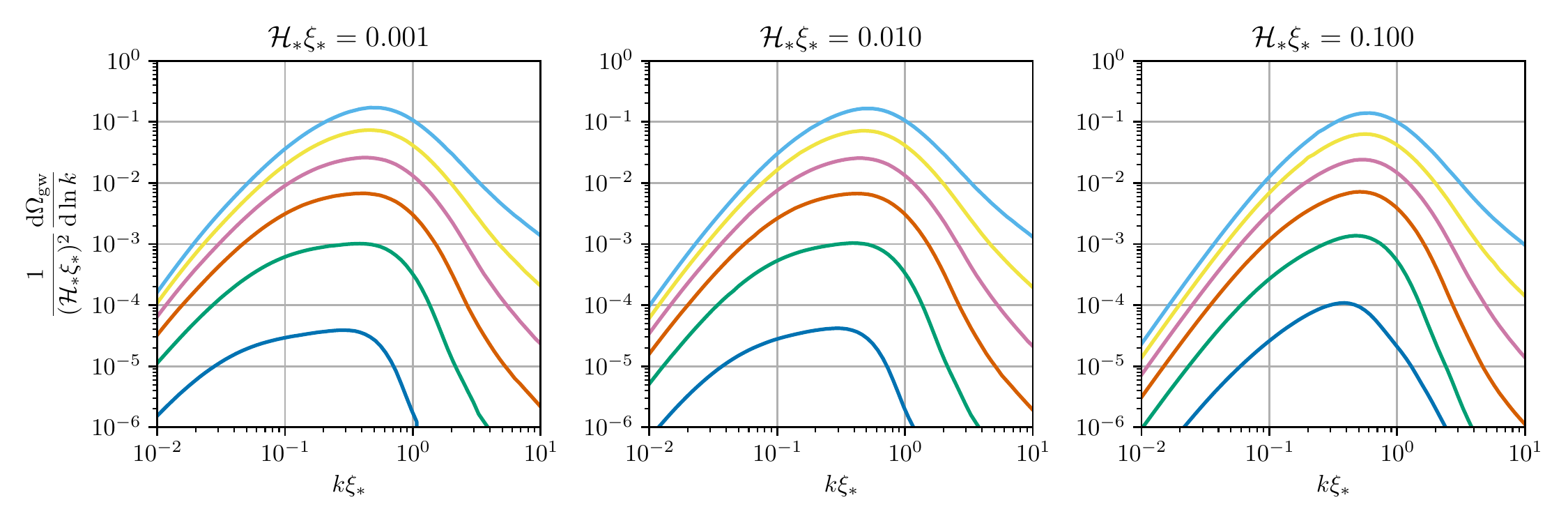}
	\caption{Gravitational wave power spectrum in the scenario with $\mathcal{C}^1$ growth phase.
		Each panel displays a different value for the initial integral scale $\mathcal{H}_* \xi_*$ (as specified in the panels titles), and each line corresponds to $\vrmsst = 0.1, 0.2, 0.3, 0.4, 0.5,$ and $0.6$ from bottom to top.
	}
	\label{fig:gwps-smooth}
\end{figure}

Different choices for the growth phase lead to different spectral shapes in the resulting SGWB.
This is demonstrated by \cref{fig:c0vsc1}, showing the SGWB with both the $\mathcal{C}^0$ and $\mathcal{C}^1$ growth phases, as well as for the instantaneous generation scenario.
While the spectra at large scales are comparable, inserting a continuous growth phase shifts the spectral peak position from the characteristic length scale of the source, $k_\mathrm{peak}\simeq \xi_*^{-1}$, to the characteristic timescale of the source, $k_\mathrm{peak}\simeq \tauxist^{-1}$: the peak position becomes therefore velocity-dependent~\cite{Caprini:2009fx}.
This behaviour is caused by the component in the velocity field UETC \cref{eq:unequal-velocity-ps} which is factorizable in time,
termed the ``coherent'' case in \Refa{Caprini:2009fx}; see the discussion below \cref{eq:constant_approx}.
Furthermore, the addition of a  growth phase which is continuous in time leads to a
reduction in the GW power, and a steeper slope at high frequencies,
as shown in \cref{fig:gwps-smooth} (in this figure we only plot the results for the $\mathcal{C}^1$ growth phase). 
Since the peak now corresponds to the inverse eddy turnover time, the $k^1$ region observed in the instantaneous growth scenario never develops, as illustrated by \cref{fig:c0vsc1}.
Note that the spectra in the $\mathcal{C}^0$ and $\mathcal{C}^1$ growth phase  are almost identical, contrary to what was observed in \Refa{Caprini:2009fx}. Indeed, we would not expect \Refa{Caprini:2009fx}
to capture all the complexity of the spectral changes due to the insertion of a continuous growth phase, since it studied a very simplified case in which the anisotropic stress UETC was separable in time.

At first, it may seem counter-intuitive that the addition of a growth phase decreases the energy in GWs, given Mercer's condition (see \cref{sec:mercer}).
We illustrate this using \cref{fig:two-time}: Mercer's condition  \cref{eq:mercer-condition} applies to intervals of the form $I\times I$, hence the contributions from the regions $1$ and $2$ are positive.
However, the contributions from the regions $3$ and $4$ can very well be negative, lowering the total GW energy.
This effect depends crucially on the correlations between the growth and the free decay phase. We leave the study of the growth of turbulence from more realistic initial conditions to a future work.

\section{Discussion}
\label{sec:discussion}

In this paper, we have studied the GW signal generated by a
phase of freely decaying vortical turbulence in a relativistic fluid.
We were motivated by the fact that thermal first order phase transitions occurring in the early Universe are expected to lead to a turbulent flow, either directly from vorticity generation
during bubble collisions, or from
the interaction of shocks following the acoustic phase.
Since the sourcing of the turbulent flow remains an active area of research, we have chosen to focus on GW production in the phase of turbulent free decay, inserting as our initial conditions a fully developed turbulent spectrum, as if the turbulence generation was instantaneous.
While this is not the most realistic assumption, it
simplified
our analysis,
allowing us to build a thorough understanding of the temporal and spatial structure of both the velocity field and the subsequent GW signal.
We thereby derived a model for the UETC of the velocity field in freely decaying turbulence.
The GW source, built in the context of this model, was then integrated to compute the GW spectrum.
The validity of our model and of the integration technique was supported with a series of simulations
using the Minkowski space relativistic hydrodynamics code SCOTTS. The simulations were prepared with equivalent initial conditions to our model, namely a fully developed turbulent spectrum.

The velocity of the turbulent fluid was assumed to follow the \vonK\ power spectrum, interpolating between a causal $k^5$ slope at low wavenumbers and a Kolmogorov $k^{-2/3}$ one at high wavenumbers.
This basic choice influences the spectral shape of the anisotropic stress power spectrum and consequently of the GW power spectrum.
Concerning the velocity UETC, motivated by the Kraichnan sweeping model,
we assumed that the velocity field  decorrelation is Gaussian in time, with a timescale set by the Eulerian eddy turnover time.
We implemented the decorrelation at larger scales, around the spectral peak, by extending the decorrelation velocity according to \Refa{kaneda_lagrangian_1993}.
We further addressed the modelling of decorrelation in the context of the theory of positive kernels, and propose to model the UETC of the turbulent velocity field as a Gibbs kernel.
This both consistently enforces time symmetry in the UETC, and guarantees that the GW power spectrum is non-negative.
We have validated our model for the velocity field decorrelation
with a series of relativistic hydrodynamic simulations performed with SCOTTS, with the initial root-mean-squared velocity, $\vrmsst$, ranging from $0.03$ to $0.3$.
We used our simulations also to study the time evolution of both the kinetic energy and
the integral scale, allowing us to fix the number of eddy turnover times that it takes for the turbulence to decay.

We have reviewed the analytical computation of the SGWB from a purely vortical fluid, arriving at an equation for the GW power spectrum in terms of an integral over the Green's function and the anisotropic stress UETC, which is provided by the turbulence model we have developed.
The computation of the GW spectrum could then be performed without any further approximation using Monte Carlo integration with importance sampling.
Varying the exponents of the turbulence decay laws had limited impact on the resulting GW spectrum.
We therefore conclude that our predictions are robust with respect to this unknown.
For instantaneously generated turbulence, we found excellent agreement between  the GW signal evaluated with direct numerical integration and the one extracted from our simulations.

To conclude our exploration of the GW signal in instantaneously generated turbulence, we studied the case of a turbulent source which is constant in time.
As the GW spectrum at high wavenumbers and around the peak is built up on short timescales in comparison to the decay of the source, the constant source approximation works well in this region.
The GW power spectrum derived within the constant source approximation is in very good agreement with both
the simulation result and the power spectrum arising from direct numerical integration of the turbulence model we have built.
We provide an analytical form for the SGWB signal from turbulence, \cref{eq:constant_approx}, which has the advantage that it can be readily used without the need to perform simulations or the challenging computation of 4-dimensional integrals.

Finally, we explored how the GW spectrum was modified by inserting a growth phase for the turbulence, therefore removing the time discontinuity inherent in the instantaneous generation assumption.
We considered in particular a $\mathcal{C}^0$ growth phase and a $\mathcal{C}^{1}$ growth phase.
These models, although simplistic, allowed us to show how the GW signal might
be modified by the turbulence generation phase.
In contrast to the decay law exponents, the form of the initial growth phase has a large effect on the GW power spectrum: flows with equivalent $\vrmsst$ can lead to GW signals with different spectral shapes and amplitudes, if the  initial condition differs.
This underlines the importance of understanding the turbulence formation mechanism in the context of first order phase transitions.

Future works will therefore focus on the generation of turbulence --- which we showed is crucial to make precise predictions for future GW experiments --- and on the inclusion of the longitudinal velocity field and a magnetic field.

\section*{Acknowledgements}

The work has been performed under the Project HPC-EUROPA3 (INFRAIA-2016-1-730897), with the support of the EC Research Innovation Action under the H2020 Programme; in particular, PA gratefully acknowledges the support of Mark Hindmarsh and the Department of Physics of the University of Helsinki and the computer resources and technical support provided by CSC.
This collaboration has been made possible thanks to the generosity of CNRS through the International Emerging Actions program 2020.
PA and CC have completed most  of this project while affiliated at the APC laboratory, Paris.
DJW (ORCID ID 0000-0001-6986-0517) was supported by Academy of
Finland grant nos. 324882 and 328958; DC (ORCID ID 0000-0002-7395-7802) was supported by Academy of
Finland grant nos. 328958 and 345070; KR (ORCID ID 0000-0003-2266-4716) was supported by Academy of Finland grants 319066, 320123 and 308791.
The work of PA (ORCID ID 0000-0002-4814-1406) was partially supported by the Wallonia-Brussels Federation Grant ARC \textnumero 19/24 - 103.
We acknowledge PRACE for awarding us access to HAWK at GCS@HLRS, Germany.

\appendix

\section{Integration variables for the evaluation of the GW power spectrum}
\label{sec:4dinteg}

\begin{figure}
	\centering
	\begin{tikzpicture}[scale=2,ultra thick, node distance=2mm]

		\coordinate(origin) at (0,0);
		\coordinate(k) at (0,3);
		\coordinate(kmid) at (0,1.5);
		\coordinate(p) at (2,1);

		\draw [->, -latex] (origin) -- (k) node [midway, left] {$\vb{k}$};
		\draw [->, thin, -latex] (origin) -- (p) node [midway, below] {$\vb{p}$};
		\draw [->, thin, -latex] (kmid) -- (p) node [midway, above] {$\vb{h}$};
		\draw [->, thin, -latex] (p) -- (k) node [midway, right] {$\vb{k-p}$};
		\draw [dashed, thin] (kmid)+(0, 0.3) arc (90:-15:0.3) node [midway, right] {$\alpha = \cos \theta$};

	\end{tikzpicture}

	\caption{Coordinate system consisting of $(h, \alpha)$.\label{fig:coord}}
\end{figure}
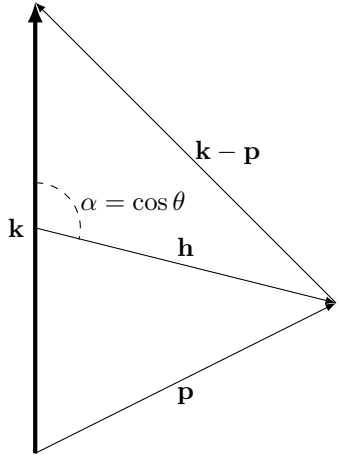

This appendix summarizes the changes of variable we perform in order to tackle the convolution integral $\dd[3]{p}$ in \cref{eq:OmPv}.
To make the symmetries of the integrand apparent, we first set $\vb{h} = \vb{p} - \vb{k}/2 $, see~\cref{fig:coord}. Denoting $\alpha = \vu{k}\vdot \vu{h}$, it follows that
\begin{align}
	p^2 = \abs{\vb{k}/2 + \vb{h}}^2                     & = k^2/4 + h k ~ \alpha + h^{2} \equiv \side^2(h,\alpha)
	\label{eq:ps}
	\\
	\abs{\vb{k} - \vb{p}}^2 = \abs{\vb{k}/2 - \vb{h}}^2 & = k^2/4 - h k ~ \alpha + h^{2}\equiv \side^2(h,-\alpha)\,.
	\label{eq:kminusps}
\end{align}
The projection factor becomes (note the symmetry $\alpha \rightarrow -\alpha$)
\begin{eqnarray}
	\qty[1 + (\vu{k}\vdot \vu{p} )^2]\qty[1+ (\vu{k}\vdot \widehat{\vb{k-p}})^2] &=& 4 \qty[1-\frac{h^2(1-\alpha^2)}{2(k^2/4 - h k ~ \alpha + h^{2})}]
	\qty[1-\frac{h^2(1-\alpha^2)}{2(k^2/4 + h k ~ \alpha + h^{2})}]
	\nonumber\\
	&\equiv& \mathrm{Proj}(h,\alpha).
	\label{eq:projection-factor}
\end{eqnarray}
For the numerical integration, it is useful to define the new variable $M \equiv {k^2}/(k^2 + h^2)$. The volume element of the spherical coordinates $(h, \theta, \phi)$ can then be written as
\begin{equation}
	\dd[3]{h} =  \frac{k \dd{M}}{2 \sqrt{1-M} M ^ {3/2}} \dd{\alpha}\dd{\phi}
\end{equation}
with $M \in ]0, 1]$ and $\alpha \in [0, 1]$. Note that the integrand in \cref{eq:OmPv} is invariant under rotations of the azimuthal angle $\phi$.

\section{Analytical expressions for the stationary assumption}
\label{sec:analytical}

Under the stationary assumption, whose implications are described in \cref{sec:stationary},  \cref{eq:OmPvAVG} can be rewritten as:
\begin{figure}
	\centering
	\includegraphics[width=\textwidth]{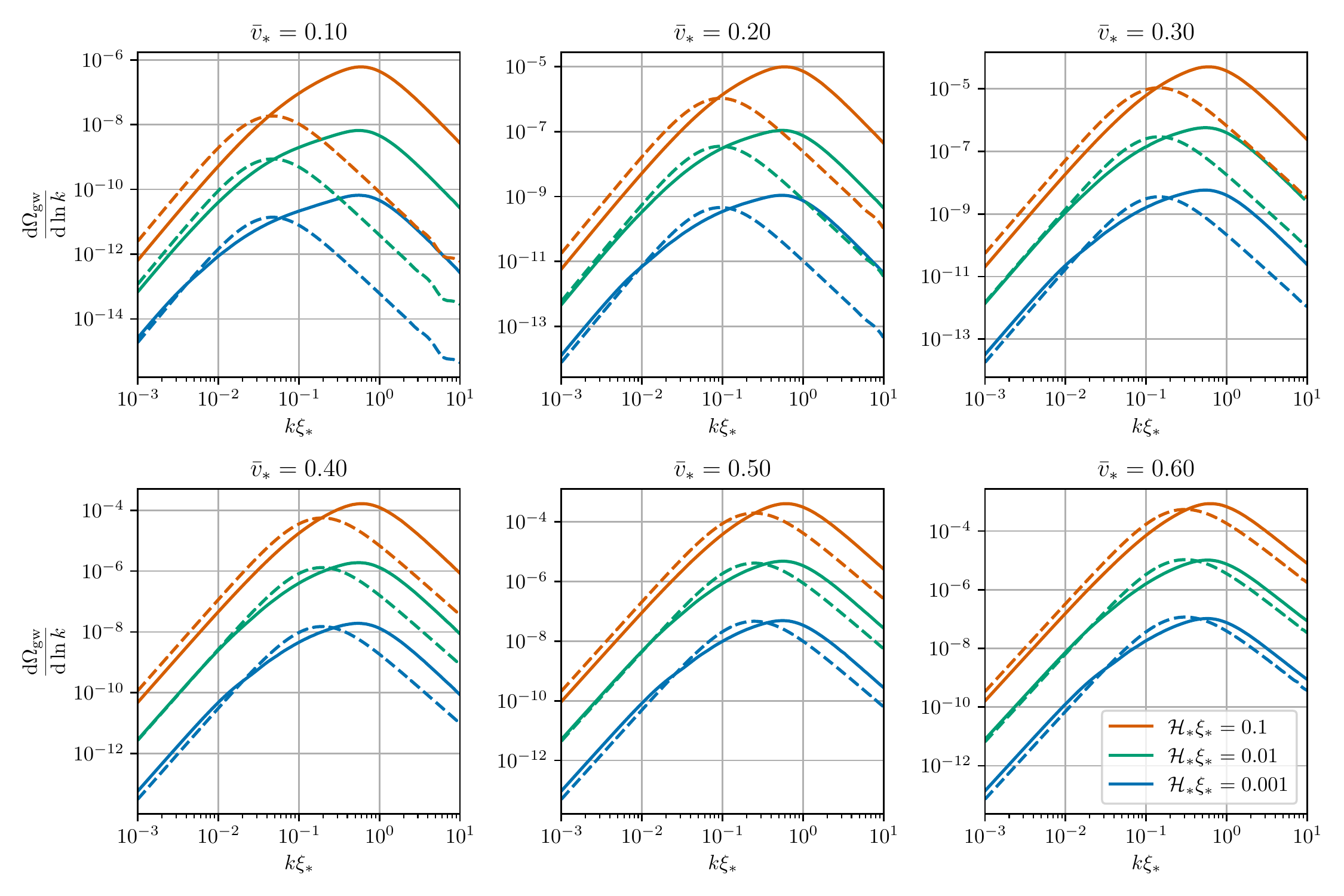}
	\caption{As in  \cref{fig:varying-beta},
		we show the SGWB spectra for several values of $\vrmsst$ and $\mathcal{H}_*\xi_*$ in the instantaneous generation scenario with $\beta=3$ (solid lines), compared with the SGWB obtained within the stationary assumption given in \cref{eq:stationary} (dashed lines). We fix $\ncut=7$ as in the constant source approximation, see \cref{eq:SGWBfinconst}.}
	\label{fig:stationary}
\end{figure}

\begin{align}
	\dv{{\Omega}_\mathrm{gw}}{\ln k} = & \frac{4}{3}k^3
	\int_0^\infty \dd h\,h^2
	\int_0^1 \dd\alpha  \,
	\frac{\mathrm{Proj}(h,\alpha)}{\side^3(h,\alpha)\side^3(h,-\alpha)}\Psv[\side(h,\alpha)]\Psv[\side(h,-\alpha)] \nonumber
	\\
	                                   & \times \Bigg\{
	\int_{\tdevel}^{\frac{\tdevel+\tfin}{2}} \dd\tmid \int_0^{2\tmid-2\tdevel}\dd \tdiff \,
	\frac{\cos(k\tdiff)}{\tmid^2-\frac{\tdiff^2}{4}} \exp\qty[-\frac{\vrmsst^2\tdiff ^2}{3}\qty(h^2+\frac{k^2}{4})]
	\nonumber                                                                                                                  \\
	                                   & + \int_{\frac{\tdevel+\tfin}{2}}^{\tfin} \dd\tmid \int_0^{2\tfin-2\tmid}\dd \tdiff \,
	\frac{\cos(k\tdiff)}{\tmid^2-\frac{\tdiff^2}{4}} \exp\qty[-\frac{\vrmsst^2\tdiff ^2}{3}\qty(h^2+\frac{k^2}{4})]\Bigg\}\,, \label{eq:Omstationary}
\end{align}
where $\side(h,\alpha)$ and $\mathrm{Proj}(h,\alpha)$ are defined in Eqs.~\eqref{eq:ps}, \eqref{eq:kminusps} and \eqref{eq:projection-factor} of \cref{sec:4dinteg}, and we have used \cref{eq:velocity_ps_mathcal} in \cref{eq:unequal-stress-final} to express the power spectral density in terms of the power spectrum $\Psv[\side(h,-\alpha)]$, defined in \cref{eq:power-spectrum}.
In order to proceed analytically, we make some extra assumptions: (i) we push the upper limit of the integration in $\dd\tdiff$ to $+\infty$; (ii)
we neglect the term $\tdiff^2/4$ with respect to $\tmid^2$ in the denominator of \cref{eq:Omstationary}.
Indeed, since $\tdiff^2 / 4 \leq \tmid^2 / 4$, we expect that the Gaussian decorrelation favours the region $\tdiff^2 \ll 1$ for the generation of GW.
Under these assumptions, the 4 dimensional  integration reduces to  a 1 dimensional one.
First, the integral over the time difference becomes
\begin{equation}
	\int_0^\infty \cos( k \tdiff) \exp[-\frac{\vrmsst^2
			\tdiff^2}{3}\qty(h^2+\frac{k^2}{4})] \dd{\tdiff} =
	\frac{\sqrt{3\pi}}{\vrmsst\sqrt{k^2+4 h^2} } \exp[ -\frac{3 k^2}{(k^2+4 h^2) \vrmsst^2} ].
\end{equation}
Additionally, the integral over absolute time starting at $\tdevel=\mathcal{H}_*^{-1}$ and lasting until $\tfin = \tdevel + \ncut\tauxist$ where $\tauxist = \xi_*/\vrmsst$, gives
\begin{equation}
	\int_{\tdevel}^{\tdevel+\ncut\tauxist} \frac{\dd{\tmid}}{\tmid^2} = \mathcal{H}_*\frac{\ncut\mathcal{H}_*\xi_*}{\vrmsst+\ncut\mathcal{H}_*\xi_*}.
\end{equation}
Finally, it is possible to integrate analytically over $\alpha = \cos \theta$, yielding the GW power spectrum in terms of the 1-dimensional integration (we define $H=\mathcal{A}\xi h$ and $K=\mathcal{A}\xi k$)
\begin{align}
	\dv{{\Omega}_\mathrm{gw}}{\ln k} = \frac{32\, 2^{1/3}}{55}\sqrt{\frac{\pi}{3}} \,\mathcal{A}\, \mathcal{B}^2\,
	\frac{N\,(\mathcal{H}_*\xi_*)^2}{\vrmsst+N\mathcal{H}_*\xi_*}\, \vrmsst^3
	\int_0^\infty \dd{H}  \frac{H^2}{K\sqrt{4H^2+K^2}} \exp[-\frac{3 K^2}{\vrms_*^2 (4 H^2+K^2)}] f(H,K),
	\label{eq:stationary}
\end{align}
where the function $f(H,K)$ is given by
\begin{align}
	f(H, K) \equiv & ~
	\frac{ f_1}{\qty(K^2+4H^2+4)^{17/3}}\, \hypergauss{\frac{1}{2}}{\frac{5}{6}}{\frac{3}{2}}{\frac{16 K^2 H^2}{K^2+4H^2+4}} \\
	               & +\frac{3 f_2}{\qty(K^2+4H^2+4)^4 \qty(K^4-8 K^2 (H^2-1)+16 (H^2+1)^2)^{11/6}}\nonumber
\end{align}
with
\begin{align}
	f_1 \equiv ~ & 16 (415 H^2+171) K^6+768 (H^2+1)^2 (5 H^2+9) K^2+6912 (H^2+1)^4                  \\
	             & +32 (305 H^4+402 H^2+225) K^4+1339 K^8 \, ,
	\nonumber                                                                                       \\
	f_2 \equiv ~ & 8 (613-247 H^2) K^{10}+18432 (H^2-3) (H^2+1)^4 K^2-36864 (H^2+1)^6+              \\
	             & 16 (-823 H^4+1042 H^2+121) K^8+256 (H^2+1)^2 (361 H^4-174 H^2-231) K^4 \nonumber \\
	             & +256 (55 H^6+73 H^4-251 H^2-141) K^6+727 K^{12} \, .
	\nonumber
\end{align}
\cref{fig:stationary} compares the SGWB obtained within the stationary assumption with the results of \cref{sec:numintegration}.
Within the stationary assumption, the GW spectrum peaks at $k\simeq 2\vrmsst/(\mathcal{A}\xi_*)$, while in the instantaneous generation scenario the peak occurs at $k\simeq 2.7/(\mathcal{A}\xi_*)$.
The stationary assumption also leads to a substantial reduction of power at small scales.
In \cref{sec:stationary} we provide the low and high wavenumber asymptotic behaviour of the SGWB \eqref{eq:stationary}, for which the prefactors are
\begin{align}
	c_1= \frac{3\sqrt{3\pi}}{110} \mathcal{A}\,\mathcal{B}^2\simeq 0.3\,,\quad
	c_2=  \frac{896\sqrt{\pi}}{405} \qty(\frac{4}{3})^{\frac{1}{6}}  \Gamma \qty(\frac{8}{3})\mathcal{A} \,\mathcal{B}^2\simeq 23\,, \quad
	c_3= \frac{24 \sqrt{6\pi}}{55} \mathcal{A}\, \mathcal{B}^2 \Upsilon \simeq 57
\end{align}
where we have fit the factor $\Upsilon\simeq 8$.

\section{Finite volume tests} \label{sec:appendix-finite-vol}

\begin{figure}
	\centering
	\includegraphics[width=0.49\textwidth]{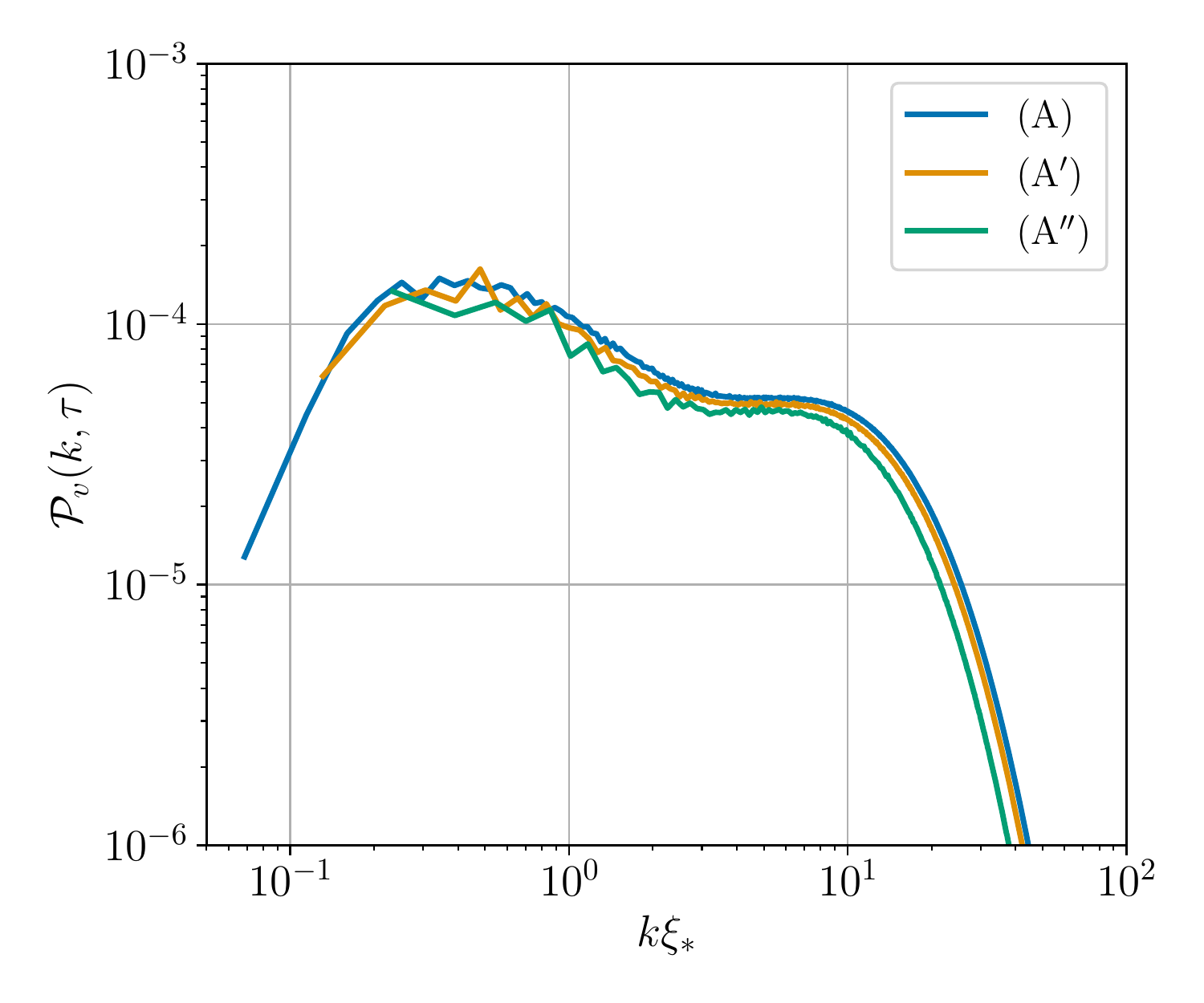}
	\caption{ Velocity power spectra at $\tau=\tend$ in simulations (A), ($\mathrm{A}'$) and ($\mathrm{A}''$).
	}
	\label{fig:finite-vol-spec}
\end{figure}

We performed two additional test simulations ($\mathrm{A}'$) and ($\mathrm{A}''$) to understand
how the low-wavenumber behaviour seen in simulation (A)
depends on the cutoff induced by the box size. As shown in
\cref{tab:list}, these simulations
have the same input power spectrum as simulation (A), but with different simulation volumes. By reducing
the number of lattice points by a factor of 2 or 4 for simulation ($\mathrm{A}'$) and ($\mathrm{A}''$) respectively,
there is less dynamic range above the
integral scale $\xi_*$. The small differences in $\vrmsst$ and $\xi_*$ are explained partially by the use of a different random seed and partially by the different simulation volume.

\begin{figure}
	\centering
	\subfloat[Evolution of $\vrms^2 \xi^{1+\beta}$]{\includegraphics[width=0.49\textwidth]{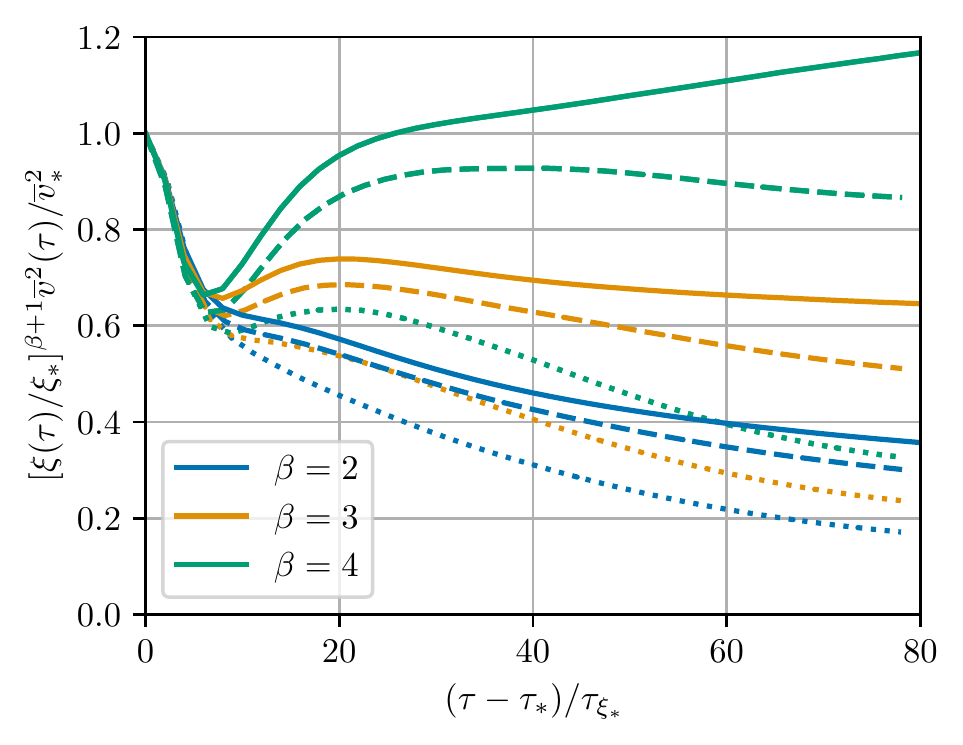}}
	\subfloat[Evolution of $\xi$ and $\vrms^2$\label{fig:xi-vrms-finite}]{	\includegraphics[width=0.49\textwidth]{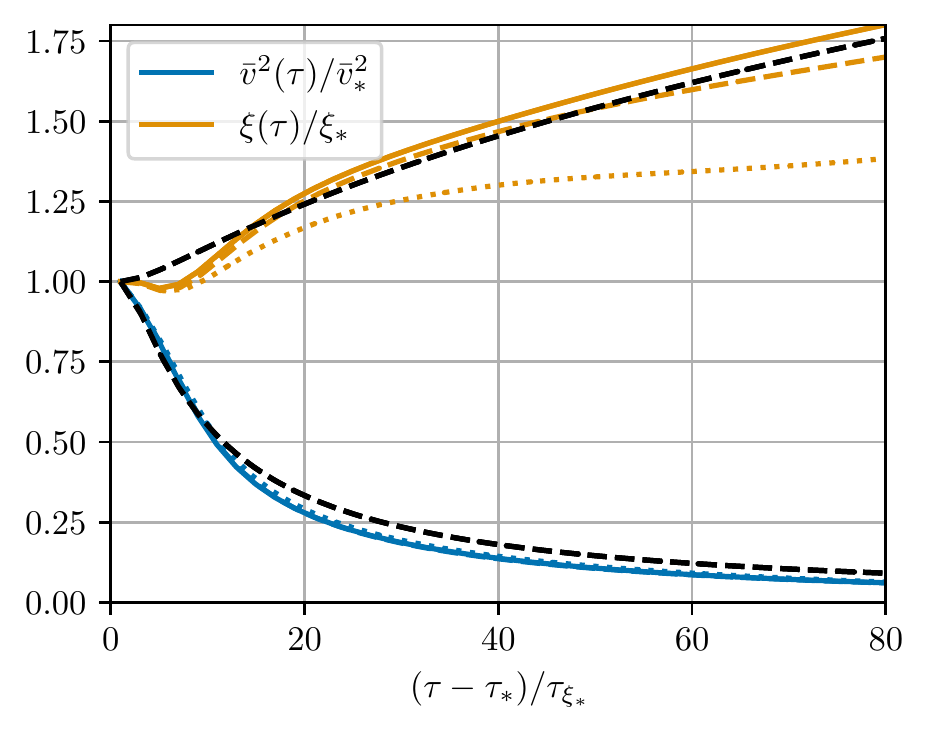}}
	\caption{
		\emph{Left panel}: Evolution of $\vrms^2 \xi^{1+\beta}$ for different values of $\beta$ in simulations (A), ($\mathrm{A}'$) and ($\mathrm{A}''$).
		As shown in the main text, this quantity should remain constant in freely decaying turbulence, thus indicating that $\beta \simeq 3$.
		\emph{Right panel}: Evolution of the velocity and integral scale in simulations (A), ($\mathrm{A}'$) and ($\mathrm{A}''$).
		The black-dashed line showcases \cref{eq:VelEvMod,eq:XiEvMod} for $\ndecay=5$, $p=4/3$ and $q=1/3$, where the values of $p$ and $q$ correspond to $\beta=3$.
		In both panels, the coloured solid lines refer to simulation (A), dashed lines to ($\mathrm{A}'$) and dotted lines to ($\mathrm{A}''$).
	}
	\label{fig:beta3-finite}
\end{figure}

In \cref{fig:finite-vol-spec} we plot the final velocity power spectra from each simulation. It can be seen that even by the end of the simulation, there is broad agreement between the simulations for the wavenumbers that they share in common.
We plot the evolution of $\xi$ and $\vrms^2$ in  each of the simulations in \cref{fig:xi-vrms-finite}. From this we can see that the evolution of the kinetic energy is relatively similar, however for ($\mathrm{A}''$) we see that the integral scale is unable to evolve to larger values due to the limited volume. Simulation $\mathrm{A}'$) has a closer agreement with ($\mathrm{A}$) at early times, but still deviates at late times as the $\xi$ continues to grow.

\begin{figure}
	\centering
	\subfloat[Instantaneous exponents $(p,q)$.]{\includegraphics[width=0.49\textwidth]{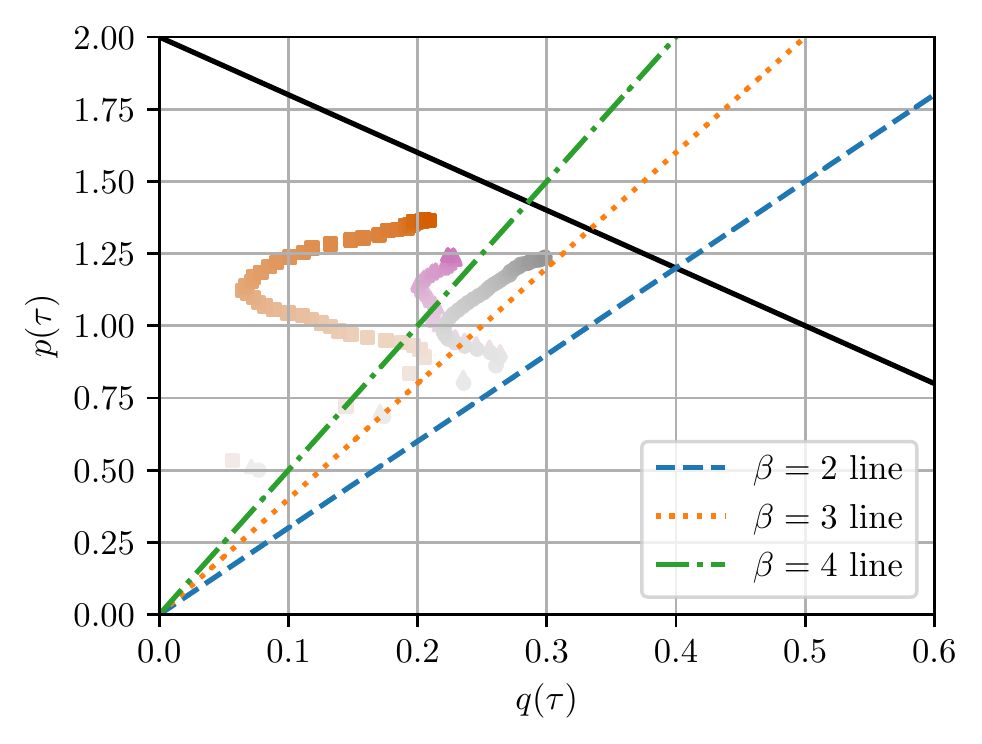}}
	\subfloat[Time evolution $p(\tau)$, $q(\tau)$.\label{fig:timeevol-finite}]{\includegraphics[width=0.46\textwidth]{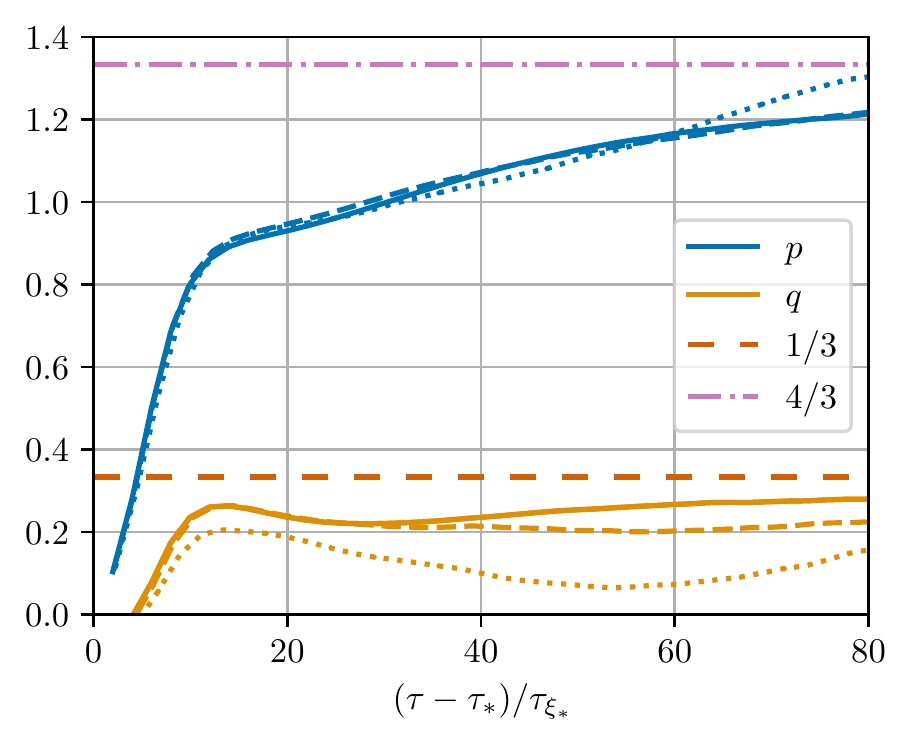}}
	\caption{\emph{Left panel}: Trajectory of the instantaneous exponents $(p,q)$ in simulation (A), ($\mathrm{A}'$) and ($\mathrm{A}''$). Simulation (A) is shown with circular gray markers, ($\mathrm{A}'$) with pink triangles, and ($\mathrm{A}''$) with orange squares. Time is represented by the colour scheme: early times are shown with lighter shades and late times with darker shades, starting at $\tau=\tdevel$, with interval $\Delta \tau \sim 2 \tauxist$. The dark solid line represents the self-similarity line $p=2(1-q)$. The coloured lines show the relation $p=(1+\beta)q$ for various choices of $\beta$. \emph{Right panel}: Evolution of the instantaneous kinetic energy and integral scale exponents $(p, q)$ as a function of time in simulation (A), ($\mathrm{A}'$) and ($\mathrm{A}''$). Solid lines refer to simulation (A), dashed lines to ($\mathrm{A}'$) and dotted lines to ($\mathrm{A}''$). The horizontal lines show the values expected for $p$ and $q$ if $\beta = 3$.}
	\label{fig:beta-finite}
\end{figure}

This effect then modifies the rest of our global plots. In \cref{fig:beta-finite}, we show the evolution of the instantaneous kinetic energy and integral scale exponents $(p,q)$,
both parametrically and as a function of time. We see that the simulations agree on $p$ and $q$ at early times, but diverge in particular for the exponent $q$ at later times. This is unsurprising, as when $q$ is positive it indicates that the integral scale is growing. Eventually $\xi$ becomes comparable to the maximum scale within the simulation and the fluid flow becomes increasingly sensitive to the size of the box. At this point it appears the growth of $\xi$ is restricted, leading to a reduction in $q$.

\FloatBarrier

\section{Unequal time correlations in other simulations}
\label{sec:appD}
\begin{figure}
	\centering
	\subfloat[Simulation (B).]{\includegraphics[width=0.49\textwidth]{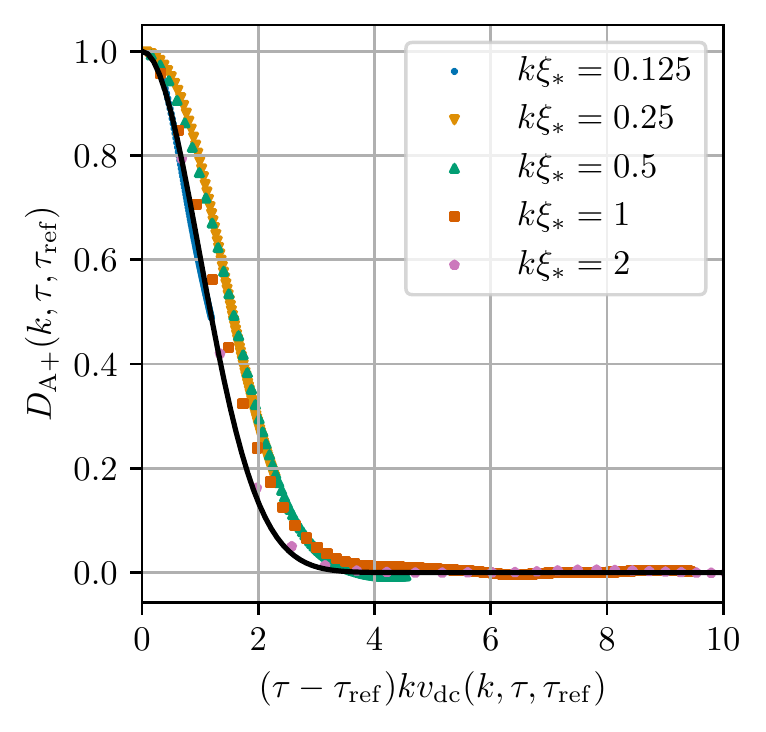}}
	\subfloat[Simulation (C).]{\includegraphics[width=0.49\textwidth]{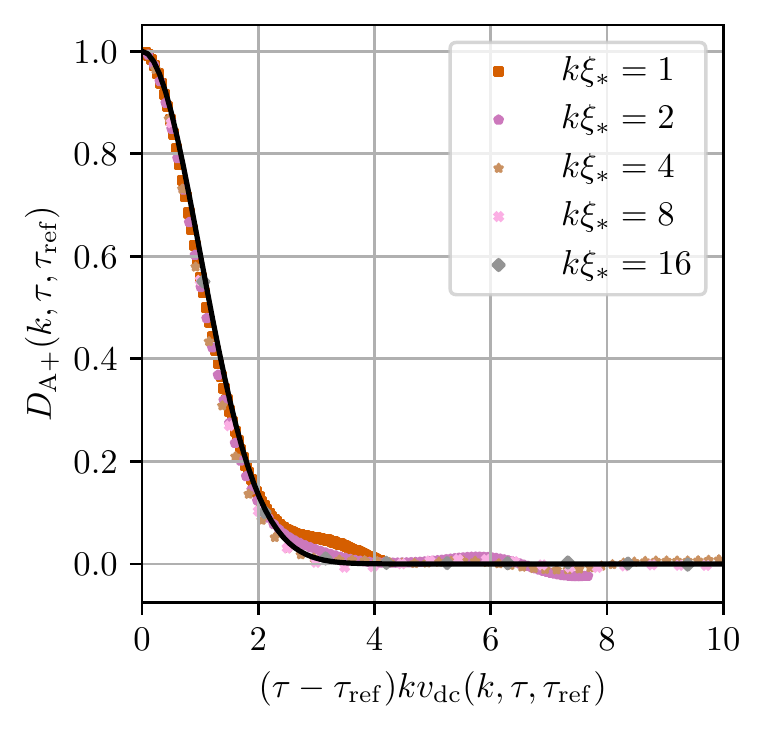}}
	\caption{Real part of the unequal time correlator measured in simulation (B) on the left panel and simulation (C) on the right panel. The $y$-axis displays $\tgauss(k, \tau, \tuetc)$. The solid dark line is the prediction of our model combining \cref{eq:expvsweep,eq:Vlarge,eq:vsweepcomplete}.}
	\label{fig:uetc-bc}
\end{figure}

\begin{figure}
	\centering
	\subfloat[Simulation (D).]{\includegraphics[width=0.49\textwidth]{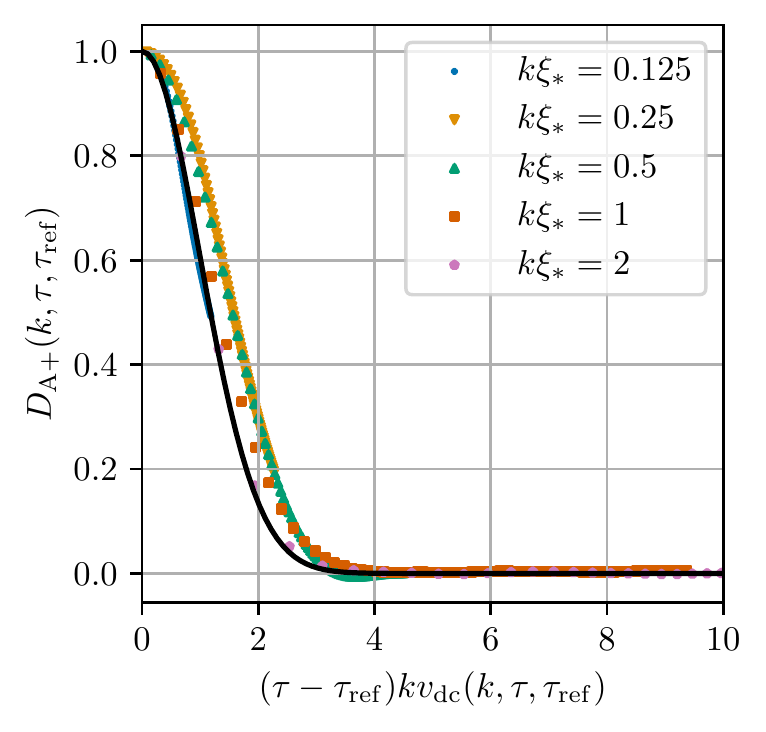}}
	\subfloat[Simulation (E).]{\includegraphics[width=0.49\textwidth]{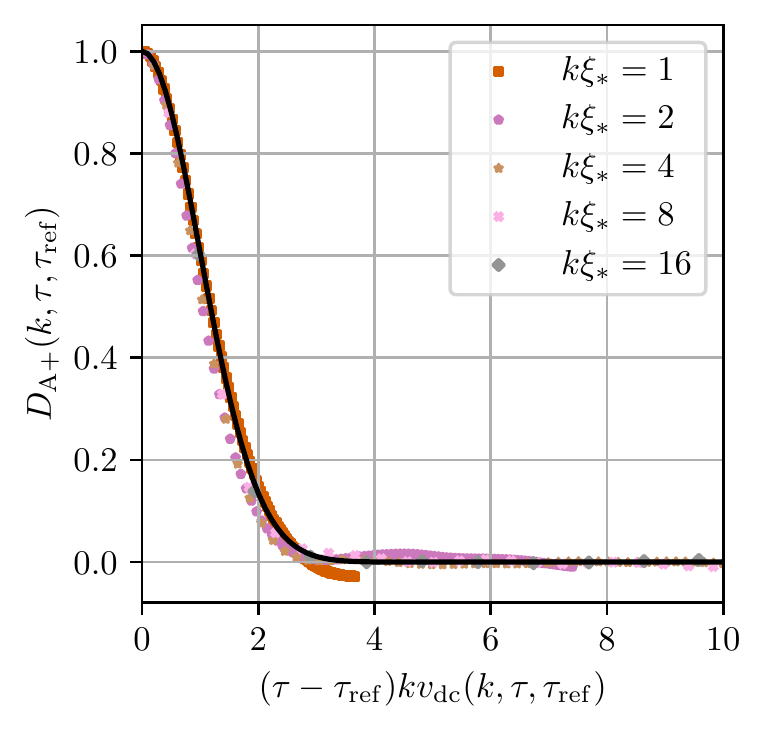}}
	\caption{Real part of the unequal time correlator measured in simulation (D) on the left panel and simulation (E) on the right panel. The $y$-axis displays $\tgauss(k, \tau, \tuetc)$. The solid dark line is the prediction of our model combining \cref{eq:expvsweep,eq:Vlarge,eq:vsweepcomplete}).}
	\label{fig:uetc-de}
\end{figure}

\begin{figure}
	\centering
	\subfloat[Simulation (F).]{\includegraphics[width=0.49\textwidth]{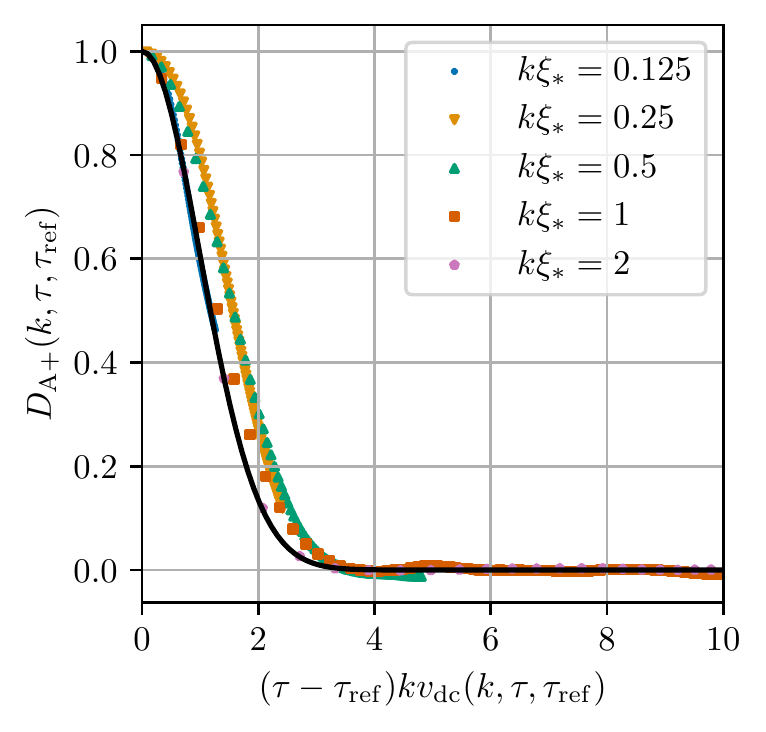}}
	\subfloat[Simulation (G).]{\includegraphics[width=0.49\textwidth]{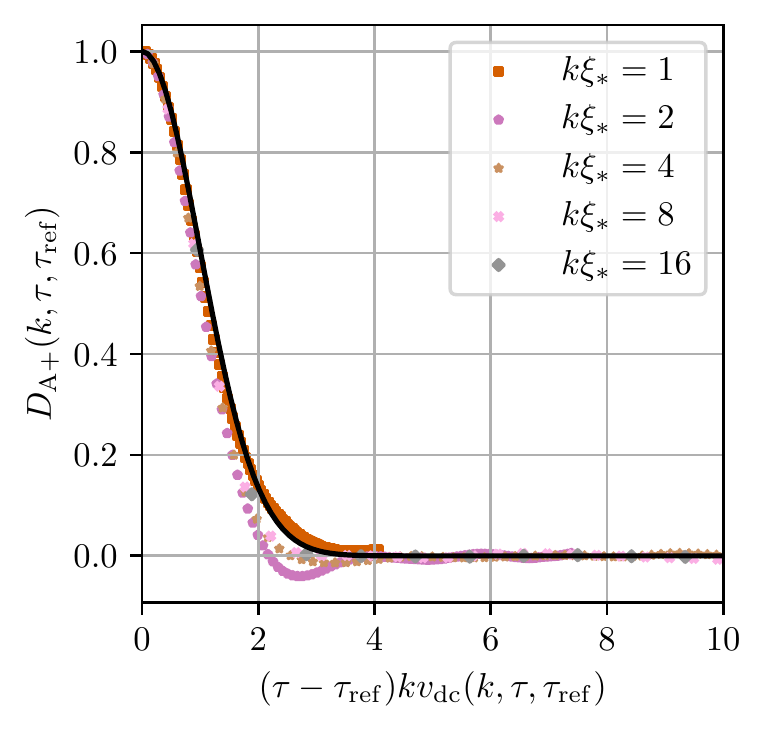}}
	\caption{Real part of the unequal time correlator measured in simulation (F) on the left panel and simulation (G) on the right panel. The $y$-axis displays $\tgauss(k, \tau, \tuetc)$. The solid dark line is the prediction of our model combining \cref{eq:expvsweep,eq:Vlarge,eq:vsweepcomplete}).}
	\label{fig:uetc-fg}
\end{figure}

In this appendix, we give the plots of the UETCs for simulations (B)-(C) in \cref{fig:uetc-bc}, simulations (D)-(E) in \cref{fig:uetc-de} and simulations (F)-(G) in \cref{fig:uetc-fg}. The $y$-axis is $\tgauss(k, \tau, \tuetc)$.
\FloatBarrier

\FloatBarrier

\section{Other GW power spectra from simulations} \label{sec:appendix-sim-gws}
\begin{figure}
	\centering
	\subfloat[Simulation (B), $\Delta\tau/\tauxist = 4.97$.]{\includegraphics[width=0.4\textwidth]{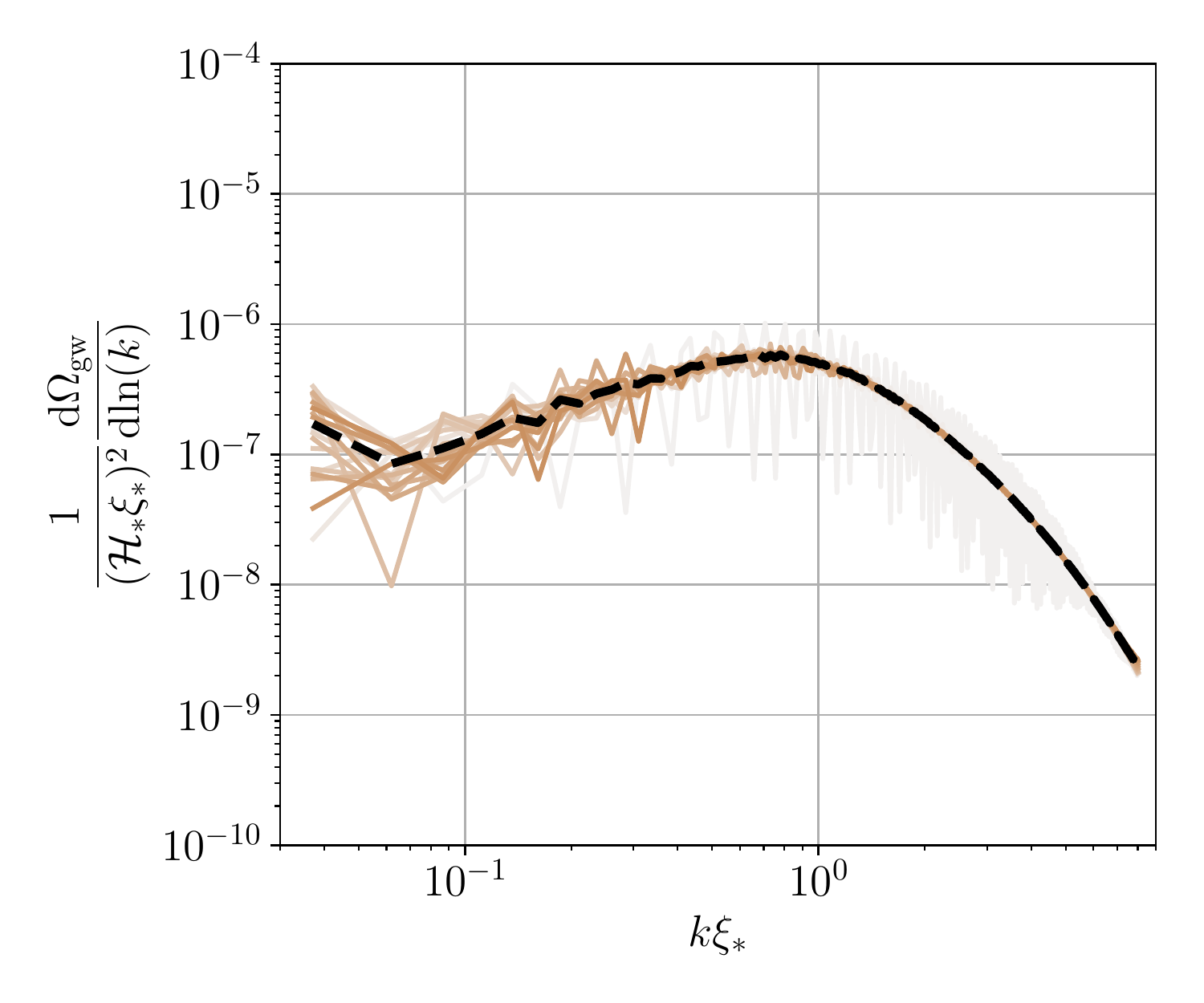}}
	\subfloat[Simulation (C), $\Delta\tau/\tauxist = 0.647$.]{\includegraphics[width=0.4\textwidth]{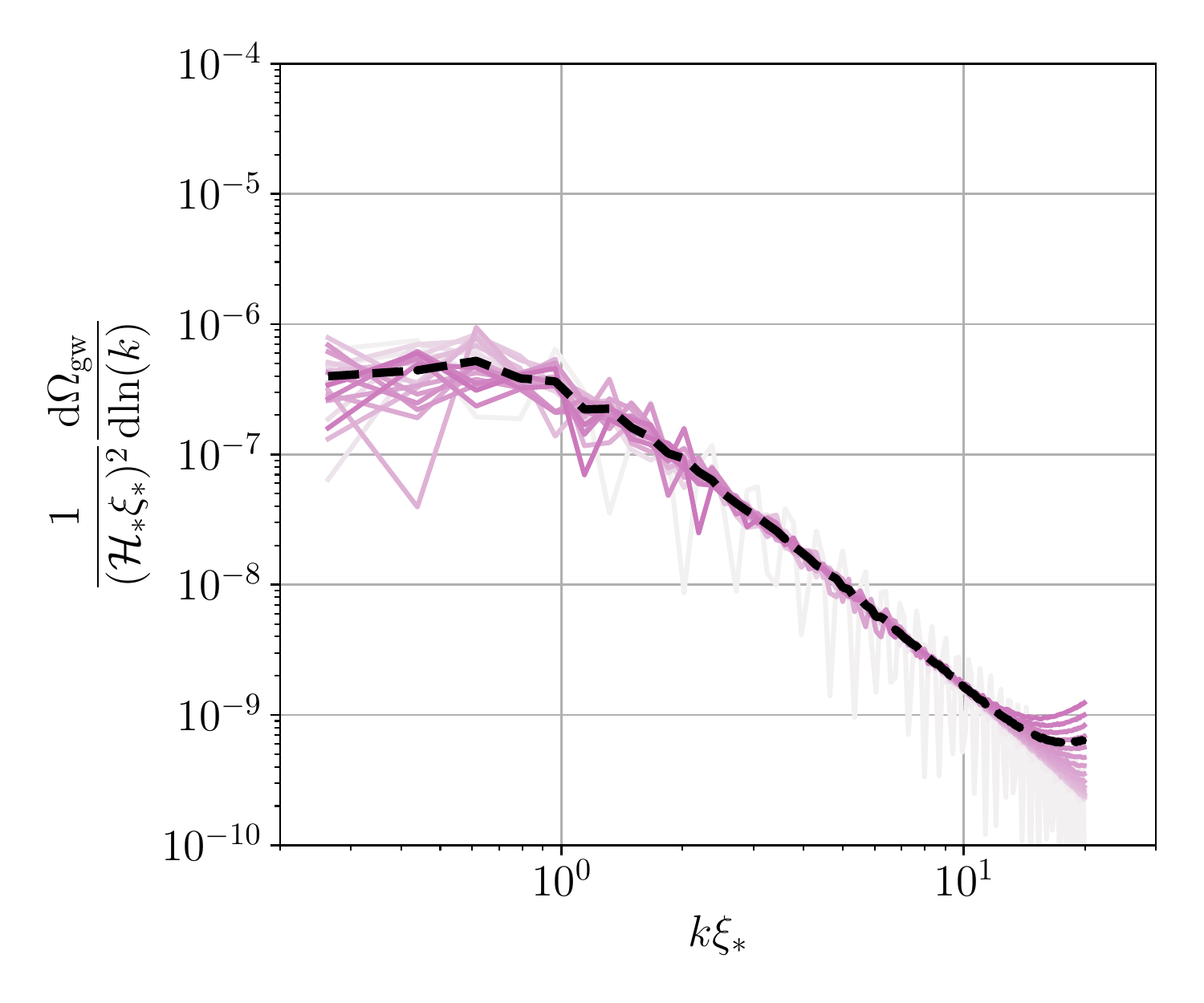}}
	\caption{GW power spectrum from  simulations with $\vrmsst \approx 0.03$. The left and right panels show simulation (B) and (C) from \cref{tab:list} respectively. The coloured lines show the GW power spectrum with interval $\Delta \tau$ as listed in the caption, with darker shades corresponding to later times. The black dashed line shows an average over the GW power spectrum in the last $50\%$ of the simulation. We have cut off the spectrum at high wavenumbers due to numerical precision noise.
	}
	\label{fig:gwps-sim-vrms-0.03}
\end{figure}

\begin{figure}
	\centering
	\subfloat[Simulation (F), $\Delta\tau/\tauxist = 5.16$.]{\includegraphics[width=0.45\textwidth]{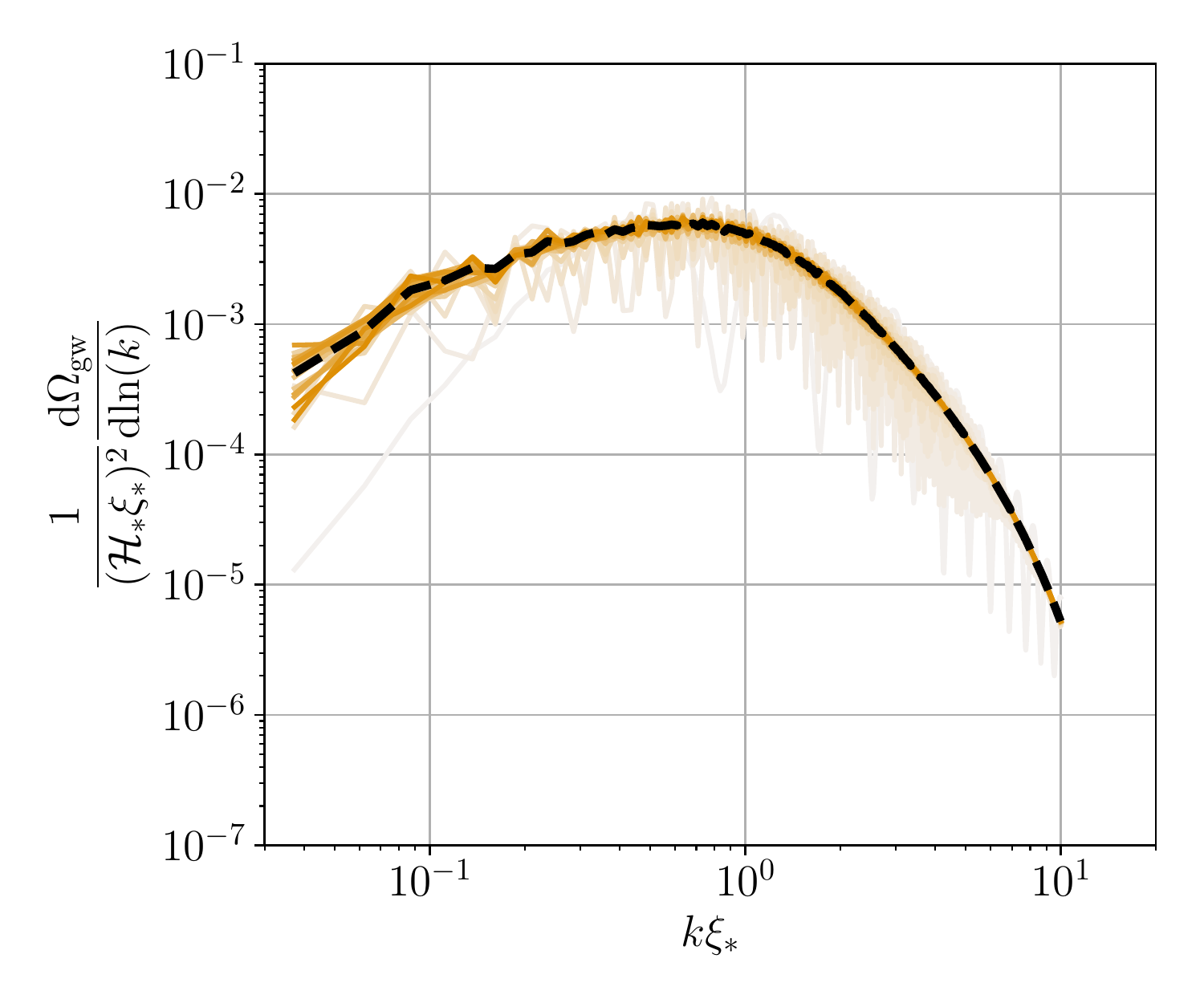}}
	\subfloat[Simulation (G), $\Delta\tau/\tauxist = 0.583$.]{\includegraphics[width=0.45\textwidth]{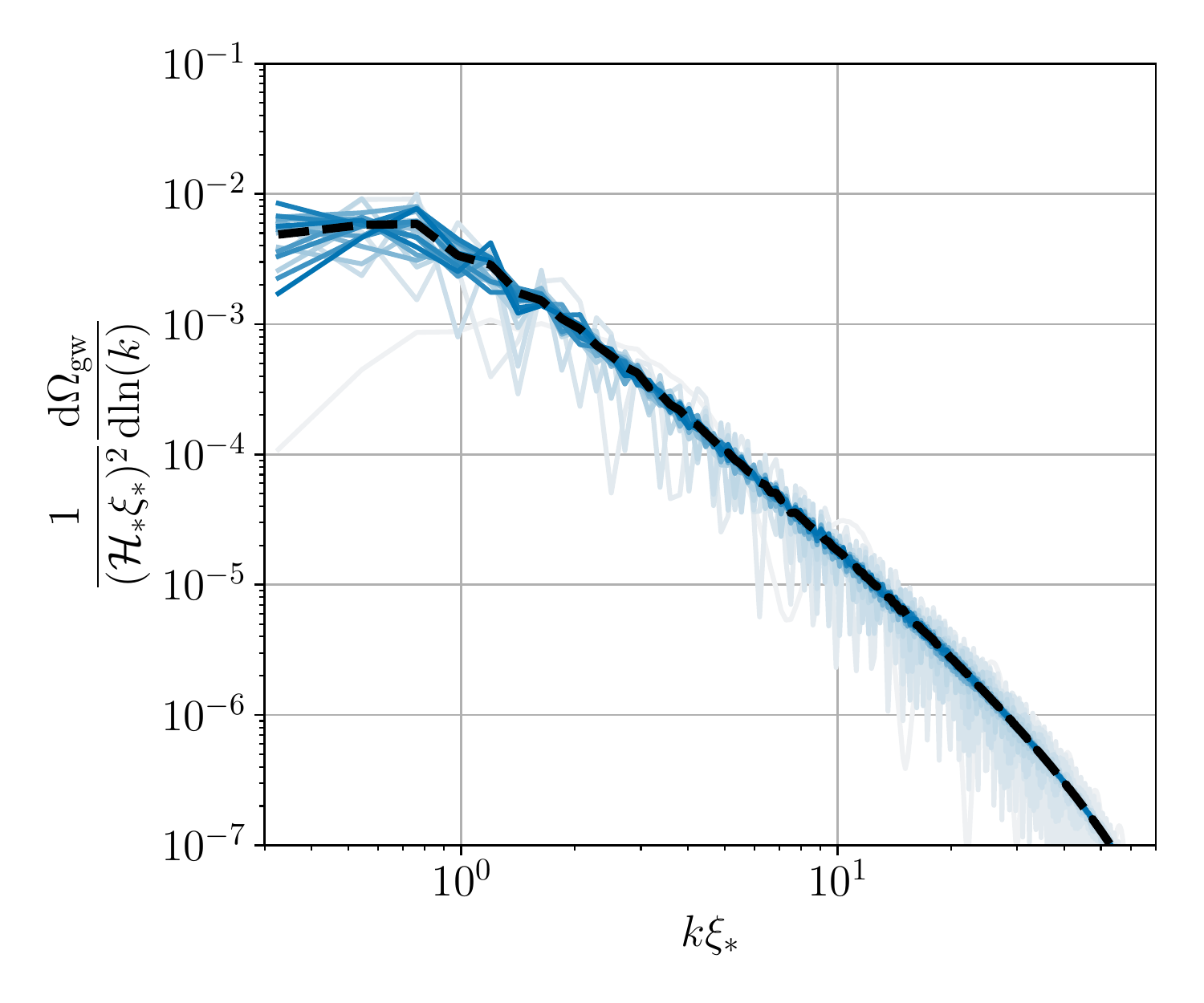}}
	\caption{GW power spectrum from  simulations with $\vrmsst \approx 0.3$. The left and right panels show simulation (F) and (G) from \cref{tab:list} respectively. The coloured lines show the GW power spectrum with interval $\Delta \tau$ as listed in the caption, with darker shades corresponding to later times. The black dashed line shows an average over the GW power spectrum in the last $50\%$ of the simulation. We have cut off the spectrum at high wavenumbers due to numerical precision noise.
	}
	\label{fig:gwps-sim-vrms-0.3}
\end{figure}

\FloatBarrier
In this appendix we present the plots of the GW power spectrum from simulations (B)-(C) with $\vrmsst \approx 0.03$ in \cref{fig:gwps-sim-vrms-0.03}, and (F)-(G) with $\vrmsst \approx 0.3$ in \cref{fig:gwps-sim-vrms-0.3}.

\FloatBarrier

\section{Vortical and longitudinal velocity in simulations}\label{sec:appendix-sim-vrms-comp}
In this work we assume that the velocity field is purely vortical. While the longitudinal component can be set to zero in the analytical modelling or during the numerical integration, in the numerical simulations a longitudinal component can in principle develop even if it is removed in the initial conditions. In the following short appendix, we present the evolution of the kinetic energy, $\vrms^2$, within the simulations, split into its vortical and longitudinal components. Note that in the following we adjust our notation to explicitly indicate both the longitudinal and vortical components of the velocity field.

We can decompose the spectral density into vortical and longitudinal components,
\begin{equation}
	\ev{v_i(\vb{k}, \tau) v_j^*(\vb{q}, \tau)} \equiv (2\pi)^3 \delta\qty(\vb{k} - \vb{q}) \left[\bot_{ij}(\vu k) P_{v_\perp}(k, \tau) + (1-\bot_{ij}(\vu k)) P_{v_\parallel}(k, \tau)\right]\text.
\end{equation}
From these we can construct the vortical and longitudinal kinetic energy analogous to \cref{eq:kinetic-energy},
\begin{equation}
	\vrms_{\perp/\parallel}^2(\tau) \equiv
	\int \frac{\dd k}{k} \mathcal{P}_{v_{\perp/\parallel}}(k, \tau)\text,\label{eq:kinetic-comp}
\end{equation}
where the vortical and longitudinal power spectra are given by
\begin{equation}
	\mathcal{P}_{v_{\perp/\parallel}}(k, \tau) = \frac{k^3}{\pi^2} P_{v_{\perp/\parallel}}(k, \tau)\text,
\end{equation}
and the total kinetic energy is
\begin{equation}
	\vrms^2(\tau) = \vrms^2_\perp(\tau) + \vrms_\parallel^2(\tau)\text.
\end{equation}

\begin{figure}
	\centering
	\includegraphics[width=0.49\textwidth]{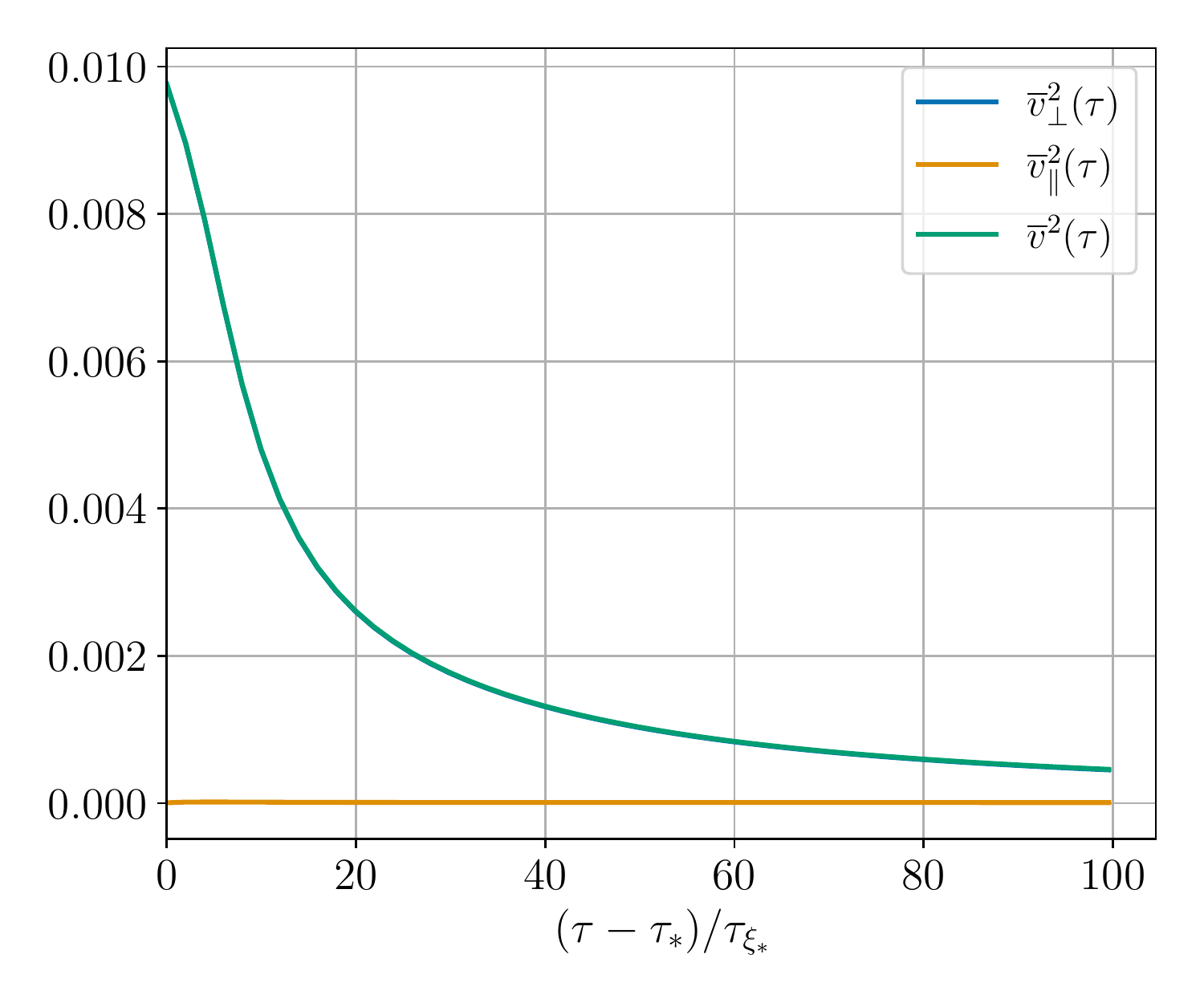}
	\caption{Evolution of the kinetic energy $\vrms^2$, decomposed into the vortical, $\vrms_\perp^2$, and longitudinal,  $\vrms_\parallel^2$, components (see \cref{eq:kinetic-comp}). The central panel shows simulation (A).}
	\label{fig:vrms-comp-a}
\end{figure}

\begin{figure}
	\centering
	\subfloat[Simulation (B).]{\includegraphics[width=0.49\textwidth]{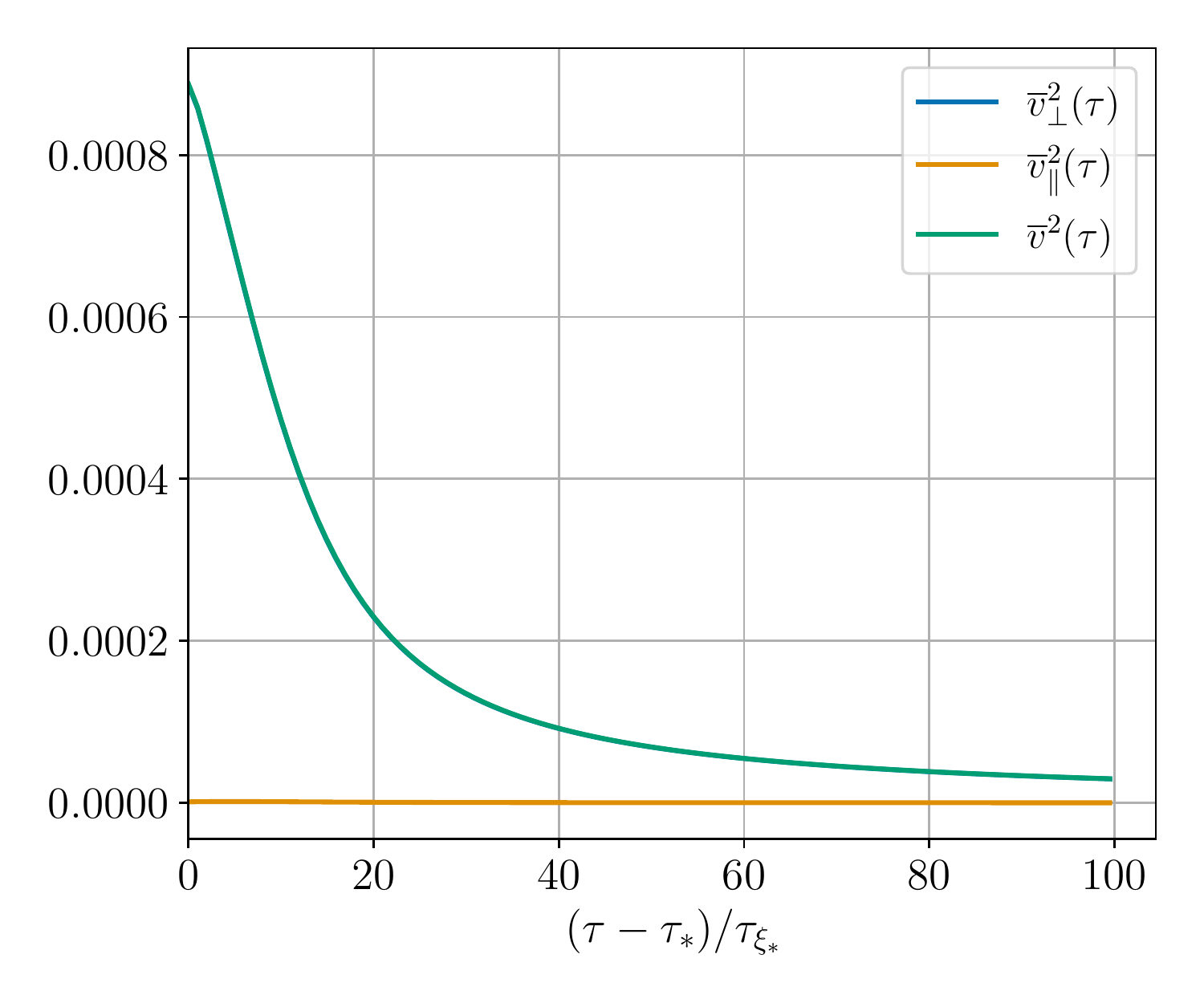}}
	\subfloat[Simulation (C).]{\includegraphics[width=0.49\textwidth]{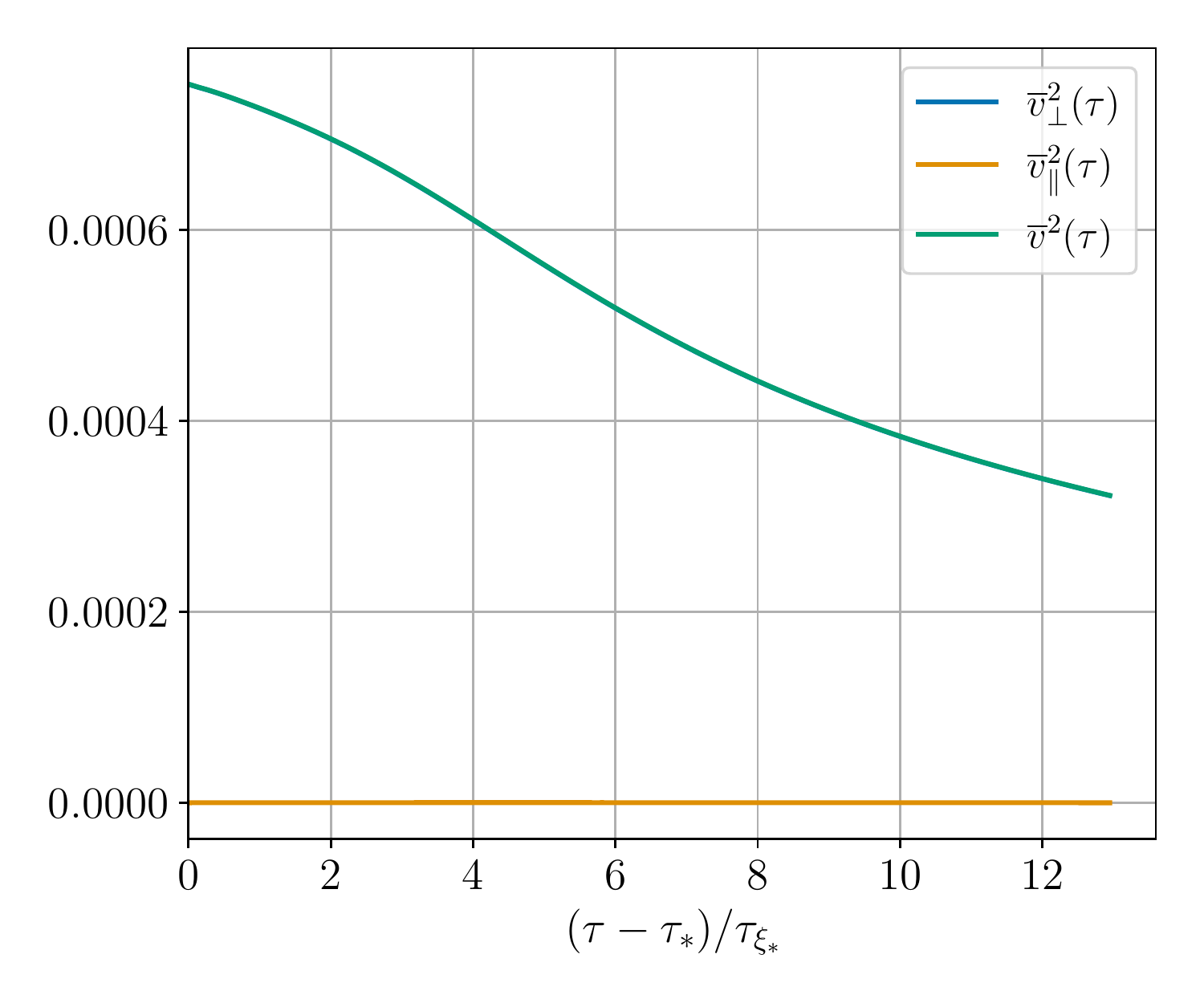}}
	\caption{Evolution of the kinetic energy $\vrms^2$, decomposed into the vortical, $\vrms_\perp^2$, and longitudinal,  $\vrms_\parallel^2$, components (see \cref{eq:kinetic-comp}). The left and right panels show simulation (B) and (C) respectively.}
	\label{fig:vrms-comp-bc}
\end{figure}

\begin{figure}
	\centering
	\subfloat[Simulation (D).]{\includegraphics[width=0.49\textwidth]{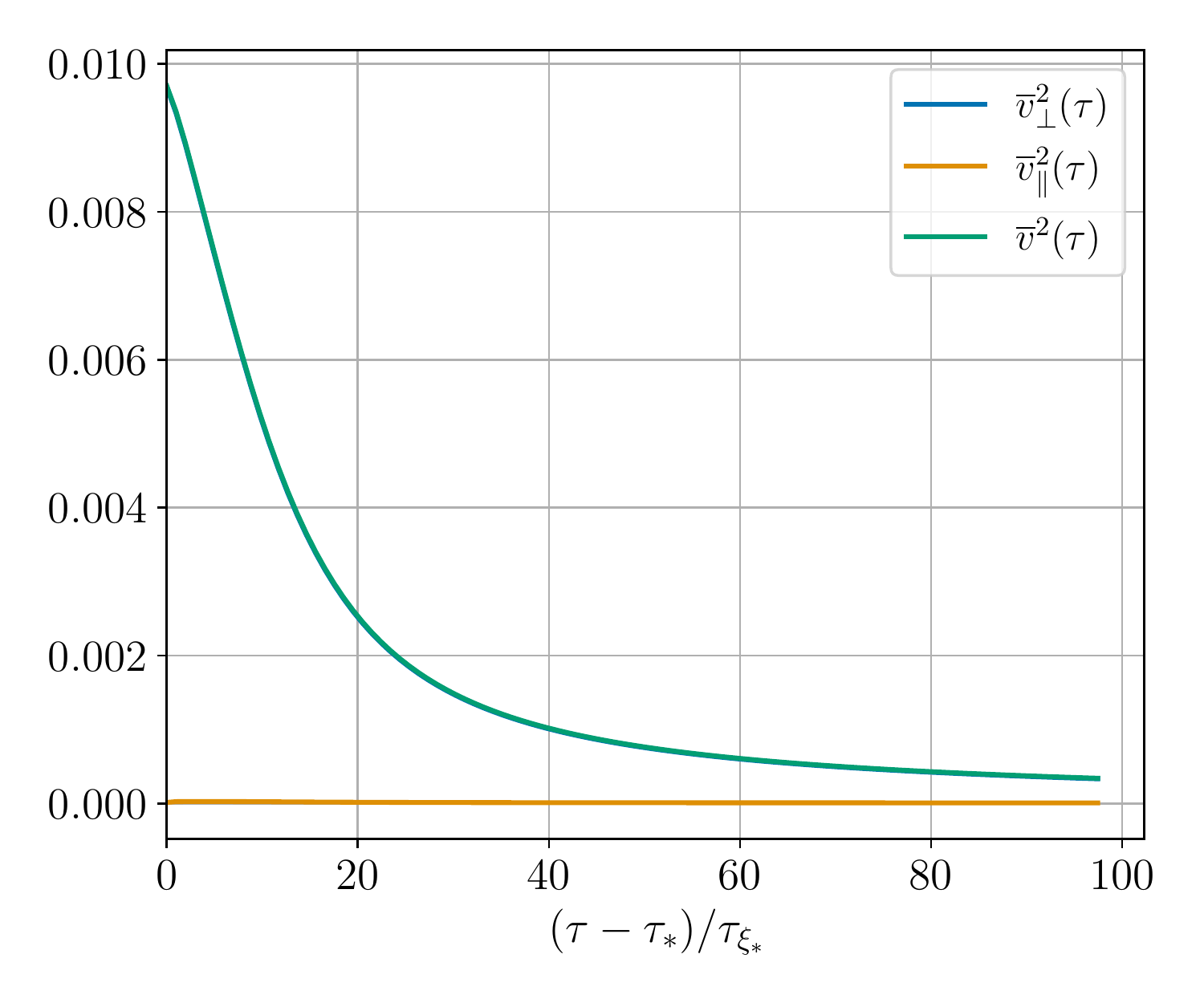}}
	\subfloat[Simulation (E).]{\includegraphics[width=0.49\textwidth]{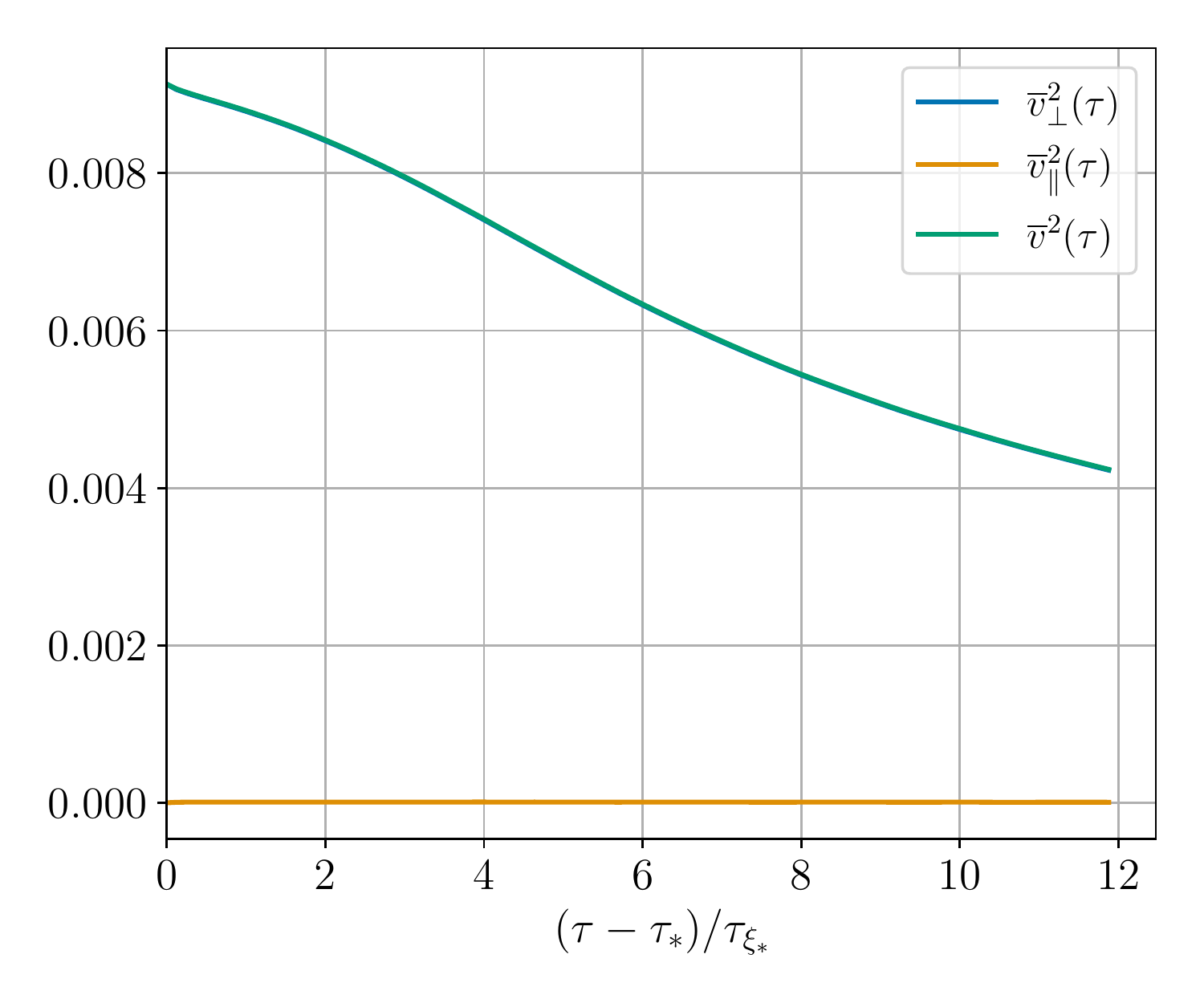}}
	\caption{Evolution of the kinetic energy $\vrms^2$, decomposed into the vortical, $\vrms_\perp^2$, and longitudinal,  $\vrms_\parallel^2$, components (see \cref{eq:kinetic-comp}). The left and right panels show simulation (D) and (E) respectively.}
	\label{fig:vrms-comp-de}
\end{figure}

\begin{figure}
	\centering
	\subfloat[Simulation (F).]{\includegraphics[width=0.49\textwidth]{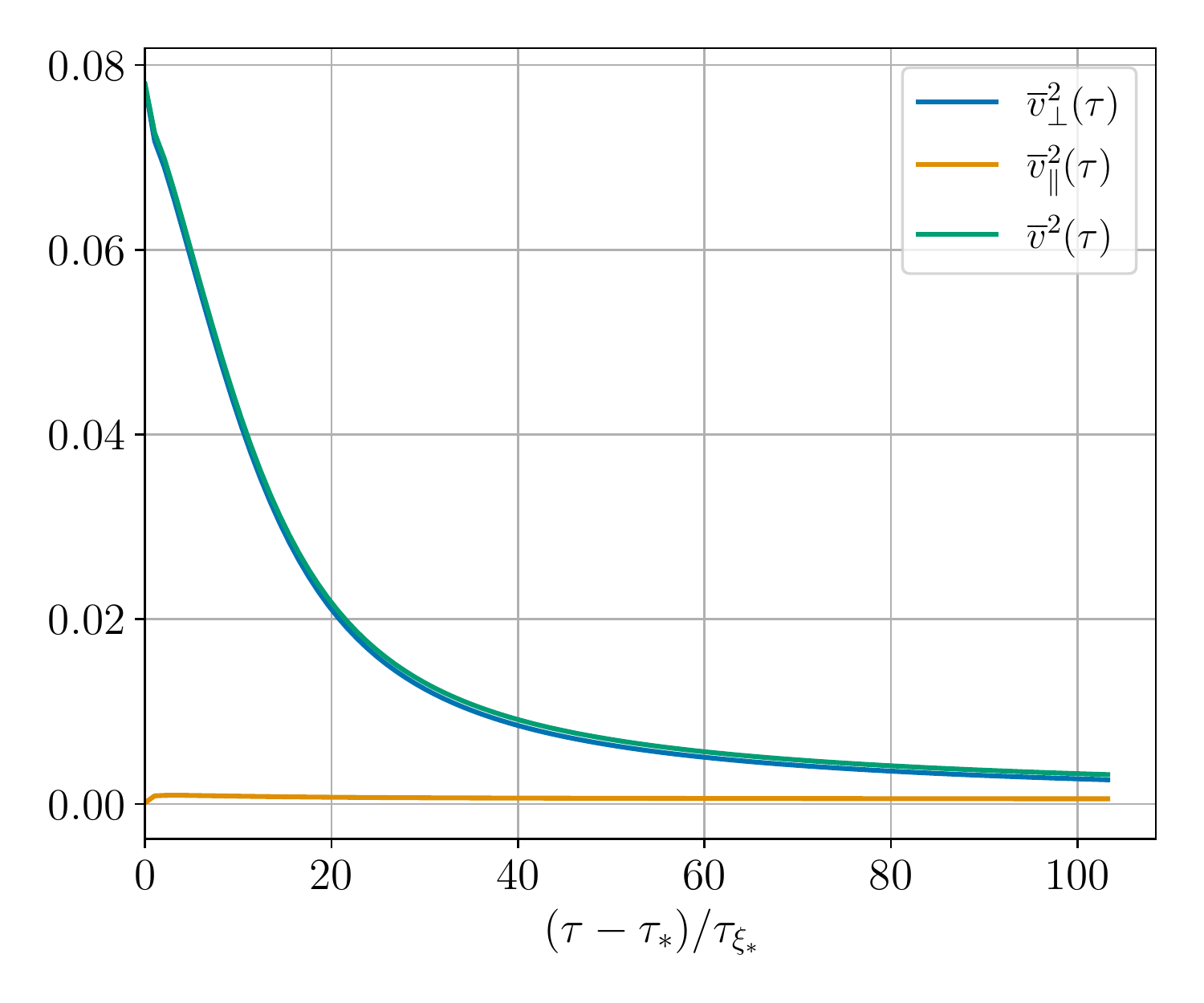}}
	\subfloat[Simulation (G).]{\includegraphics[width=0.49\textwidth]{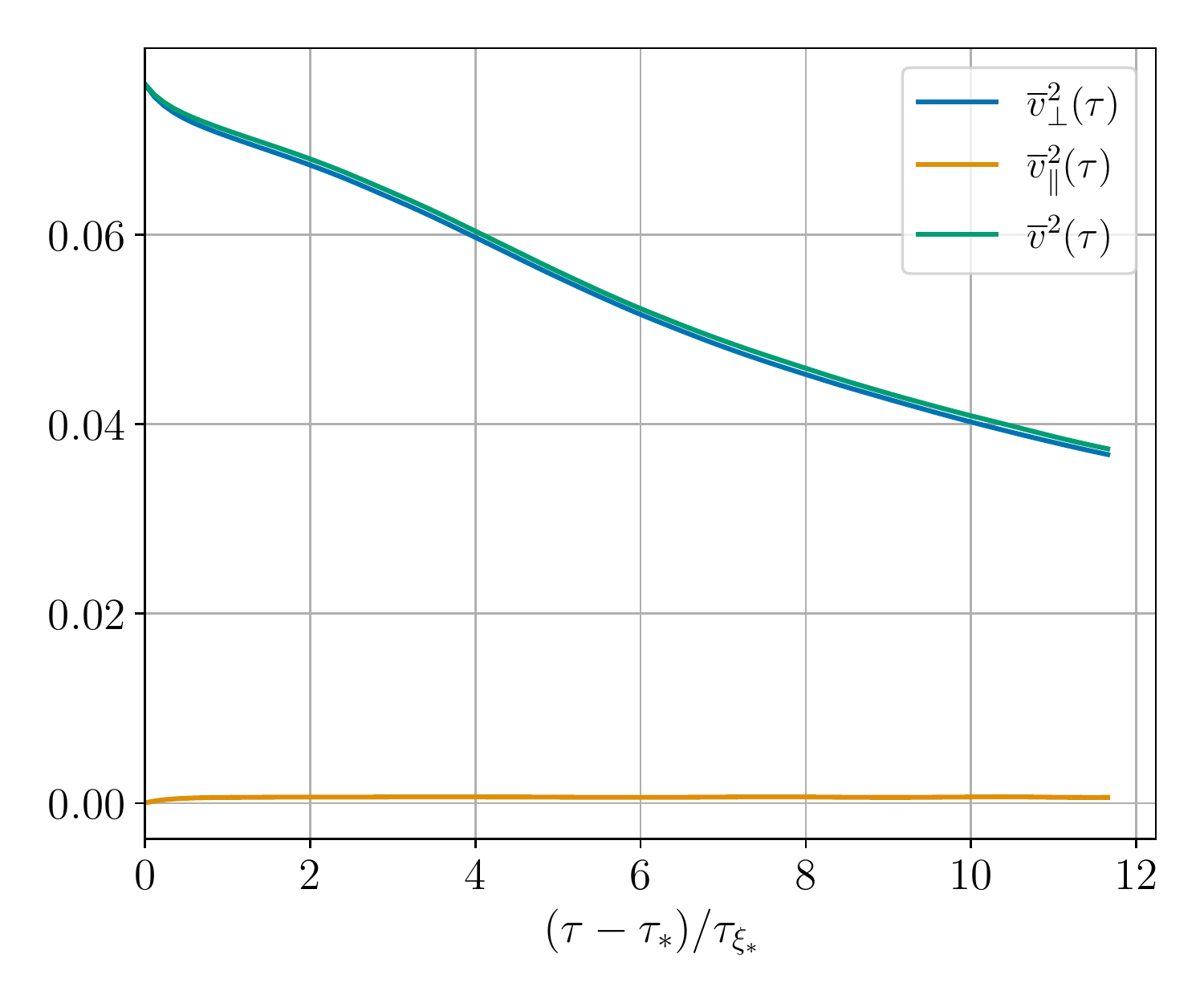}}
	\caption{Evolution of the kinetic energy $\vrms^2$, decomposed into the vortical, $\vrms_\perp^2$, and longitudinal,  $\vrms_\parallel^2$, components (see \cref{eq:kinetic-comp}). The left and right panels show simulation (F) and (G) respectively.}
	\label{fig:vrms-comp-fg}
\end{figure}

In the following \cref{fig:vrms-comp-a,fig:vrms-comp-bc,fig:vrms-comp-de,fig:vrms-comp-fg}, we show the evolution of $\vrms^2$, $\vrms^2_\perp$ and $\vrms_\parallel^2$ in simulations (A)-(G). The longitudinal component of the kinetic energy stays small throughout each simulation, with the total kinetic energy being dominated by the vortical component. The simulations (F) and (G) with $\vrmsst\simeq 0.3$ have the largest increase of $\vrms_\parallel^2$ at the start of the simulation, and interestingly this remains relatively constant throughout the simulation. By the end of simulation (F) the ratio of the longitudinal to total kinetic energy is $\vrms_\parallel^2/\vrms^2=0.176$, but by this time the total kinetic energy has decayed by more than a factor 20.

\FloatBarrier

\bibliographystyle{JHEP}
\bibliography{biblio}

\end{document}